\documentclass[aps,twocolumn,pra,superscriptaddress,showpacs]{revtex4}

\usepackage{mathrsfs}
\usepackage{amssymb}
\usepackage{amsmath}
\usepackage{graphicx}
\usepackage{epsfig}
\usepackage{txfonts}
\usepackage{subfigure}
\usepackage{amsfonts}
\usepackage{color}
\usepackage[colorlinks,citecolor=blue]{hyperref}
\usepackage{bm}
\usepackage{xcolor}

\begin{document}
\title{Quantum-enhanced distributed network sensing using multiple quantum resources}

\author{Rui Zhang}\thanks{Co-first authors with equal contribution}
\affiliation{Key Laboratory of Low-Dimensional Quantum Structures and Quantum Control of Ministry of Education, Hunan Research Center of the Basic Discipline for Quantum Effects and Quantum Technologies, XJ-Laboratory and Department of Physics, Hunan Normal University, Changsha 410081, China}

\author{Zi-Yu Zhou}\thanks{Co-first authors with equal contribution}
\affiliation{Key Laboratory of Low-Dimensional Quantum Structures and Quantum Control of Ministry of Education, Hunan Research Center of the Basic Discipline for Quantum Effects and Quantum Technologies, XJ-Laboratory and Department of Physics, Hunan Normal University, Changsha 410081, China}

\author{Wen-Quan Yang}
\affiliation{Key Laboratory of Low-Dimensional Quantum Structures and Quantum Control of Ministry of Education, Hunan Research Center of the Basic Discipline for Quantum Effects and Quantum Technologies, XJ-Laboratory and Department of Physics, Hunan Normal University, Changsha 410081, China}

\author{Ya-Feng Jiao}
\affiliation{Academy for Quantum Science and Technology, Zhengzhou University of Light Industry, Zhengzhou 450002, China}

\author{Xun-Wei Xu}
\email{xwxu@hunnu.edu.cn}
\affiliation{Key Laboratory of Low-Dimensional Quantum Structures and Quantum Control of Ministry of Education, Hunan Research Center of the Basic Discipline for Quantum Effects and Quantum Technologies, XJ-Laboratory and Department of Physics, Hunan Normal University, Changsha 410081, China}

\author{Le-Man Kuang}
\email{lmkuang@hunnu.edu.cn}
\affiliation{Key Laboratory of Low-Dimensional Quantum Structures and Quantum Control of Ministry of Education, Hunan Research Center of the Basic Discipline for Quantum Effects and Quantum Technologies, XJ-Laboratory and Department of Physics, Hunan Normal University, Changsha 410081, China}
\affiliation{Academy for Quantum Science and Technology, Zhengzhou University of Light Industry, Zhengzhou 450002, China}

\date{\today}

\begin{abstract}
We propose a theoretical scheme for quantum-enhanced distributed network sensing, targeting multiphase estimation by leveraging multiple quantum resources. 
Specifically, we investigate the performance advantage in a distributed quantum network (DQN) for multiphase sensing by integrating three types of quantum resources (TQRs): quantum catalysis, entanglement, and squeezing. 
Our results reveal that employing all three TQRs leads to better sensing performance than using only two TQRs under both lossless and lossy conditions, with precision approaching the Heisenberg limit. 
We further demonstrate that partial quantum catalysis provides a stronger precision advantage than global catalysis in both ideal and noisy regimes. 
We identify a practical homodyne  measurement scheme for globally and partially catalyzed multimode W-type coherent states, whose measurement sensitivity can approach the corresponding quantum Cram\'{e}r-Rao bound.  In this practical setting, partial catalysis also yields better measurement sensitivity than global catalysis.
Moreover, under photon loss, both global and partial catalysis of multimode W-type coherent states exhibit a loss–catalysis dual-enhanced sensitivity region.
These findings highlight the quantum-enhanced advantages conferred by hybrid quantum resources for practical DQN sensing applications. Our work opens a way for realizing quantum-enhanced DQN sensing.
\end{abstract}

\maketitle

\section{Introduction}
Quantum metrology seeks to harness non-classical resources, such as  quantum entanglement and squeezing, to achieve high-precision estimation of physical quantities and parameters beyond classical limit ~\cite{Giovannetti2006quantum}, which has been widely applied in various fields such as biological detection~\cite{taylor2013biological}, navigation~\cite{lai2019observation}, and gravitational wave detection~\cite{schnabel2010quantum,abbott2016observation}.  In general, the precision of parameter estimation depends on the properties of the probe quantum state, the interaction between the system and the parameter of interest, and the measurement strategy employed. 
The main tool for quantum parameter estimation is the quantum Fisher information (QFI) ~\cite{fisher1925theory,holevo2011probabilistic,helstrom1969quantum,braunstein1996generalized} to measures the sensitivity of a quantum state concerning changes in the parameters encoded in it, where its inverse is known as the quantum Cramér-Rao bound (QCRB), scaling as $1/\sqrt{\mathcal{F}_{Q}}$, which sets the ultimate theoretical limit on precision for estimating parameters of a quantum system~\cite{cramer1999mathematical,helstrom1967minimum}.
Consequently, the central objective of quantum parameter estimation is to minimize the QCRB, thereby achieving ultrasensitive measurement performance. The precision of the parameter estimation, using only classical resources, is limited by the standard quantum limit (SQL) that scales as $1/\sqrt{N}$, but  with quantum resources can beat SQL and can even achieve the Heisenberg limit (HL), which scales as $1/N$, where $N$ is the number of resources. Single-parameter estimation has been extensively studied both theoretically and experimentally~\cite{gerry2001generation,nagata2007beating,dowling2008quantum,joo2011quantum,xiang2011entanglement,sanders2012review,tan2014enhanced,lee2015quantum,chen2024asymmetry,zhang2025enhancing,dorner2009optimal,joo2012quantum,krischek2011useful}. However, many practical physical systems involve the simultaneous estimation of multiple correlated parameters, rendering the single-parameter framework insufficient. This limitation has naturally driven the extension of quantum metrology into the domain of multiparameter estimation.

Multiparameter quantum metrology is essential for a wide range of practical applications involving the simultaneous estimation of multiple correlated parameters. A key emerging direction in this field is distributed quantum network (DQN) sensing, which aims to exploit quantum resources shared among 
spatially separated sensors to enhance the estimation sensitivity of global properties, such as weighted linear functionals of local parameters~\cite{qian2019heisenberg,rubio2020quantum,bringewatt2021protocols,ehrenberg2023minimum,Pezze2017Optimal,Ge2018Distributed,Proctor2018Multiparameter,Gatto2019Distributed,Marco2021Distributed,Maleki2022Distributed,wang2025exact}. This paradigm has applications in global-scale clock synchronization~\cite{K2014A}, phase imaging~\cite{Zhang2017Quantum}, and ultrasensitive positioning~\cite{xia2020demonstration,sun2022quantum}.
%Multiparameter quantum metrology is essential for a wide range of practical applications, among which distributed quantum network (DQN) sensing aims to exploit quantum resources to enhance the estimation sensitivity of global functions of multiple distributed parameters.  A key emerging direction in multiparameter quantum metrology is the estimation of global properties, for example weighted linear functionals of local parameters, across a network of sensors prepared in distributed entangled states~\cite{qian2019heisenberg,rubio2020quantum,bringewatt2021protocols,ehrenberg2023minimum}. This paradigm, known as distributed quantum network (DQN)~\cite{Pezze2017Optimal,Ge2018Distributed,Proctor2018Multiparameter,Gatto2019Distributed,Marco2021Distributed,Maleki2022Distributed,wang2025exact}, and has applications to global-scale clock synchronization~\cite{K2014A}, phase imaging~\cite{Zhang2017Quantum}, and ultrasensitive positioning ~\cite{xia2020demonstration,sun2022quantum}. 
The theoretical foundation of multiparameter quantum metrology is built upon the quantum Fisher information matrix  (QFIM) \cite{fujiwara1995quantum,petz1996geometries,petz1996monotone,Jie2014Quantum,J2015Quantum,Gagatsos2016Gaussian,liu2020quantum,albarelli2020perspective}, which sets the fundamental precision limit for jointly estimating multiple parameters. Furthermore, the corresponding quantum Cramér-Rao bound, given by the inverse of the QFIM, is generally not saturable due to incompatibility among measurement observables\cite{matsumoto2002new,ragy2016compatibility}. Nevertheless, multiparameter quantum metrology seeks to exploit entanglement and other quantum resources to enhance the joint precision. It was demonstrated that entanglement is a necessary conditionin for achieving quantum-enhanced precision in distributed metrology.~\cite{Ge2018Distributed,Humphreys2013Quantum,Liu2016Quantum}. 
Experimental progress in distributed quantum sensing has also been made. Hong $et~al.$ demonstrated four-node distributed phase sensing using two-photon N00N states~\cite{Hong2021Quantum}.  Pan's group achieved three-mode sensing with six-photon entangled states, surpassing the SQL~\cite{liu2021distributed}. That same year, Zhao $et~al.$ implemented distributed phase sensing over $10$km fiber with $2$ modes, also exceeding the SQL~\cite{zhao2021field}. More recently, Kim $et~al.$~\cite{kim2024distributed} used two-photon polarized entangled states to perform four-node sensing with precision approaching the HL.

Quantum squeezing is also a fundamental quantum resource for surpassing classical precision limits~\cite{caves1981quantum,wineland1992spin,ma2011quantum,toth2014quantum,pezze2018quantum}. Their application in distributed phase sensing has been extensively explored~\cite{zhuang2018distributed,guo2020distributed,oh2020optimal,triggiani2021heisenberg,Gessner2020Multiparameter,malitesta2023distributed,ge2025heisenberg}. It has been shown that distributed single-mode squeezed vacuum states can enable phase estimation with HL scaling~\cite{Gatto2019Distributed}.  Combining quantum squeezing with entanglement one experimentally demonstrated high-precision DQN sensing across four nodes~\cite{guo2020distributed}. 
Within Gaussian strategies, squeezed states emerge as optimal resources for estimating weighted sums of displacements and phases, aligning naturally with multiparameter tasks~\cite{oh2020optimal}. Crucially, N00N-like entangled states excited by squeezed vacuum further enhance precision, outperforming counterparts with other excitation types under the same number of input resources~\cite{Zhang2017Quantum}. Recently, Peng‘s group~\cite{ma2026high} have experimentally realized a high-sensitivity DQN sensing of the multimodal parameters by mixing two kinds of quantum resources, quantum squeezing and entanglement.
These results collectively highlight the critical role of squeezing and entanglement in advancing multiparameter quantum metrology.

Recently,  quantum catalysis~\cite{lvovsky2002quantum,lipka2024catalysis,datta2023catalysis} is found  as a new kind of quantum resources. It has demonstrated that quantum catalysis can enhance  improve the   performance of quantum sensing~\cite{Li2016Multiphoton,Hu2017Continuous,jia2018comparison,Zhang2021Improved,zhao2024phase,zhang2023quantum}. Moreover, quantum catalytic states have been successfully prepared in experiments~\cite{lvovsky2002quantum,Bartley2012Multiphoton}. However, the role of quantum catalysis in multiparameter estimation remains unexplored.  It is  an interesting question how to synergistically utilize quantum entanglement, squeezing, and catalysis to enhance multiparameter quantum sensing. In particular, pushing distributed quantum sensing of multiparameters from the single to multiple quantum resources is crucial for a complete and deep understanding of the nature of the quantum resources at the root of the quantum-enhanced advantagy in multiparameter quantum metrology.

In this paper, we propose a theoretical scheme of a quantum-enhanced distributed network sensing for multiple-phase estimation by mixing multiple quantum resources.  We introduce quantum catalysis as a quantum resource into the framework of DQN sensing and investigate the potential performance enhancement enabled by  quantum  entanglement, squeezing, and catalysis within the theory of multiparameter estimation. We show that quantum probes equipped with multiple resources achieve significantly enhanced sensitivity in distributed phase estimation under both lossless and lossy conditions, with precision approaching the HL. Furthermore, we demonstrate that partial catalysis consistently outperforms global catalysis in both theoretical precision bounds and practical measurement sensitivity, providing additional enhancement in both lossless and lossy regimes.

The organization of this paper is as follows: In Sec.~\ref{section 1}, we propose a DQN sensing scheme based on multiphoton quantum catalysis, and present general expressions to calculate the success probability of multiphoton catalysis and QFIM. In Sec.~\ref{section 2}, we investigate the improvement of distributed phase sensing performance in the DQN sensing by mixing two quantum resources—quantum catalysis and entanglement, and  explores the advantages of partial catalysis over global catalysis on the sensitivity.
In Sec.~\ref{section 3} we make an extension of the DQN sensing to mixing three types of quantum resources by introducing squeezing as an additional quantum resource, and further explores the advantages of partial catalysis over global catalysis on the sensitivity.
In Sec.~\ref{section 4} we  propose a practical measurement scheme using the homodyne measurement method  to obtain the actual measurement sensitivity of distributed phase in DQN.
In Sec.~\ref{section 5}, we evaluate the robustness of the proposed scheme under photon loss. The last section provides a summary of the main results and discussion.

%%%%%%%%%%%%------------------------------------------%%%%%%%%%%%%
\begin{figure}[t!]
	\centering
	\includegraphics[width=1.0\columnwidth]{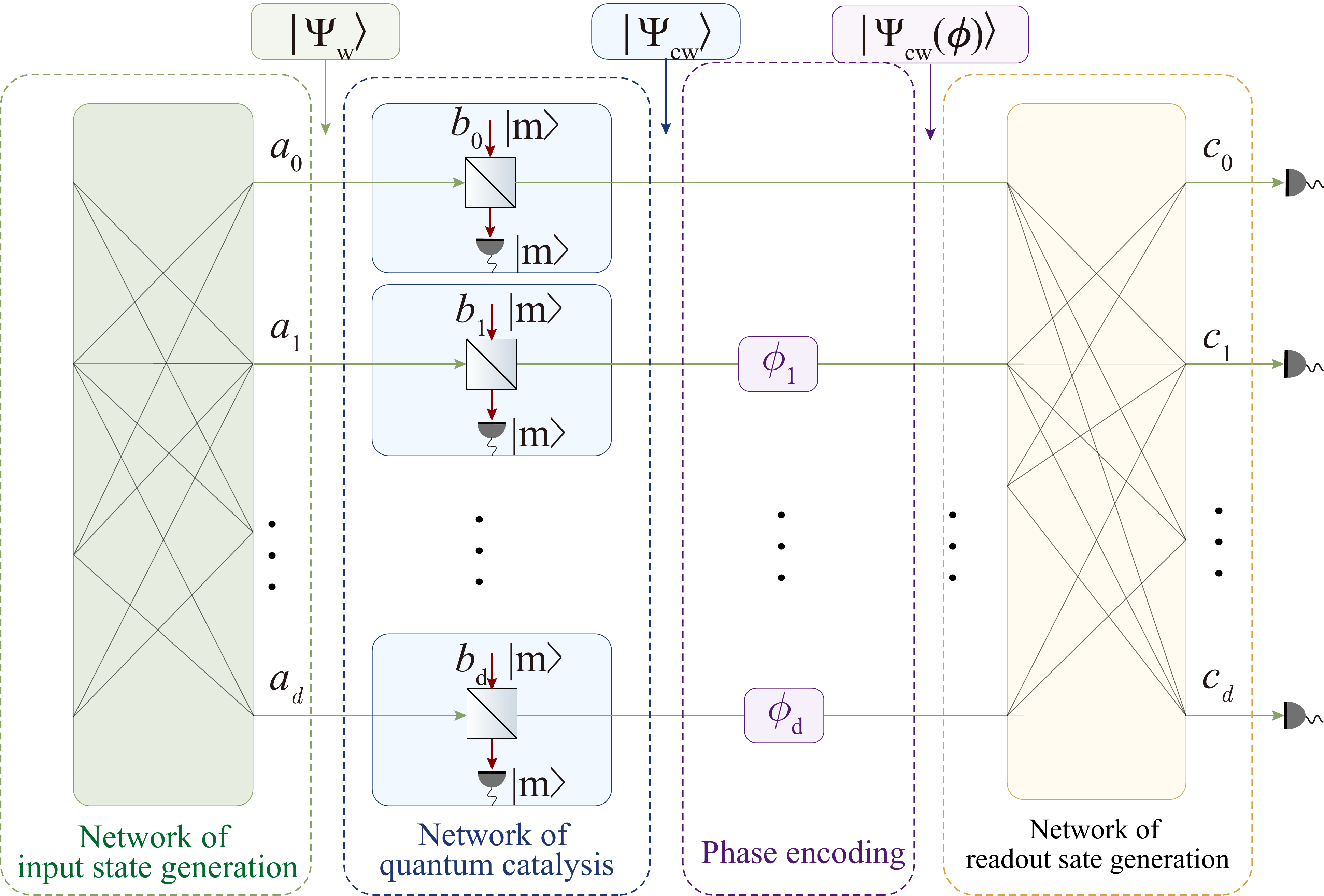}
	\caption{Schematic of DQN sensing of multiple phases with quantum catalysis. A $(d+1)$-mode W-type entangled state $\left|\Psi_{\mathrm{w}}\right\rangle$ is prepared by a $d+1$-port network of input state generation consisting of beam splitters, Kerr nonlinear media, and photon number resolving detectors.
	The probe modes are then sent into a network of quantum catalysis with $d+1$ local quantum catalysis modules, where the $a_j$ modes are locally catalyzed with the aid of the auxiliary catalysis modes $b_j$ prepared in the Fock state $|m\rangle$, producing the catalyzed W-type entangled state $\left|\Psi_{\mathrm{cw}}\right\rangle$. 
	The mode $a_0$ serves as a reference, while the remaining $d$ modes encode the local phase shifts $\phi_i$ $(i=1,\ldots,d)$. 
	After recombination by a $d+1$-port network of readout state generation composed of beam splitters and detectors, the phase information is obtained via local homodyne measurement on the output mode $c_j$.\label{fig:1}}
\end{figure}
%%%%%%%%%%%%------------------------------------------%%%%%%%%%%%%

\section{DQN sensing scheme for multiple phases}\label{section 1}

In this section, we propose a DQN sensing scheme for multiphase estimation and derive the corresponding success probability of multiphoton catalysis and the QFIM. As illustrated in Fig.~\ref{fig:1}, the scheme consists of four functional modules. First, a $d+1$-mode W-type entangled state $\left|\Psi_{\mathrm{w}}\right\rangle$ is prepared as the initial shared probe by a $d+1$-port network of  input state generation consisting of beam splitters (BSs), Kerr nonlinear media, and photon number resolving detectors (see details in Appendix \ref{C_HEP}) . 
Second, the W-type probe state $\left|\Psi_{\mathrm{w}}\right\rangle$ is injected into a catalytic network of $d+1$ local quantum-catalysis modules, each consisting of an auxiliary catalysis mode $b_i$ prepared in the Fock state $|m\rangle$, a BS, and a conditional photon number detector. 
In the $i$th module, the probe mode $a_i$ $(i=0,1,\ldots,d)$ is mixed with $b_i$ via the BS, and local $m$-photon quantum catalysis is realized by conditionally detecting $m$ photons in the auxiliary output mode, thereby producing the $d+1$-mode catalyzed W-type entangled state $\left|\Psi_{\mathrm{cw}}\right\rangle$.
Third, the $a_0$ mode is used as a reference, while the remaining $d$ modes encode local phase shifts $\phi_i$ $(i=1,\ldots,d)$, leading to the phase encoded state $\left|\Psi_{\mathrm{cw}}(\boldsymbol{\phi})\right\rangle$. Finally, the $d+1$ modes are recombined by a $d+1$-port network of readout state genenration composed of BSs and detectors, and the phase information is obtained via local homodyne measurement. The goal of this DQN scheme is to estimate a linear combination of the distributed phases.

We consider the case that the input state of $(d+1)$ catalyzed modes is the generalized W state  ~\cite{Zhang2017Quantum,Maleki2022Distributed,Hong2021Quantum}
\begin{equation}\label{eq:1}
	\begin{aligned}
		\left|\Psi_{\mathrm{w}}\right\rangle=\mathcal{N}\sum_{m=0}^{d}\left|0\right\rangle_{0}\left|0\right\rangle_{1}\left|0\right\rangle_{2}...\left|\psi\right\rangle_{m}...\left|0\right\rangle_{d},
	\end{aligned}
\end{equation}
where $\mathcal{N}=\left[\left(d+1\right)\left(1+d\left|\left\langle 0|\psi\right\rangle\right|^{2}\right)\right]^{-\frac{1}{2}}$ is the normalization coefficient of the multimode quantum state. The $m$-th mode is prepared in a particular state given by $\left|\psi\right\rangle_{m}$,  and the other modes are vacuum state. 

The $m$-photon catalytic procedure includes the catalytic  network action represented by $\otimes_{j=0}^{d}\hat{B}_{j}(\theta)$ and  local conditional measurements expressed by $\hat{M}=\otimes_{j=0}^{d}\left|m\right\rangle_{b_{j}}\left\langle m\right|_{b_{j}}$.  This process generates a multi-photon catalyzed entanglement state with $d+1$ modes, which serves as the quantum probe state of the DQN sensing (see details in the Appendix \ref{A_HEP})
\begin{equation}
	\begin{aligned}
		\left|\Psi_{\mathrm{cw}}\right\rangle
		&=\otimes_{j=0}^{d}\hat{C}_{j}(\theta)\left|\Psi_{\mathrm{w}}\right\rangle,\\
	\end{aligned}
\end{equation}
where $\hat{C}_{j}(\theta)=\left\langle m\right|_{b_{j}}\hat{B}_{j}\left(\theta\right)\left|m\right\rangle _{b_{j}}=:\mathcal{L}_{m}\left(\hat{a}_{j}^{\dagger}\hat{a}_{j}\tan^{2}\theta\right):\left(\cos\theta\right)^{\hat{a}_{j}^{\dagger}\hat{a}_{j}+m}$ is the $j$-th mode multiphoton catalytic operator with 
Laguerre's Polynomials $\mathcal{L}_{m}\left(.\right)$.  $\theta $ is the parameter of the BS used to generate the catalytic state. $j$-th mode generation (annihilation) operator is  $\hat{a}_{j}^{\dagger}$ $\left(\hat{a}_{j}\right)$, the number of network nodes  $d$ and the catalytic photon number $m$. According to the catalysis model, the prerequisite for generating an $m$-photon catalyzed quantum state at the $a_{j}$ output port  is the detection of $m$ photons at the reference output port $b_{j}$. For a ($d+1$)-mode entangled state, we implement $d+1$ independent $m$-photon catalysis processes, ensuring each mode undergoes an individual catalysis operation conditioned on detecting $m$ photons in the corresponding reference mode. Therefore, the generation of catalytic states is probabilistic and its success probability can be solved by the following expression (the detailed calculation process is shown in Appendix~\ref{B_HEP})
\begin{equation}\label{eq:2}
	\begin{aligned}
		P_{c}=\left\langle \Psi_{\mathrm{w}}\right|\otimes_{j=0}^{d}\hat{C}_{j}^{\dagger}\left(\theta\right)\hat{C}_{j}\left(\theta\right)\left|\Psi_{\mathrm{w}}\right\rangle,
	\end{aligned}
\end{equation}
where $\left|\Psi_{\mathrm{w}}\right\rangle$ is the quantum probe state input into the quantum network.
Then the quantum probe state undergoes local phase shifts $\phi_{j}$   $\left(j=1,...,d\right)$. To extract relative phase information, we encode phases only in $d$ out of the total $d+1$ modes. The encoding is implemented via the unitary operation
\begin{equation}
	\begin{aligned}
		\hat{U}(\boldsymbol{\phi})&=\exp\left(-i\sum_{j=1}^{d}\phi_{j}\hat{H}_{j}\right)
	\end{aligned}
\end{equation}
where $\boldsymbol{\phi}=\left(\phi_{1},...,\phi_{d}\right)^{T}$ represents $d$ independent phases and $\hat{H}_{j}=\hat{n}_{j}$ denotes the phase shift generator associated with mode $j$, Applying this transformation maps the initial catalyzed state $\left|\Psi_{\mathrm{cw}}\right\rangle$ to the phase-dependent state $\left|\Psi_{\mathrm{cw}}\left(\boldsymbol{\phi}\right)\right\rangle=\hat{U}(\boldsymbol{\phi})\left|\Psi_{\mathrm{cw}}\right\rangle$. %~\cite{Pezz`e2017Optimal,Ge2018Distributed,Proctor2018Multiparameter,Gessner2020Multiparameter,Marco2021Distributed}. 
% In an optical network, the phase shift generation operator can be represented by the boson operator, i.e., $\hat{H}_{j}=\hat{n}_{j}$. 
Each network node contributes an independent phase shift $\phi_{j}$, and thus DQN sensing involves multi-parameter quantum metrology to estimate a family of unknown parameters $\boldsymbol{\phi}$.
%~\cite{Gessner2020Multiparameter,Jie2014Quantum,Marco2021Distributed}. 
The goal of DQN sensing is often to estimate not each $\phi_{j}$ individually, but rather a specific linear combination of multiple distributed parameters as $f(\phi)=\mathbf{w}^{T} \boldsymbol{\phi}=\sum_{j=1}^{d}w_{j}\phi_{j}$, where $\mathbf{w}=\left(w_{1},...,w_{d}\right)^{T}$ is a vector of non-negative weights~\cite{Proctor2018Multiparameter}. Without loss of generality, we assume normalization $\sum_{j=1}^{d}w_{j}=1$, such that $w_{j}$ can be interpreted as a probability weight. In the case of balanced distribution $w_{j}=1/d$, we obtain the average of the local phases
%~\cite{Gatto2019Distributed,Proctor2018Multiparameter,Ge2018Distributed,Maleki2022Distributed,Marco2021Distributed}
\begin{align}\label{eq:5}
	\bar{\phi}=\frac{1}{d}\sum_{j=1}^{d}\phi_{j}.
\end{align}
%where $\phi_{j}$ is the local phase shift of the $j$-th mode.

%We aim to infer the linear combination of multiple phases. 
The performance of the multiparameter estimation is quantified by the covariance matrix $Cov\left(\boldsymbol{\Phi}\right)$, where $\boldsymbol{\Phi}$ denotes a vector of unbiased estimators for the true parameters $\boldsymbol{\phi}$. The ultimate precision limits are governed by the multiparameter quantum Cramér-Rao bound (QCRB), which sets a lower bound on the covariance matrix as
%~\cite{Ge2018Distributed,Proctor2018Multiparameter,Jie2014Quantum}:
\begin{equation}\label{eq:3}
	Cov\left(\boldsymbol{\Phi}\right)\ge\mathcal{F}^{-1},
\end{equation}
where the elements of covariance matrix are defined as $Cov\left(\boldsymbol{\Phi}\right)_{ij}=E\left[\left(\Phi_{i}-\phi_{i}\right)\left(\Phi_{j}-\phi_{j}\right)\right]$ with $E\left[X\right]$ being the expected value of the quantity $X$. Moreover, the QFIM is given by $\mathcal{F}$,  its elements for pure states are given by %~\cite{J2015Quantum,Liu2016Quantum}%
\begin{equation}
	\begin{aligned}
		\mathcal{F}_{ij}=&4\left\langle \partial_{i}\Psi\left(\boldsymbol{\phi}\right)|\partial_{j}\Psi\left(\boldsymbol{\phi}\right)\right\rangle \\
		&-4\left\langle \partial_{i}\Psi\left(\boldsymbol{\phi}\right)|\Psi\left(\boldsymbol{\phi}\right)\right\rangle\left\langle \Psi\left(\boldsymbol{\phi}\right)|\partial_{j}\Psi\left(\boldsymbol{\phi}\right)\right\rangle,
	\end{aligned}
\end{equation}
with $\partial_{i}\Psi\left(\boldsymbol{\phi}\right)=\left(\partial\Psi\left(\boldsymbol\phi\right) /\partial\phi_{i}\right)$. For the unbiased estimation of $f(\Phi)=\sum_{j}w_{j}\Phi_{j}$ as the parameter $f(\phi)$, the uncertainty $\bigtriangleup^{2} f(\phi)\equiv E\left[(f(\Phi)-f(\phi))^{2}\right]=\sum_{i,j}w_{i}Cov\left(\boldsymbol{\Phi}\right)_{ij}w_{j}$  is given by the following expression  
\begin{equation}
	\bigtriangleup^{2} f(\phi)\ge\sum_{ij}w_{i}\mathcal{F}_{ij}^{-1}w_{j},
\end{equation}
however, the QFIM $\mathcal{F}$ is real, symmetric, positive semidefinite 
%~\cite{Ge2018Distributed,Proctor2018Multiparameter,Gagatsos2016Gaussian}
, and is thus generally not invertible. Therefore, solving the inverse matrix of QFIM is a difficult task. To overcome this, we seek an alternative bound that captures the metrological performance of the quantum probe via the Cauchy-Schwarz inequality
%\begin{equation}
%\sum_{kl}w_{k}\mathcal{F}_{kl}^{-1}w_{l}\sum_{ij}w_{i}\mathcal{F}_{ij}w_{j}\ge\left|\mathbf{w}\right|^{4},
%\end{equation} 
and obtain the weak form of QCRB~\cite{wang2025exact,kim2024distributed}
\begin{equation}
	\bigtriangleup^{2} f(\phi)\ge\frac{\left|\mathbf{w}\right|^{4}}{\mathcal{F}_{w}},
\end{equation}
which $\mathcal{F}_{w}=\sum_{ij}w_{i}\mathcal{F}_{ij}w_{j}$ denotes the weighted QFI. In the case of a balanced distribution $f(\phi)=\bar{\phi}$, the bound of the distributed phase estimation reduce to
% If all of the phases $\phi_{j}$ contribute meaningfully to $q$, a notion captured formally by the requirement $\left|\mathbf{w}\right|^{2}\sim\frac{1}{d}$, then we will say that the weights are well-distributed~\cite{Ge2018Distributed}. 
\begin{equation}\label{Eq.12}
	\bigtriangleup^{2} \bar{\phi}\ge\frac{1}{H},
\end{equation}
where $H=\sum_{ij}\mathcal{F}_{ij}$ called the effective QFI, is the sum of all matrix elements in the QFIM, which can be used to describe the performance of quantum sensing to estimate the global parameter \textcolor{red}{$\bar{\phi}$} in DQN sensing. 

Thus the expression for the effective QFI for the multimode W-type quantum state of distributed phase estimation in the DQN can be obtained
\begin{equation}
	\begin{aligned}\label{eq:13}
		H&=4\mathcal{N}^{2}\left(d\left\langle \hat{n}^{2}\right\rangle-\mathcal{N}^{2}d^{2}\left\langle \hat{n}\right\rangle^{2}\right),
	\end{aligned}
\end{equation}
where the mean values of the operator of photon number and its squared  operator in each mode are taken in the  state  $|\psi\rangle$.

As for DQN sensing problems, one quantity to be focused on is the number of quantum resources used which reflects the energy characteristics of a light field. For the  W-type quantum probe state given by Eq.~(\ref{eq:1}), the available resources of the quantum state are determined by the average total photon number
\begin{equation}\label{Eq.13}
	\begin{aligned}
		\bar{N}&=\mathcal{N}^{2}\left(d+1\right)\left\langle\hat{n}\right\rangle. \\
	\end{aligned}
\end{equation}

%Next, to determine whether the QFI of multimode W-type entangled states exhibits the Heisenberg scaling, we rewrite the effective QFI as $H=\frac{4d\bar{N}}{d+1}\left(\frac{\left\langle \hat{n}^{2}\right\rangle}{\left\langle \hat{n}\right\rangle}-\frac{\bar{N}d}{d+1}\right)$.
% To ensure that the effective QFI satisfies $H>\bar{N}$, the normalized second-order moment must satisfy the inequality
% \begin{equation}
	% 	\begin{aligned}
		% 		\frac{\left\langle \hat{n}^{2}\right\rangle}{\left\langle \hat{n}\right\rangle}>\frac{1}{4}(1+\frac{1}{d})+\frac{\bar{N}d}{d+1}
		% 	\end{aligned}
	% \end{equation}
%  In the asymptotic limit $d\to\infty$, this condition reduces to $\frac{\left\langle \hat{n}^{2}\right\rangle}{\left\langle \hat{n}\right\rangle}>\frac{1}{4}+\bar{N}$. This defines the threshold for achieving a metrological advantage beyond the SQL.
% To ensure that the effective QFI exceeds the square of the average photon number, $H>\bar{N}^2$, the normalized second-order moment must satisfy the condition
%   \begin{equation}
	%  	\begin{aligned}
		%  		\frac{\left\langle \hat{n}^{2}\right\rangle}{\left\langle \hat{n}\right\rangle}>\bar{N}[\frac{1}{4}(1+\frac{1}{d})+\frac{d}{d+1}]
		%  	\end{aligned}
	%  \end{equation}
%  In the asymptotic limit $d\to\infty$, this condition reduces to $\frac{\left\langle \hat{n}^{2}\right\rangle}{\left\langle \hat{n}\right\rangle}>\frac{5\bar{N}}{4}$, indicating that a super-Heisenberg enhancement requires the second-order moment to grow faster than linear in $\bar{N}$.

%%%%%%%%%%%%%%%
\section{Enhancing sensitivity using quantum catalysis and entanglement}\label{section 2}
%%%%%%%%%%%%%%

In this section, we investigate the influence of quantum catalysis and entanglement on the sensitivity of DQN-based phase sensing. Specifically, we apply global and partial quantum catalytic operations to multimode W-type coherent states, thereby constructing a hybrid quantum probe that concurrently harnesses entanglement and quantum catalysis. Under global quantum catalysis, all $d+1$ modes of the quantum probe undergo catalytic action; under partial quantum catalysis, only a subset of the modes are catalyzed. Our results demonstrate that combining these two quantum resources yields a distinct advantage over probes utilizing only a single resource in DQN phase sensing. Furthermore, partial catalysis of the  ($d+1$)-mode quantum probe can further enhance distributed phase sensing performance and increase the cooperation factor.

%%%%%%%%%%%%------------------------------------------%%%%%%%%%%%%
\begin{figure}[t!]
	\centering
	\includegraphics[width=1.0\columnwidth]{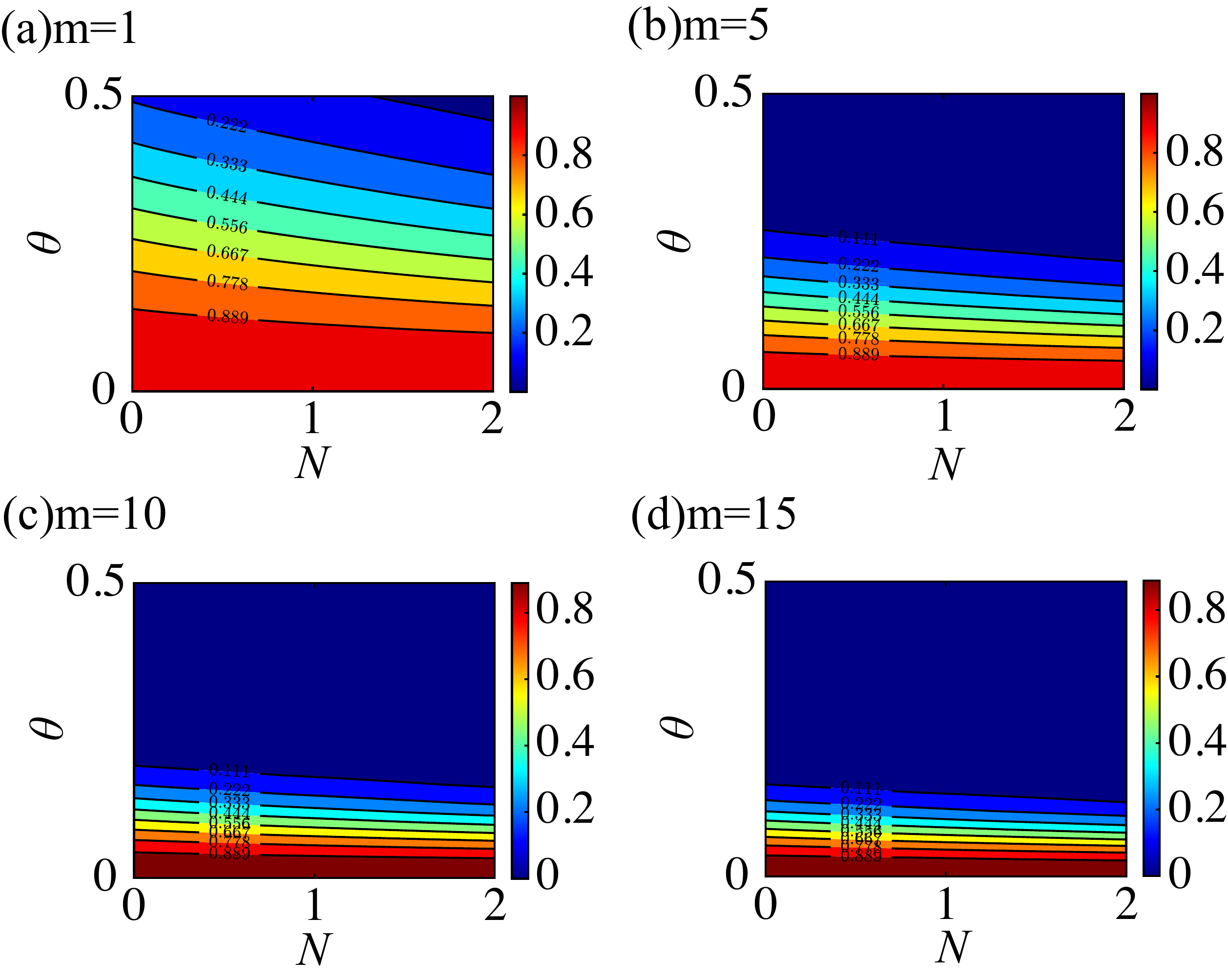}
	\caption{(a)-(d) represents the relationship between the success probability $P_{\mathrm{cwc}}$ with input quantum resources $N$ and the transmissivity parameter $\theta$ for the catalysis photon numbers $m = 1, 5, 10,15$, with a fixed number of quantum network nodes $d=5$.}\label{fig:2}
\end{figure}
%%%%%%%%%%%%------------------------------------------%%%%%%%%%%%%

%%%%%%%%%%%%
\subsection{The case of global quantum catalysis}
%%%%%%%%%%%%
In this subsection we study  the performance of the DQN phase sensing under the condition of global quantum catalysis in which all ($d+1$) modes of the quantum probe are catalyzed. 
We consider the case that the input state of the quantum probe before quantum catalytic operations is the multimode W-type  coherent state,   i.e., we take $\left|\psi\right\rangle=\left|\alpha\right\rangle$.  Then Eq. (\ref{eq:1}) becomes
\begin{equation}
	\left|\Psi_{\mathrm{wc}}\right\rangle=\mathcal{N}_{1}\sum_{m=0}^{d}\left|0\right\rangle_{0}\left|0\right\rangle_{1}\left|0\right\rangle_{2}...\left|\alpha\right\rangle_{m}...\left|0\right\rangle_{d},
\end{equation}
where the normalization constant is given by $\mathcal{N}_{1}=\left[\left(d+1\right)\left(1+d e^{-\left|\alpha^{2}\right|}\right)\right]^{-\frac{1}{2}}$.  

The multimode W-type coherent state is globally catalyzed to produce a multiphoton globally catalyzed multimode W-type coherent state given by 
\begin{equation}\label{eq:10}
	\begin{aligned}
		\left|\Psi_{\mathrm{cwc}}\right\rangle&\equiv\mathcal{N}_{2}\sum_{m=0}^{d}\left|0\right\rangle_{0}\left|0\right\rangle_{1}\left|0\right\rangle_{2}...\left|\psi'\right\rangle_{m}...\left|0\right\rangle_{d},
	\end{aligned}
\end{equation}
This state serves as a shared quantum probe that mixes two essential quantum resources: quantum catalysis and entanglement.
After the multiphoton catalytic operation,  the vacuum state remains the vacuum state, while the coherent state becomes catalytic coherent state  (see Eq.(\ref{eq:A5}) in Appendix \ref{A_HEP})
\begin{equation}\label{eq:15}
	\left|\psi'\right\rangle=\bar{N}_{m}\mathcal{L}_{m}\left(\hat{a}^{\dagger}\mu\right)\left|\alpha_{\theta}\right\rangle,
\end{equation}
where we have set $\alpha_{\theta}=\alpha\cos\theta$. The coefficients associated with the normalization constant of catalytic coherent state is given by $\bar{N}_{m}^{-2}=\sum_{n,k=0}^{m}\Pi_{n,k}^{m}\left(-\mu,\mu^{\ast}\right)H_{nk}\left(\alpha_{\theta}^{\ast},-\alpha_{\theta}\right)$, where $H_{n,k}$ is two-variable Hermite polynomials, the parameter $\mu=\alpha\cos\theta\tan^{2}\theta$, and the multinomial expression for $\Pi_{n,k}^{m}\left(x,y \right)=\binom{m}{n}\binom{m}{k}\frac{x^{n}y^{k}}{n!k!}$.
From this we can obtain the normalization constant $\mathcal{N}_{2}$ of $\left|\Psi_{\mathrm{cwc}}\right\rangle$, as
\begin{equation}
	\begin{aligned}
		\mathcal{N}_{2}&=\left[\left(d+1\right)\left(1+d\bar{N}_{m}^{2}\exp(-\left|\alpha_{\theta}\right|^{2})\right)\right]^{-\frac{1}{2}}.
	\end{aligned}
\end{equation}

From  Eq.~(\ref{eq:2}) and Eq.~(\ref{eq:10}), we can obtain the expression of successful probability for multiphoton globally catalyzed multimode W-type coherent state $\left|\Psi_{\mathrm{cwc}}\right\rangle$ (see Eq. (\ref{eq:B5}) in Appendix \ref{B_HEP})
\begin{equation}\label{eq:17}
	\begin{aligned}
		P_{\mathrm{cwc}}&=\mathcal{N}_{1}^{2}N_{m}^{-2}\mathcal{N}_{2}^{-2}\left(\cos\theta\right)^{2md}.
	\end{aligned}
\end{equation}
To investigate the parameter conditions that lead to a higher success probability, Fig.~\ref{fig:2} illustrates the relationship between the success probability, the number of input resources $N$, and the transmissivity parameter $\theta$ when the number of quantum network nodes $d=5$. It can be concluded that the success probability decreases with the increase of $ \theta$ and the number of catalytic photons $m$, but for the case of multiphoton catalysis, the success probability can still remain at a high level in the region of high transmittance.

%%%%%%%%%%%%------------------------------------------%%%%%%%%
\begin{figure}[t!]
	\centering
	\includegraphics[width=1.0\columnwidth]{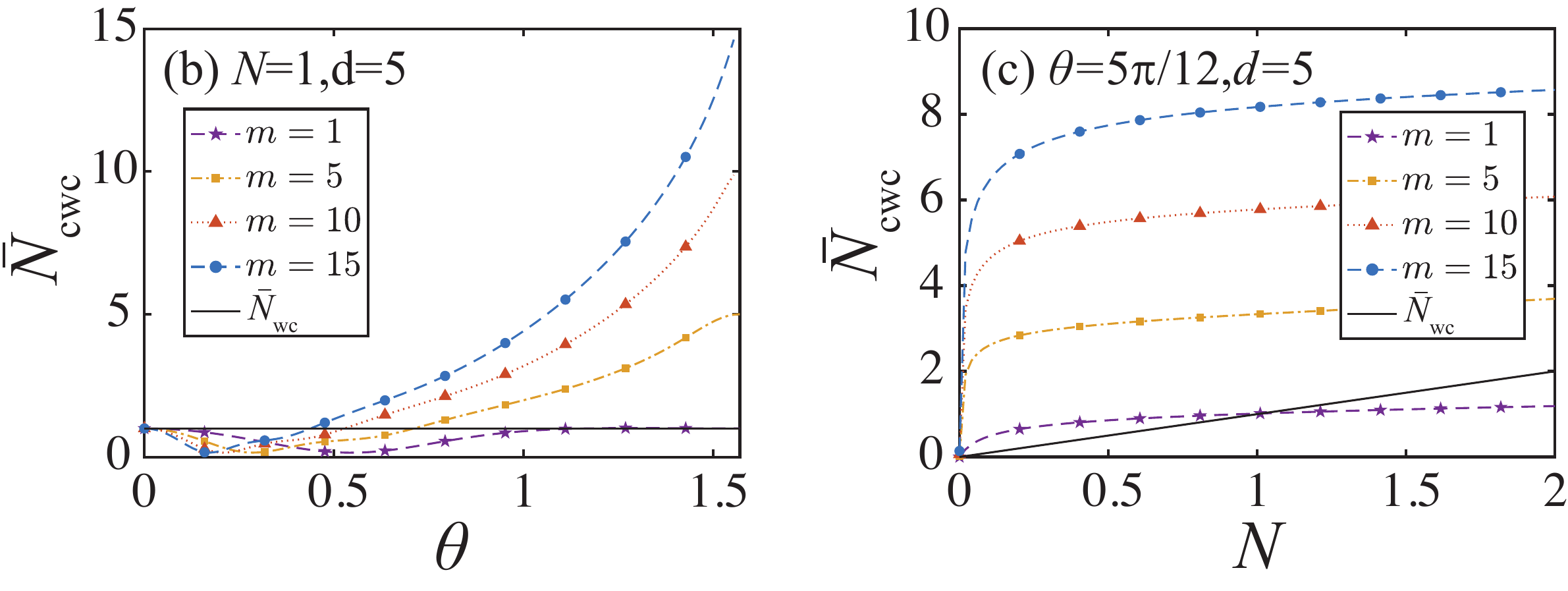}
	\caption{For the catalysis photon numbers $m = 1, 5 ,10, 20$, (a) the average photon number $\bar{N}_{\mathrm{cwc}}$of the transmissivity parameter $\theta$ for a given the input quantum resources $N=1$ and the network nodes $d=5$, and (b) $\bar{N}_{\mathrm{cws}}$ as a function of $N$ for a given $\theta=5\pi/12$ and $d = 5$, and its comparison with $\bar{N}_{\mathrm{wc}}$. }\label{fig:3}
\end{figure}
%%%%%%%%%%%%------------------------------------------%%%%%%%

Then we discuss the variation of the number of quantum resources. According to Eq.~(\ref{Eq.13}), the available resources of mutimode W-type coherent state $\left|\Psi_{\mathrm{wc}}\right\rangle$ are $\bar{N}_{\mathrm{wc}}=\mathcal{N}_{1}^{2}\left(d+1\right)\left|\alpha\right|^{2}$. However, for multiphoton globally catalyzed multimode W-type coherent state $\left|\Psi_{\mathrm{cwc}}\right\rangle$, the average number of photons can represent as
\begin{equation}
	\begin{aligned}
		\bar{N}_{\mathrm{cwc}}=&-\mathcal{N}_{2}^{2}\left(d+1\right)\bar{N}_{m}^{2}\sum_{n,k}^{m}\Pi_{n,k}^{m}(-\mu,\mu^{*}) \\
		&\times H_{n+1,k+1}(\alpha^{*}_{\theta}, -\alpha_{\theta})-1, \\
	\end{aligned}
\end{equation}

Since quantum catalysis is a probabilistic event, the average photon number $\bar{N}_{\mathrm{cwc}}$ varies under different catalytic conditions. To characterize this behavior, we examine how $\bar{N}_{\mathrm{cwc}}$ varies with different catalysis photon numbers $m = 1, 5, 10, 15$.
Fig.~\ref{fig:3}(a) illustrates its dependence on $\theta$ for a given $N=1$ (corresponding to $\alpha=1.4686$), while Fig.~\ref{fig:3}(b) illustrates the dependence of $\bar{N}_{\mathrm{cwc}}$ on the input resource $N$ for a fixed BS transmissivity parameter $\theta=5\pi/12$.
 From Fig.~\ref{fig:3}(a), it is observed that $\bar{N}_{\mathrm{cwc}}$ increases with the catalysis photon number $m$ increasing in the space of the larger transmissivity parameter $\theta$. A comparison indicates that the $\bar{N}_{\mathrm{cwc}}$ corresponding to $m>1$ is greater than the $\bar{N}_{\mathrm{wc}}$ when $\theta$ takes larger values and as $m$ increasing, the region of $\bar{N}_{\mathrm{cwc}}>\bar{N}_{\mathrm{wc}}$ expanding. As demonstrated in Fig.~\ref{fig:3}(b), at $\theta=5\pi/12$, $\bar{N}_{\mathrm{cwc}}$ increases with both $m$ and $N$, gradually tends to saturate as $N$ becomes large, and the $\bar{N}_{\mathrm{cwc}}$ surpasses the $\bar{N}_{\mathrm{w}}$ within the whole space of $N$ for multiphoton catalysis ($m>1$). 

%%%%%
%%%%%%%%%%%%%%%%%%%%%%%%%%%%%
%Enhancement of Quantum Probes through catalytic operations
%Enhancing sensitivity via introducing quantum catalysis as a new quantum resource
%Quantum Catalysis as a new Quantum resource: Enhancing sensitivity via hybrid two Quantum resources

%\subsection{Quantum entanglement and quantum catalysis collaborate to enhance senstivity}
%Enhanced sensing performance via two hybridQuantum resources
%\subsubsection{Advantage of  global Quantum Catalysis: Comparison between  $\left|\Psi_{cwc}\right\rangle$ with $\left|\Psi_{wc}\right\rangle$}

We now study the performance of the DQN phase sensing by calculating effective QFI of  quantum probe state. On the one hand, for multimode W-type coherent state $\left|\Psi_{\mathrm{wc}}\right\rangle$, we can calculate the effective QFI of this multimode quantum probe according to Eq.(\ref{eq:13})
\begin{equation}\label{eq:19}
	\begin{aligned}
		H_{\mathrm{wc}}&=4\mathcal{N}_{1}^{2}\left|\alpha\right|^{2}d\left[1+\left(1-d\mathcal{N}_{1}^{2}\right)\left|\alpha\right|^{2}\right],
	\end{aligned}
\end{equation}
where we have used $\left\langle \hat{n}\right\rangle=\left|\alpha\right|^{2}$, and $\left\langle \hat{n}^{2}\right\rangle=\left|\alpha\right|^{2}\left(1+\left|\alpha\right|^{2}\right)$. 

%In order to discuss whether the effective QFI of $\left|\Psi_{wc}\right\rangle$ can reach or surpass the HL, we analyze the behavior of $H_{wc}$ in two asymptotic regimes. 
%In the week-fied regime $\left|\alpha\right|^2\ll1$, the average photon number is approximated as $\bar{N}_{wc}=\frac{\left|\alpha\right|^2}{1+d(1-\left|\alpha\right|^2)}$, leading to  $H_{wc}=\frac{4d}{d+1}[\bar{N}_{wc}+(d+\frac{1}{d+1})\bar{N}_{wc}^2]$. Since $d\ge1$, this expression guarantees that $H_{wc}>\bar{N}_{wc}^2$,  thus surpassing the HL.
%In the strong-field regime $\left|\alpha\right|^2\gg1$, the approximation $\bar{N}_{wc}=\left|\alpha\right|^2$ yields $H_{wc}=\frac{4d\bar{N}_{cw}}{d+1}(1+\frac{\bar{N}_{wc}}{d+1})$, confirming that $H_{wc}>\bar{N}_{wc}$ is always satisfied, but $H_{wc}>\bar{N}_{wc}^{2}$  is satisfied only $\frac{4d}{(d+1)^2}\ge1$, which restricts the validity of HL surpassing to small $d$ (specifially $d\le1.7$).

On the other hand, we can also calculate  the effective QFI of multiphoton globally catalyzed multimode W-type coherent state $\left|\Psi_{\mathrm{cwc}}\right\rangle$ (See Eq. (\ref{eq:C5}) in Appendix \ref{C_HEP})
\begin{equation}\label{eq:20}
	\begin{aligned}
		H_{\mathrm{cwc}}=&4d\mathcal{N}_{2}^{2}\bar{N}_{m}^{2}\Big[\sum_{n,k}^{m}\Pi_{n,k}^{m}\left(-\mu,\mu^{*}\right)H_{n+2,k+2}\left(\alpha^{*}_{\theta},-\alpha_{\theta}\right) \\
		&+\left(3-2d\mathcal{N}_{2}^{2}\right)\sum_{n,k}^{m}\Pi_{n,k}^{m}\left(-\mu,\mu^{*}\right)H_{n+1,k+1}\left(\alpha^{*}_{\theta},-\alpha_{\theta}\right) \\
		&-\mathcal{N}_{2}^{2}\bar{N}_{m}^{2}d\left(\sum_{n,k}^{m}\Pi_{n,k}^{m}\left(-\mu,\mu^{*}\right)H_{n+1,k+1}\left(\alpha^{*}_{\theta},-\alpha_{\theta}\right)\right)^{2}\Big] \\
		&+4d\mathcal{N}_{2}^{2}\left(1-d\mathcal{N}_{2}^{2}\right).  \\
	\end{aligned}
\end{equation}

From Eq.~(\ref{eq:20}) we can see  that the effective QFI of $\left|\Psi_{\mathrm{cwc}}\right\rangle$ not only depends on the amplitude of the coherent state $\alpha$ and the number of nodes in the network $d$, but also on the catalytic correlation coefficient BS parameter $\theta$ and the number of catalytic photons $m$. Therefore, we can improve estimation precision through adjust these parameters. As expected, when $m=0$ and $\theta =0$, this relation recovers the phase estimation with $\left|\Psi_{\mathrm{wc}}\right\rangle$ given by Eq.~(\ref{eq:19}).

%%%%%%%%%%%%------------------------------------------%%%%%%%%%%%%
\begin{figure}[t!]
	\centering
	\includegraphics[width=1.0\columnwidth]{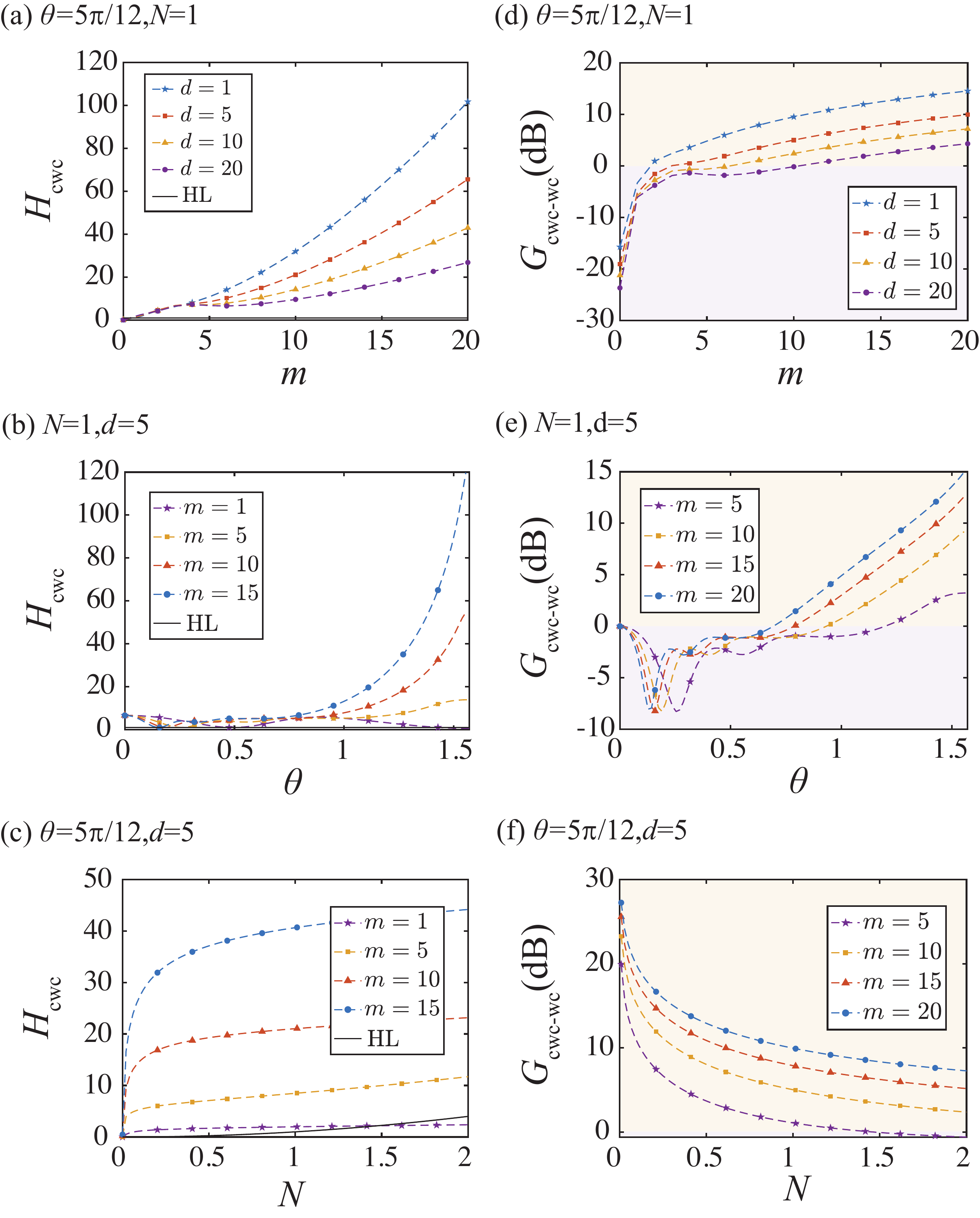}
	\caption{Effective QFI $H_{\mathrm{cwc}}$
		(left panels, a–c) and corresponding gain factor $G_{\mathrm{cwc-wc}}$(right panels, d–f) as functions of key system parameters. (a) $H_{\mathrm{cwc}}$ and (d) $G_{\mathrm{cwc-wc}}$ dependence on the number of catalytic photons $m$ for different network nodes $d=1,5,10,20$, with fixed input resource $N=1$ and beam splitter transmissivity $\theta=\frac{5\pi}{12}$. (b) $H_{\mathrm{cwc}}$ and (e) $G_{\mathrm{cwc-wc}}$ dependence on beam splitter paramete $\theta$ for $m=1,5,10,15$ and for $m=5,10,15,20$ respectively, with fixed $d=5$ and $N=1$.  (c)$H_{\mathrm{cwc}}$ and (f) $G_{\mathrm{cwc-wc}}$ dependence on input resource $N$ for $m=1,5,10,15$ and for $m=5,10,15,20$ respectively, with fixed $d=5$ and $\theta=\frac{5\pi}{12}$. The black solid line in (a–c) indicates the Heisenberg limit.  The yellow shaded area indicates $\left|\Psi_{\mathrm{cwc}}\right\rangle$ exhibits better estimation precision than $\left|\Psi_{\mathrm{wc}}\right\rangle$. \label{fig:4}}
\end{figure}
%%%%%%%%%%%%------------------------------------------%%%%%%%%%%%%

In Figure~\ref{fig:4}, we plot the effective \textcolor{red}{QFI} in the case of global quantum catalysis, and its gain factor with respect to that without quantum catalysis.
From Figure~\ref{fig:4} (a-c) we can observe that quantum catalysis can significantly enhance the effective QFI of the probe over a broad parameter range. For fixed $N=1$ and $\theta=5\pi/12$, increasing the catalytic photon number $m$ improves the sensitivity for all considered network sizes, and for sufficiently large $m$ the sensitivity loss caused by increasing $d$ can be partially mitigated. For fixed $N=1$ and $d=5$, despite a dip in the small-$\theta$ region, $H_{\mathrm{cwc}}$ increases rapidly for larger $\theta$, with higher catalysis photon numbers $m$ further enhancing the QFI and broadening the above-HL parameter region.
 In addition, for fixed $d=5$ and $\theta=5\pi/12$, the effective QFI increases with both $N$ and $m$, with stronger catalysis yielding a more pronounced gain in the low-$N$ region and a saturation tendency at larger $N$. Overall, $\left|\Psi_{\mathrm{cwc}}\right\rangle$ exhibits clear metrological superiority, and its effective QFI exceeds the HL in nearly all considered cases.

A comparison between $H_{\mathrm{cwc}}$ and $\bar{N}_{\mathrm{cwc}}$ shows that their dependences on the input resource $N$, the beam-splitter parameter $\theta$, and the catalytic photon number $m$ are largely consistent. This indicates that the behavior of $H_{\mathrm{cwc}}$ is mainly governed by the variation of $\bar{N}_{\mathrm{cwc}}$, so that $\bar{N}_{\mathrm{cwc}}$ serves as the dominant factor determining the overall features of $H_{\mathrm{cwc}}$. At the same time, slight discrepancies can still be observed between the two quantities, suggesting that $H_{\mathrm{cwc}}$ is not determined by $\bar{N}_{\mathrm{cwc}}$ alone, but is also modified by the corresponding weighting factor.

According to the weak QCRB, a higher effective QFI implies better sensitivity. To evaluate the impact of quantum catalysis on the performance of DQN sensing. We compare the effective QFI of quantum probes before and after the catalytic operation under the condition of same input resources. To this end, we define a gain factor as $G=10\log(\frac{H_{\mathrm{c}}}{H})$ to quantize the enhancement in metrological performance. If $G$ is greater than $0$, we can say the quantum state after catalysis $\left|\Psi_{\mathrm{c}}\right\rangle$ performs better than before $\left|\Psi\right\rangle$. As for $\left|\Psi_{\mathrm{wc}}\right\rangle$, we have 

%Next, we examine the effect of the catalytic operation on the performance of the multimode W-type quantum probe, quantified by the value of $G$ for $\left|\Psi_{c}\right\rangle=\left|\Psi_{cwc}\right\rangle$  and $\left|\Psi\right\rangle=\left|\Psi_{wc}\right\rangle$ (Note that, to ensure the equivalent utilization of resources, the same coherent state is employed in this case, except that the former has been catalyzed), the resulting expression is as follows
%\begin{widetext}
\begin{equation}
	G_{\mathrm{cwc-wc}}=10\log\left(\frac{H_{\mathrm{cwc}}}{H_{\mathrm{wc}}}\right). 
\end{equation}
Fig.~\ref{fig:4} (d)–(f) show the gain factor $G_{\mathrm{cwc-wc}}$ as functions of the catalytic photon number $m$, the beam-splitter parameter $\theta$, and the input resource $N$, respectively. As shown in Fig.~\ref{fig:4} (d), $G_{\mathrm{cwc-wc}}$ increases with the number of catalytic photons $m$ but decreases with the number of network nodes $d$. Even so, for sufficiently large $m$, the gain remains positive in large-scale networks; for example, when $d=20$, it can still reach $4.32$ dB at $m=20$. Fig.~\ref{fig:4} (e) shows that, $G_{\mathrm{cwc-wc}}$ increases with $\theta$ in the larger $\theta$ region, reaching a maximum near $\theta=\pi/2$. In Fig.~\ref{fig:4} (f), with fixed $d=5$ and $\theta=5\pi/12$, $G_{\mathrm{cwc-wc}}$ is positive in the low $N$ regime for all considered $m$, indicating that $\left|\Psi_{\mathrm{cwc}}\right\rangle$  outperforms $\left|\Psi_{\mathrm{wc}}\right\rangle$ under the same initial resources. As $N$ increases, however, the gain gradually decreases and may even become negative for small $m$. The largest gains, up to $27.67$~dB, appear in the low $N$ regime for large $m$.

These results show that quantum catalysis can significantly enhance the metrological performance of $|\Psi_{\mathrm{wc}}\rangle$, with the enhancement being most pronounced for large $m$, an appropriately large $\theta$, and low input resource $N$. 
Following the above discussion, the enhancement with increasing $m$ and $\theta$ mainly appears in the larger-$\theta$ region originates from the stronger catalytic modification of the coherent component, which amplifies the effective photon-number resource and the non-Gaussian contribution to phase sensitivity. The decrease of the gain with increasing $d$ reflects that the available quantum resources are distributed over more network nodes, which weakens the relative catalytic advantage. As $N$ increases, however, $\bar{N}_{\mathrm{cwc}}$ gradually saturates, whereas $\bar{N}_{\mathrm{wc}}$ grows nearly linearly, reducing the relative resource advantage introduced by catalysis. Therefore, the decrease of $G_{\mathrm{cwc-wc}}$ with increasing $N$ can be understood as a crossover from catalysis-induced non-Gaussian enhancement in the low-$N$ regime to the intrinsic entanglement of the original probe in the high-$N$ regime.
\begin{figure}[t!]
	\centering
	\includegraphics[width=1.0\columnwidth]{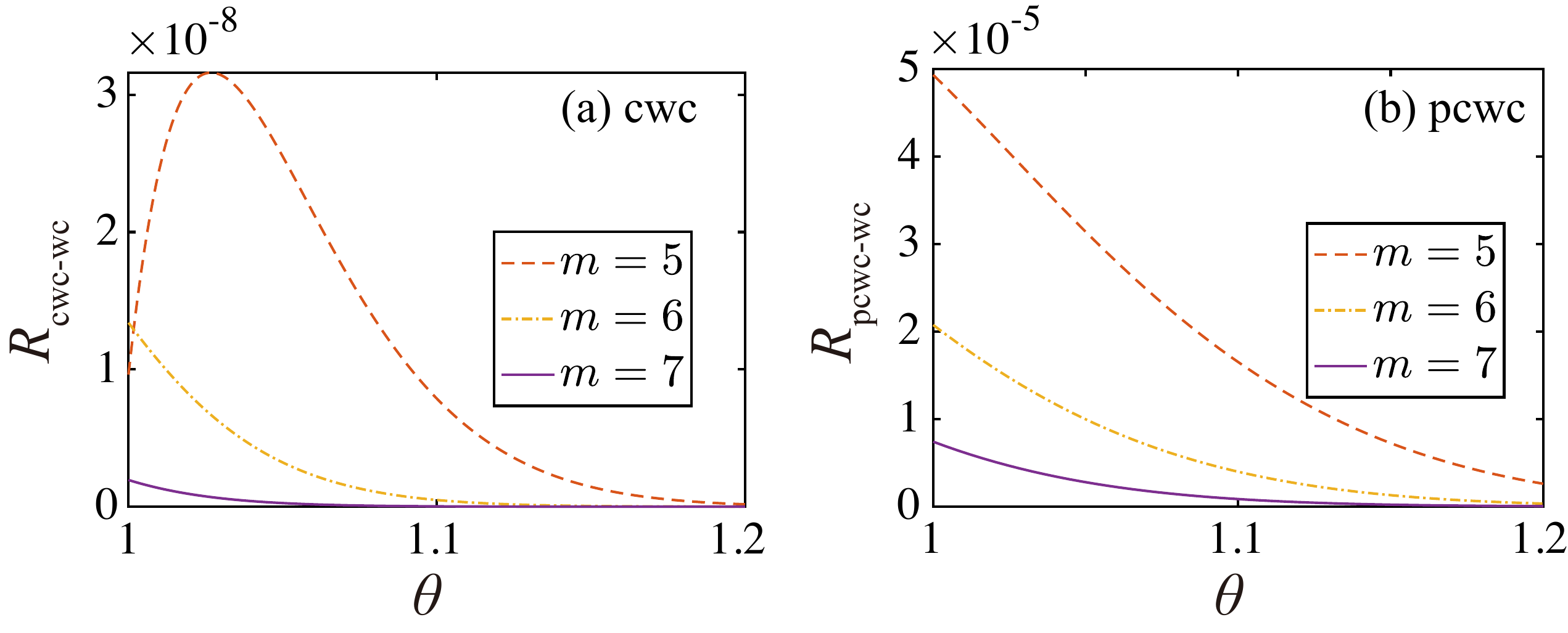}
	\caption{(a) The cooperation factor for global catalysis $R_{\mathrm{cwc-wc}}$  as a function of the transmissivity coefficient $\theta$ for different catalytic photons $m=5,6,7$ for given $N=1$ and $d=2$ and (b) The cooperation factor for partial catalysis $R_{\mathrm{pcwc-wc}}$ as a function of $\theta$  for $m=5,6,7$ when $N=1$, $d=2$, and $s=1$. \label{fig:5}}
\end{figure}
%%%%%%%%%%%%------------------------------------------%%%%%%%%%%%%

Accurately, what we want is the presence of a high gain region that also exhibits a high success probability. Therefore we accounts for two factors: the overlap between the positive gain region $(H_{\mathrm{cwc}}>H_{\mathrm{wc}})$ and  the high success probability region. To quantitatively capture this trade-off, we define a cooperation factor
\begin{equation}
	\begin{aligned}
		R_{\mathrm{cwc-wc}}=\bigtriangleup H_{\mathrm{cwc-wc}}\times P_{\mathrm{cwc}},
	\end{aligned}
\end{equation}
where $\bigtriangleup H_{\mathrm{cwc-wc}}=H_{\mathrm{cwc}}-H_{\mathrm{wc}}$ denotes the sensitivity gain with respect to that without quantum catalysis.

Now we numerically investigate the cooperation coefficient $R_{\mathrm{cwc-wc}}$ for global catalysis scheme by  adjusting the transmission coefficient $\theta$ under fixed $N=1$ and $d=2$, with catalytic photon numbers $m=5,6,7$.  Fig.~\ref{fig:5} (a) shows that $R_{\mathrm{cwc-wc}}$ remains positive over the plotted range, indicating that $\left|\Psi_{\mathrm{cwc}}\right\rangle$ still retains a positive advantage over $\left|\Psi_{\mathrm{wc}}\right\rangle$ after the success probability is taken into account. Moreover, as the number of catalytic photons \textcolor{red}{$m$} increases, the cooperation coefficient decreases. It indicates that although increasing $m$ may strengthen the sensing enhancement, it also causes a faster reduction in the success probability, thereby suppressing the $R_{\mathrm{cwc-wc}}$ performance. This implies that only a limited parameter region allows a reasonable trade-off between sensitivity enhancement and success probability. Nevertheless, the success probability remains very low in the regime where a positive gain is observed, which limits the practical applicability of the present scheme.
%The optimal operating point for $m$-photon catalysis renders this cooperation coefficient maximum.
%%%%%%
%From the Fig.~\ref{fig:5}, it can be observed that for each catalytic photon number $m$, there exists an optimal $\theta$ that maximizes $R_{cwc-wc}$, for $m=1,5,10,20$, the optimal $\theta$ respectively is 0.2631, 0.1421, 0.1018, and 0.0716. Additionally, as the number of catalytic photons increases, the maximum cooperation factor also increases, for $m=1$, the maximal $R_{cwc-wc}$ is 0.4844 with $P_{cwc}=0.4937, \bigtriangleup H_{cwc-wc}=0.9811$, for $m=5$, the maximal $R_{cwc-wc}$ is 1.2496 with $P_{cwc}=0.3987, \bigtriangleup H_{cwc-wc}=3.1340$, for $m=10$, the maximal $R_{cwc-wc}$ is 1.3723 with $P_{cwc}=0.3958, \bigtriangleup H_{cwc-wc}=3.4675$, and for $m=20$, the maximal $R_{cwc-wc}$ is 1.4353 with $P_{cwc}=0.1644, \bigtriangleup H_{cwc-wc}=6.5652$. In other words, for a given number of catalytic photons, network nodes, and input resources, an optimal transmission coefficient can always be identified to achieve the best operating point, in which both the phase estimation sensitivity and success probability have a high value.  Furthermore, the phase estimation sensitivity can be further improved by increasing the number of catalytic photons, although at the cost of a reduced success probability. 

\subsection{The case  of partial quantum catalysis} 
%Comparison between  $\left|\Psi_{pcwc}\right\rangle$ with $\left|\Psi_{wc}\right\rangle$}\label{Section 4}

We now study  the performance of the DQN phase sensing under the condition of partial quantum catalysis in which only partial modes in the ($d+1$) modes of the quantum probe are catalyzed. Assume that ($s+1$) modes are catalyzed  in the ($d+1$) modes of the quantum probe.  The probe state of the partial catalyzed quantum probe can be written as follows
\begin{equation}\label{eq:23}
	\begin{aligned}
		\left|\Psi_{\mathrm{pcwc}}\right\rangle=&\mathcal{N}_{1}'\Big(\sum_{m=0}^{s}\left|0\right\rangle_{0}\left|0\right\rangle_{1}...\left|\psi'\right\rangle_{m}\left|0\right\rangle_{d} \\
		&+\sum_{m=s+1}^{d}\left|0\right\rangle_{0}\left|0\right\rangle_{1}...\left|\alpha\right\rangle_{m}\left|0\right\rangle_{d}\Big), \\
	\end{aligned}
\end{equation}
where the first $s+1$ modes are catalyzed, their excited states are catalyzed coherent state $\left|\psi'\right\rangle$ defined by Eq.~(\ref{eq:15}), the remaining modes are not catalyzed and their excited states are coherent state $\left|\alpha\right\rangle$. $\mathcal{N}_{1}'$ is the normalization given by
\begin{equation}
	\begin{aligned}
		\mathcal{N}_{1}^{'-2}=&\left(s+1\right)\left[1+s\left|\left\langle 0|\psi'\right\rangle\right|^{2}+\left(d-s\right)e^{-\frac{\left|\alpha\right|^{2}}{2}}\left\langle 0|\psi'\right\rangle\right] \\
		&+\left(d-s\right)\left[1+\left(d-s-1\right)e^{-\left|\alpha\right|^{2}}+\left(s+1\right)e^{-\frac{\left|\alpha\right|^{2}}{2}}\left\langle \psi'|0\right\rangle\right], 
	\end{aligned}
\end{equation}
where we have $\left\langle 0|\psi'\right\rangle=N_{m}\cos^{m}\left(\theta\right)e^{-\frac{\left|\alpha\right|^{2}}{2}}$.

The number of quantum resources in the partial quantum catalysis scheme is the average photon number  in the probe state of the partial catalyzed quantum probe given by Eq. (\ref{eq:23}) with the following expression
\begin{equation}\label{eq:27}
	\begin{aligned}
		\bar{N}_{\mathrm{pcwc}}&=\mathcal{N}_{1}^{'2}\left[\left(s+1\right)\left\langle \hat{n}'\right\rangle+\left(d-s\right)\left\langle\hat{n}\right\rangle\right]. \\
	\end{aligned}
\end{equation}
where $\left\langle\hat{n}'\right\rangle$ is the average photon number of catalytic coherent state   $\left|\psi'\right\rangle$, and $\left\langle\hat{n}\right\rangle$ is the average photon number of coherent state $|\alpha\rangle$. 

In order to discuss the sensitivity of the DQN phase sensing under the partial quantum catalysis, we need to calculate  the effective QFI of the partial quantum catalysis state $\left|\Psi_{\mathrm{pcwc}}\right\rangle$.
Based on Eq.~(\ref{Eq.12}), we can obtain the value of $H$ for this case as
% (the detailed calculation process is shown in Appendix~\ref{D_HEP})
\begin{equation}
	\begin{aligned}
		H_{\mathrm{pcwc}}=&4\mathcal{N}_{1}^{'2}\left[s\left\langle \hat{n}'^{2}\right\rangle+\left(d-s\right)\left\langle \hat{n}^{2}\right\rangle\right], \\
		&-4\mathcal{N}_{1}^{'4}\left[s\left\langle \hat{n}'\right\rangle+\left(d-s\right)\left\langle \hat{n}\right\rangle\right]^{2}. \\
	\end{aligned}
\end{equation}
Moreover, we want to discuss the sensitivity advantage of quantum probes with a certain number of catalyzed modes over quantum probes that are not catalyzed. Therefore, we compare the gain factors of the two schemes, i.e., $G_{\mathrm{pcwc-wc}}=10\log\left(\frac{H_{\mathrm{pcwc}}}{H_{\mathrm{wc}}}\right)$. If the ratio is greater than zero, it means that only $s$ modes catalyzed can obtain a better sensitivity than that in the case without catalyzing. Since we want to make sure that the both schemes of the partial catalysis and without catalysis use the same quantum resources when we make the comparison, we are using the same coherent state here.

\begin{figure}[t!]
	\centering
	\includegraphics[width=0.78\columnwidth]{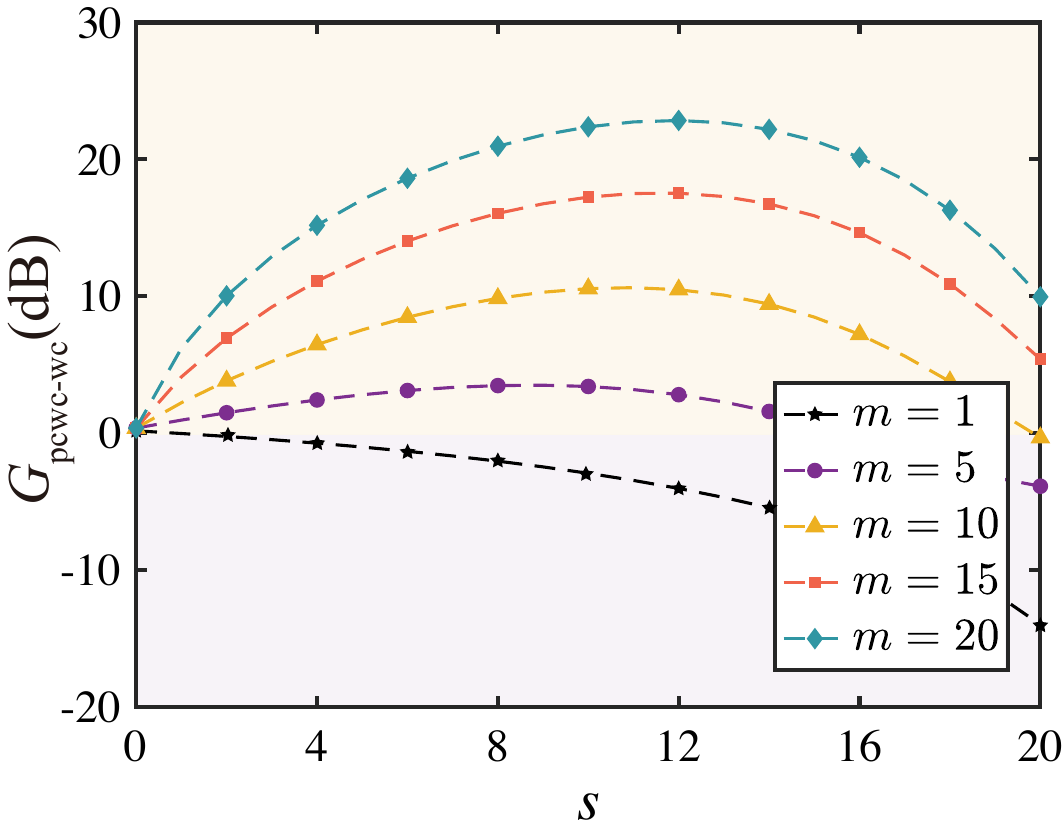}
	\caption{Sensitivity gain factor $G_{\mathrm{pcwc-wc}}$ as a funcation of the number of catalytic modes $s$ for different catalytic photons $m=1,5,10,15,20$ when the BS transmittance $\theta=5\pi/12$, the input resource number $N=1$ and the network node number $d=20$. The yellow shaded area indicates $\left|\Psi_{\mathrm{pcwc}}\right\rangle$ exhibits better estimation precision than $\left|\Psi_{\mathrm{wc}}\right\rangle$.}\label{fig:6}
\end{figure}

To verify the effect of the catalyzed modes on the sensitivity of the DQN phase sensing in the partial catalysis scheme, we plot the gain factor $G_{\mathrm{pcwc-wc}}$ for $d=20$ with respect to the modes of catalysis $s$ with the condition of $N=1$ and $\theta=5\pi/12$, as shown in Fig.~\ref{fig:6}. We find that for each case of multiphoton catalysis ($m>1$), there exists an optimal catalytic-mode number that leads to the most significant improvement in phase estimation sensitivity  of DQN sensing. Specifically, for $m = 5, 10, 15, 20$, the optimal catalytic-mode numbers are $9, 11, 12$, and $12$, respectively. This means that applying partial catalysis only up to the optimal number of catalyzed modes can achieve the maximum sensitivity enhancement of the DQN phase sensing while significantly improving the experimental success probability of the postselection process. Furthermore, as the number of catalyzing photons increases, the sensitivity gain obtained at the optimal catalyzed-mode number also increases. Specially, a $22.84$~dB improvement is observed at $m=20$ in the optimal mode. 

By contrast, when $s=d$, partial catalysis reduces to global catalysis, but its sensitivity gain is lower than that of the optimal partial-catalysis case. This behavior can be understood from the perspective of resource allocation in distributed quantum metrology. Ref~\cite{Ge2018Distributed}  have shown that, even with the same total photon number, metrological enhancement depends not only on the total amount of resources, but also on how these resources are distributed among different modes. Resources uniformly spread over many modes are not always converted into useful metrological enhancement, whereas a more concentrated resource distribution can lead to Heisenberg-scaling advantages. Similarly, in our $d+1$-mode W-type entangled scheme, the initial resource before catalysis is fixed. Global catalysis acts on all modes and preserves their equivalence, leading to a relatively uniform distribution of catalysis-induced non-Gaussian resources. In contrast, partial catalysis acts only on the first s modes, dividing the system into catalyzed and uncatalyzed subspaces. This breaks the uniform photon-number distribution and concentrates the catalysis-induced non-Gaussian resources in the catalyzed modes. Therefore, in certain parameter regimes, partial catalysis can yield a higher QFI or better effective quantum enhancement.

%%%%%%%%%%%%------------------------------------------%%%%%%%%%%%%
%%%%%%%%%%%%------------------------------------------%%%%%%%%%%%%

It is interesting to note that partial catalysis not only achieves the high sensitivity but also significantly improves the success rate of preparing catalytic quantum probes. This point can be observed by analyzing the  cooperation factor
\begin{equation}
	\begin{aligned}
		R_{\mathrm{pcwc-wc}}&=\Delta{H}_{\mathrm{pcwc-wc}}\times P_{\mathrm{pcwc}} ,
		\end{aligned}
\end{equation}
where $\Delta{H}_{\mathrm{pcwc-wc}}=H_{\mathrm{pcwc}}-H_{\mathrm{wc}}$ represents the sensitivity gain with $H_{\mathrm{wc}}$ and $H_{\mathrm{pcwc}}$ being given by Eqs. (\ref{eq:19}) and (\ref{eq:28}), respectively.  $P_{\mathrm{pcwc}}$ represents the successful probability of preparation which can be obtain as followed (see Eq. (\ref{eq:B6}) in Appendix \ref{B_HEP})
\begin{equation}\label{eq:28}
	\begin{aligned}
		P_{\mathrm{pcwc}}=&\mathcal{N}_{1}^{2}\left(\cos\theta\right)^{2ms}\big\{(s+1)\left[\left(N_{m}\right)^{-2}+d(\cos\theta)^{2m}\exp(-|\alpha|^{2})\right]\\&+(d-s)(\cos\theta)^{2m}\left[1+d\exp(-|\alpha|^{2})\right]\big\}.
	\end{aligned}
\end{equation}

In Figure~\ref{fig:5}(b), we plot the cooperation factor for partial catalysis $R_{\mathrm{pcwc-wc}}$ as a function of $\theta$  for $m=5,6,7$ when $N=1$, $d=2$, and $s=1$.  The results show that, for any fixed catalytic photon number $m$, the cooperation factor  $R_{\mathrm{pcwc-wc}}$ decreases monotonically with increasing $\theta$. Comparing with the case of the global quantum catalysis,   partial catalysis yields a cooperation factor four orders of magnitude higher than that of global catalysis under the same parameter conditions. Therefore, we can conclude that partial catalysis has more advantages than the global catalysis in the DQN phase sensing.

%%%%%%%%
\section{Enhancing Sensitivity using quantum catalysis, entanglement, and squeezing }\label{section 3}
%%%%%%%%%%%

In this section, we examine the DQN-based phase sensing scheme incorporating three types of quantum resources—quantum catalysis, entanglement, and squeezing, and demonstrate that the DQN sensing protocol utilizing all three resources exhibits a stronger quantum-enhanced advantage compared to the protocol using only two types. Moreover, we find that larger network systems depend more critically on the cooperative interplay among multiple quantum resources. In what follows, we analyze the phase sensing performance of the DQN scheme with three quantum resources, considering both global and local quantum catalysis.

%%%%%%%%%%%%
\subsection{The case of global quantum catalysis}
%%%%%%%%%%%%
In this subsection we study  the sensing performance of the DQN phase sensing scheme with three quantum resources under the condition of global quantum catalysis.  The quantum squeezing is introduced to  the DQN phase sensing scheme by taking the excited state $\left|\psi\right\rangle$ as the squeezed state,  i.e., $\left|\psi\right\rangle=\left|r\right\rangle=S(r)|0\rangle$ with $S(r)$ being the simple-mode squeezing operator. Then the input state of the quantum probe is a $\left(d+1\right)$-mode W-type squeezed state $\left|\Psi_{\mathrm{ws}}\right\rangle$ with the following expression
\begin{equation}\label{eq:29}
	\left|\Psi_{\mathrm{ws}}\right\rangle=\mathcal{N}_{3}\sum_{m=0}^{d}\left|0\right\rangle_{0}\left|0\right\rangle_{1}\left|0\right\rangle_{2}...\left|r\right\rangle_{m}...\left|0\right\rangle_{d},
\end{equation}
where $\mathcal{N}_{3}=\left[\left(d+1\right)\left(1+d \operatorname{sech}  r\right)\right]^{-\frac{1}{2}}$ is the normalization constant.

%%%%%%%%%%%%------------------------------------------%%%%%%%%%%%%
\begin{figure}[t!]
	%\textwidth
	\centering
	\includegraphics[width=1.0\columnwidth]{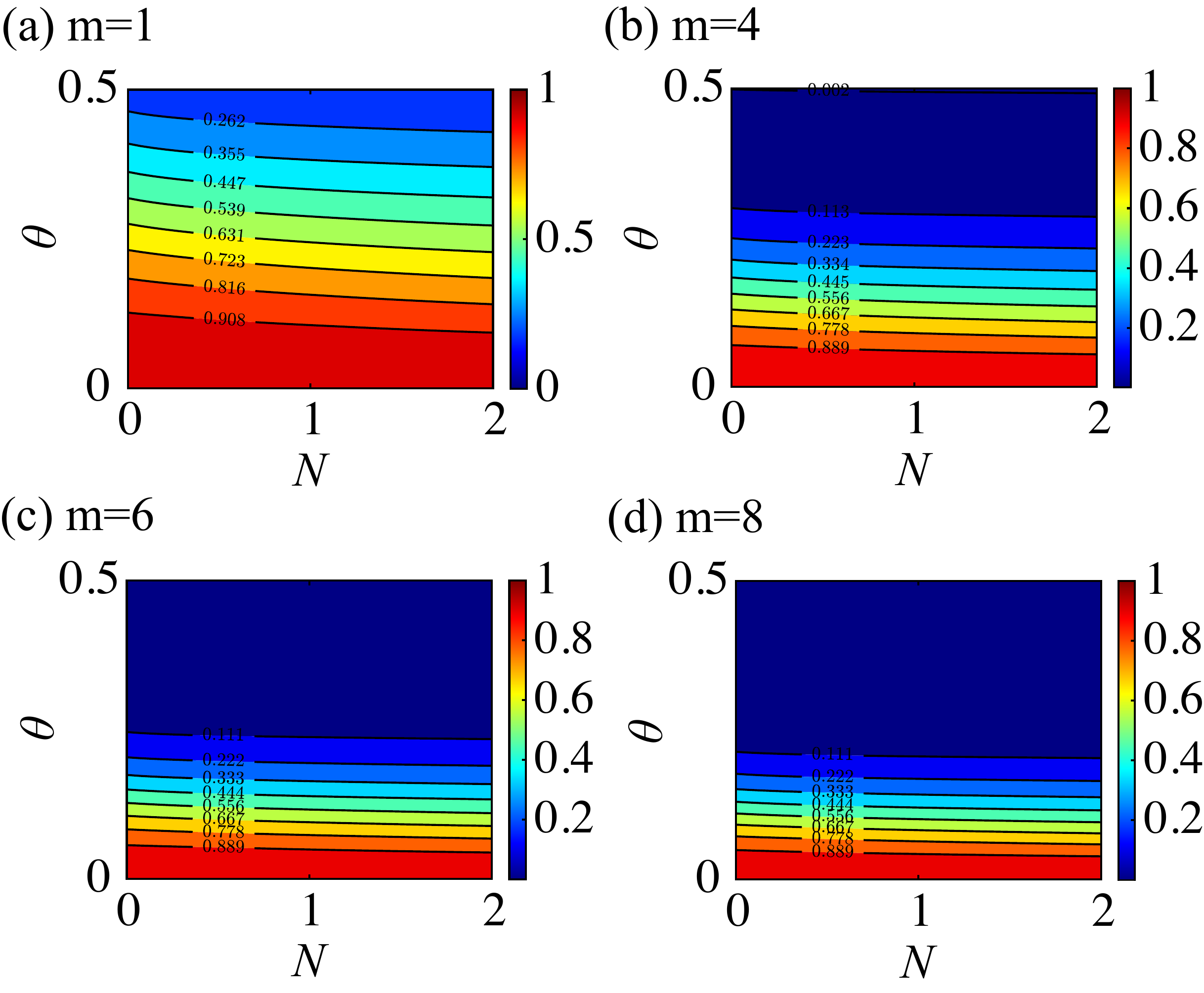}
	\caption{(a)-(d) represents the relationship between the success probability $P_{\mathrm{cws}}$ with input quantum resources $N$ and the transmissivity parameter $\theta$ for the catalysis photon numbers $m = 1, 4, 6, 8$, with a fixed number of quantum network nodes $d=5$.}\label{fig:7}
\end{figure}
%%%%%%%%%%%%------------------------------------------%%%%%%%%%%

To compare the sensing performance of quantum probes having three resources (catalysis, entanglement, and squeezing) with two resources (entanglement, and squeezing), we globally catalyze the W-type squeezed state $\left|\Psi_{\mathrm{ws}}\right\rangle$, then obtain a $(d+1)$-mode globally catalyzed W-type squeezed state $\left|\Psi_{\mathrm{cws}}\right\rangle$  which can be expressed as
\begin{equation}\label{eq:30}
	\begin{aligned}
		\left|\Psi_{\mathrm{cws}}\right\rangle&\equiv\mathcal{N}_{4}\sum_{m=0}^{d}\left|0\right\rangle_{0}\left|0\right\rangle_{1}\left|0\right\rangle_{2}...\left|\psi_{s}'\right\rangle_{m}...\left|0\right\rangle_{d},
	\end{aligned}
\end{equation}
%%%%%%%%%%%%%%%%%%%%%%%%
\begin{figure}[t]
	\centering
	\includegraphics[width=1.0\columnwidth]{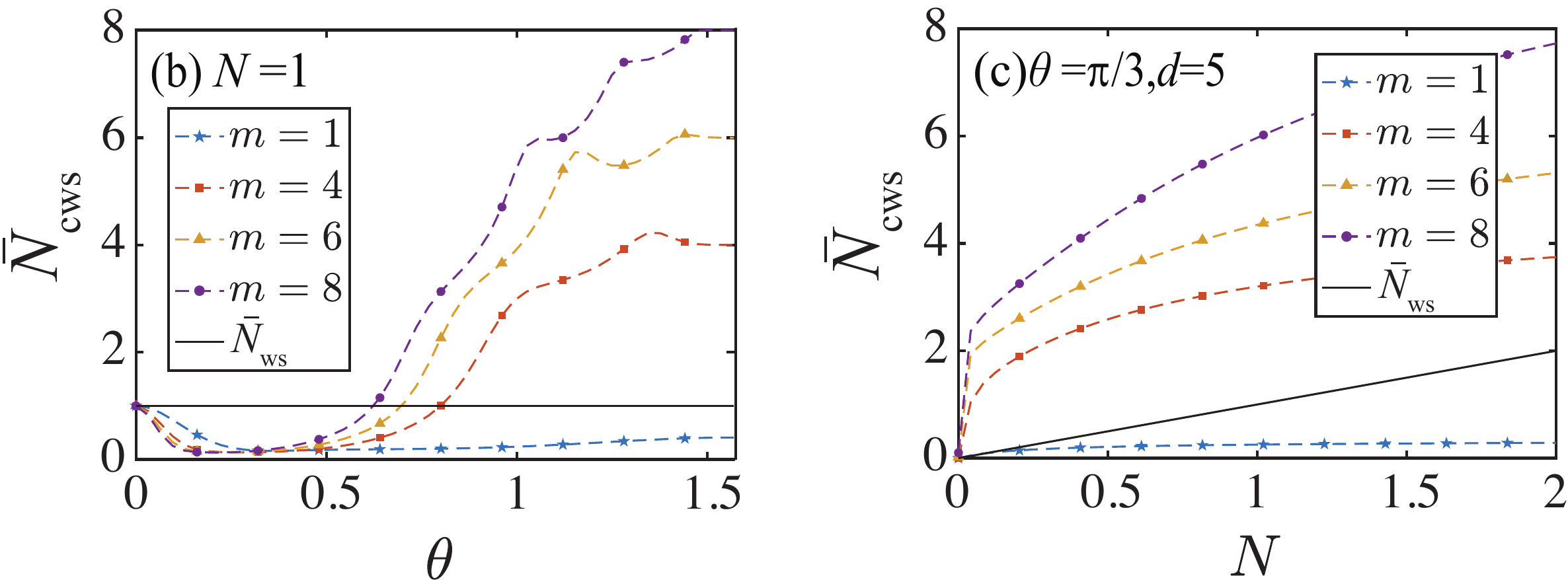}
	\caption{For the catalysis photon numbers $m = 1,4,6,8$, 
		(a) the average photon number $\bar{N}_{\mathrm{cws}}$ as a function of 
		he transmissivity parameter $\theta$ for a given the input quantum resourcethe $N=1$ and the networke nodes $d=5$, and (b) $\bar{N}_{\mathrm{cws}}$ as a function of $N$ for a given $\theta=\pi/3$ and $d=5$, and its comparison with $\bar{N}_{\mathrm{ws}}$.\label{fig:8}}
\end{figure}
%%%%%%%%%%%%%%%%%%%%%%%%%
which is utilized as the shared quantum probe state for our DQN phase sensing with three quantum resources. after the multiphoton catalytic operation,  the vacuum state remains the vacuum state, while the squeezed state becomes the following catalytic squeezed state  (see Eq. (\ref{eq:A8}) in the Appendix \ref{A_HEP})
\begin{equation}\label{eq:31}
	\left|\psi_{s}'\right\rangle=\widetilde{N}_{m}\frac{\partial^{m}}{\partial\tau^{m}}\frac{1}{1-\tau}\exp\left[\frac{1}{2}\cos^{2}\theta \tanh rC_{\tau}^{2}\hat{a}^{\dagger2}\right]\left|0\right\rangle |_{\tau=0},
\end{equation}
where $\widetilde{N}_{m}^{-2}=\mathcal{D}_{m}\left(\frac{\triangle_{t,\tau}}{\sqrt{1-B^{2}A_{t}^{2}C_{\tau}^{2}}}\right)|_{t,\tau=0}$ is coefficients associated with the normalization constant of catalytic squeezed state with $A_{t}=\left(\frac{1-t\sec^{2}\theta}{1-t}\right)$, $B=\cos^{2}\theta \tanh r$, $C_{\tau}=\frac{1-\tau \sec^{2}\theta}{1-\tau}$, $\mathcal{D}_{m}=\frac{\partial^{m}}{\partial t^{m}}\frac{\partial^{m}}{\partial\tau^{m}}\left(\text{.}\right)|_{t=\tau=0}$ and $\triangle_{t,\tau}=\frac{1}{1-t}\frac{1}{1-\tau}$.
Meanwhile, we can obtain the normalization constant of $\left|\Psi_{cws}\right\rangle$ as
\begin{equation}
	\begin{aligned}
		\mathcal{N}_{4}&=\left[\left(d+1\right)\left(1+d\widetilde{N}_{m}^{2}m!^{2}\right)\right]^{-\frac{1}{2}}.
	\end{aligned}
\end{equation}

Following the same catalytic protocol in the previous section, we implement quantum catalysis on the $\left|\Psi_{\mathrm{ws}}\right\rangle$ given by Eq. (\ref{eq:29}). Then  the target state $\left|\Psi_{\mathrm{cws}}\right\rangle$ can be produced with the following success probability (see Eq. (\ref{eq:B8}) in Appendix \ref{B_HEP})
\begin{equation}\label{eq:33}
	\begin{aligned}
		P_{\mathrm{cws}}&=\mathcal{N}_{3}^{2}N_{m}^{'-2}\mathcal{N}_{4}^{-2}\left(\cos\theta\right)^{2md}.
	\end{aligned}
\end{equation}
To investigate the parameter conditions that lead to a higher success probability, Fig.~\ref{fig:7} illustrates the relationship between the success probability, the number of input resources $N$, and the projection coefficient $\theta$ when the number of quantum network nodes $d=5$. It can be concluded that the success probability decreases with the increase of $ \theta$ and the number of catalytic photons $m$. 
%Furthermore, we find that the success probability of this scheme is lower than that of the multimode catalyzed W-type coherent state scheme under the same parameter conditions.

In addition, the total number of resources input to the DQN can also be calculated by the Eq.~(\ref{Eq.13}). On the one hand, for multimode W-type squeezed state $\left|\Psi_{\mathrm{ws}}\right\rangle$, the excited state is $\left|\psi\right\rangle=\left|r\right\rangle$, we get $\bar{N}_{\mathrm{ws}}=\mathcal{N}_{3}^{2}\left(d+1\right)\sinh^{2}r$ . On the other hand, for multiphoton globally catalyzed multimode W-type squeezed state $\left|\Psi_{\mathrm{cws}}\right\rangle$, the excited state is $\left|\psi\right\rangle=\left|\psi_{s}'\right\rangle$, the average number of photons can be represent as
\begin{equation}
	\begin{aligned}
		\bar{N}_{\mathrm{cws}}&=\mathcal{N}_{4}^{2}\left(d+1\right)\left[\widetilde{N}_{m}^{2}\mathcal{D}_{m}\left(\frac{\triangle_{t,\tau}}{\left(1-B^{2}A_{t}^{2}C_{\tau}^{2}\right)^{3/2}}\right)-1\right].
	\end{aligned}
\end{equation}
However, the average photon number after the catalytic operation varies under different catalytic conditions, including the catalytic photon number $m$, the projection coefficient $\theta$, and the total input resources $N$. 
Fig.~\ref{fig:8} (a) illustrates the dependence of $\bar{N}_{\mathrm{cws}}$ on $\theta $ for given $N=1$ (corresponding  to a squeezing parameter $r=1.5501$), i.e., a squezzed level is $13.47$~dB. Notably, squeezed states with a squeezing level up to $15$~dB have already been demonstrated in experiments~\cite{vahlbruch2016detection}, confirming the experimental feasibility of our setting. From Fig.~\ref{fig:8} (a), it is observed that $\bar{N}_{\mathrm{cws}}$ oscillates with 
$\theta$ while generally increasing with the catalysis photon number $m$ across most $\theta$ values. A comparison indicates that the $\bar{N}_{\mathrm{cws}}$ corresponding to $m > 1$ is greater than the $\bar{N}_{\mathrm{ws}}$ when $\theta$ takes most values and as $m$ increasing, the region of $\bar{N}_{\mathrm{cws}}>\bar{N}_{\mathrm{ws}}$ expanding. 
Fig.~\ref{fig:8} (b) presents the average photon number $\bar{N}_{\mathrm{cws}}$ as a funcation of $N$ for fixed $\theta=\frac{\pi}{3}$.  We find that $\bar{N}_{\mathrm{cws}}$ grows with the catalysis photon number $m$ and the input photon number $N$, gradually approaches saturation at large $N$, and for multiphoton catalysis ($m>1$), it surpasses $\bar{N}_{\mathrm{ws}}$ across the entire $N$-range. 

%%%%%%%%%%%%%%%%%%%%%%%%%%%
\begin{figure}[t]
	\centering
	\includegraphics[width=1.0\columnwidth]{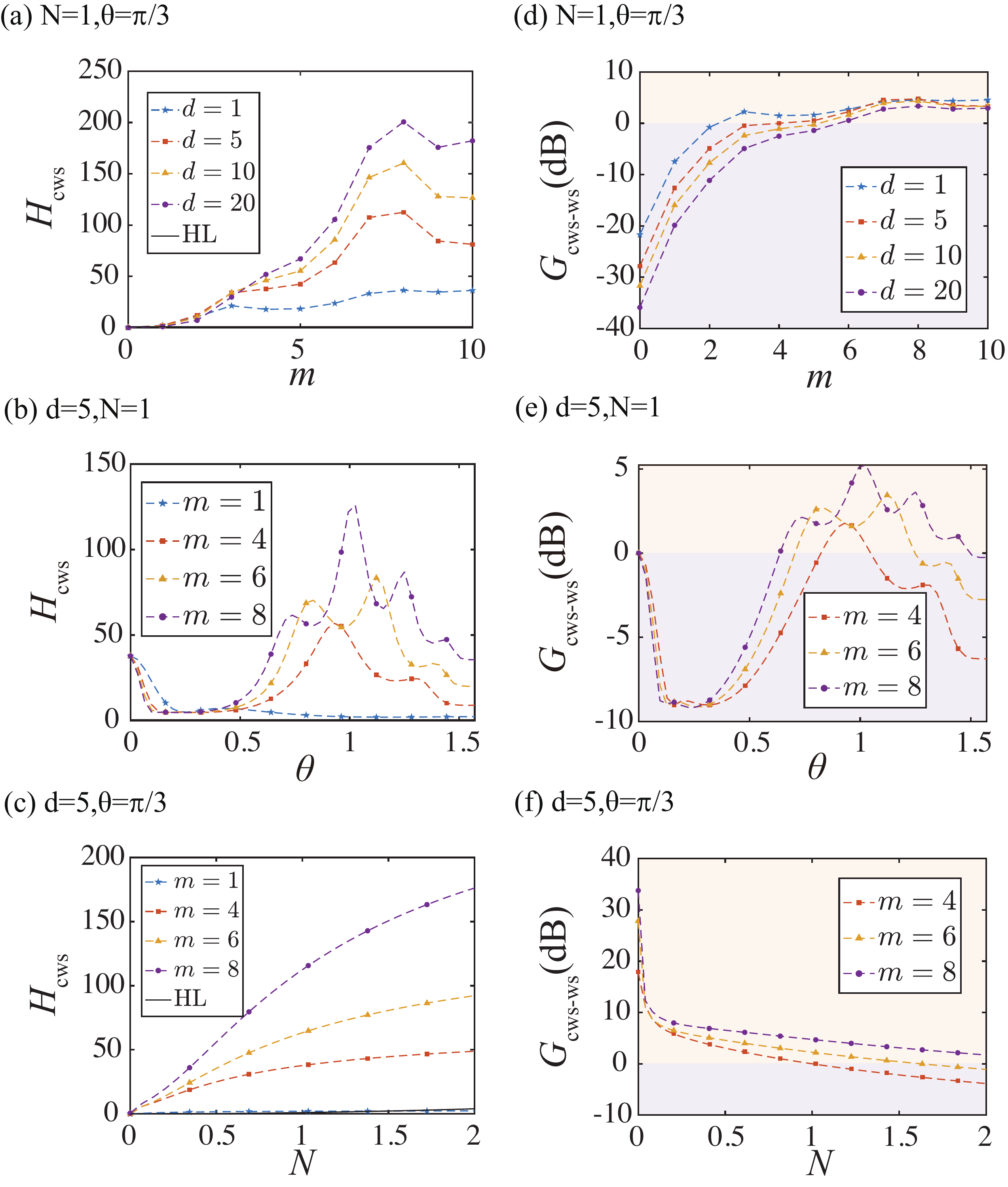}
	\caption{Effective QFI $H_{\mathrm{cws}}$
		(left panels, a–c) and corresponding gain factor $G_{\mathrm{cws-ws}}$(right panels, d–f) as functions of key system parameters. (a) $H_{\mathrm{cws}}$ and (d) $G_{\mathrm{cws-ws}}$ dependence on the number of catalytic photons $m$ for different network nodes $d=1,5,10,20$, with fixed input resource $N=1$ and beam splitter transmissivity $\theta=\frac{\pi}{3}$. (b)$H_{\mathrm{cws}}$ and (e) $G_{\mathrm{cws-ws}}$ dependence on beam splitter paramete $\theta$ for $m=1,4,6,8$ and  $m=4,6,8$, respectively, with fixed $d=5$ and $N=1$. (c) $H_{\mathrm{cws}}$ and (f) $G_{\mathrm{cws-ws}}$ dependence on input resource $N$ for $m=1,4,6,8$ and for $m=4,6,8$, with fixed $d=5$ and $\theta=\frac{\pi}{3}$. The black solid line in (a–c) indicates the Heisenberg limit. The yellow shaded area indicates $\left|\Psi_{\mathrm{cws}}\right\rangle$ exhibits better estimation precision than $\left|\Psi_{\mathrm{ws}}\right\rangle$.\label{fig:9}}
\end{figure}
%%%%%%%%%%%%%%%%%%%%%%%%%%%%%%%

We turn to explore the sensing performance of the DQN phase sensing scheme with three types of quantum resources for the case of the global quantum catalysis. Making use of Eq. (\ref{eq:13}) we can obtain the effective QFI of  the W-type squeezed state $\left|\Psi_{\mathrm{ws}}\right\rangle$ in Eq. (\ref{eq:29}) with the expression
\begin{equation}\label{eq.34}
	\begin{aligned}
		H_{\mathrm{ws}}&=4\mathcal{N}_{3}^{2}\left[\left(3d-\mathcal{N}_{3}^{2}d^{2}\right)\sinh^{4}r+2d\sinh^{2}r\right],
	\end{aligned}
\end{equation}
where we have used meanvalues $\left\langle \hat{n}\right\rangle=\sinh^{2}r$ and  $\left\langle \hat{n}^{2}\right\rangle =\sinh^{2}r\left(3\sinh^{2}r+2\right)$. 
%In the small-squeezing regime $(r\ll1)$ holds, the relation $\bar{N}_{ws}\approx \frac{r^{2}}{(d+1)}$ holds, leading to an effective QFI of $H_{ws}\approx \frac{4dN_{ws}}{d+1}[\frac{3(d+1)^2-d}{d+1}N_{ws}+2]$. Since $d\ge1$, this expression confirms that $H_{ws}$ surpasses the HL. Conversely, in the strong-squeezing regime $(r\gg1)$, we have $\bar{N}_{ws}\approx\frac{1}{4}e^{-2r}$, which again leads to $H_{ws}=\frac{4d}{d+1}[\frac{(2d+3)\bar{N}_{ws}}{d+1}+2]$. Thus, regardless of the squeezing strength, the effective QFI maintains a scaling that surpasses the HL, provided the network nodes satisfies $d\ge1$.
%when $d\to\infty$, we find that the inequality $\frac{\left\langle \hat{n}^{2}\right\rangle}{\left\langle \hat{n}\right\rangle}>\frac{1}{4}+\bar{N}$ holds only for $r>0.288$, implying that the effective QFI exceeds the Heisenberg scaling when squeezing exceeds $2.5$dB. In fact, squeezed states with a squeezing level of $15$dB have been successfully generated experimentally~\cite{vahlbruch2016detection}, demonstrating that this condition is experimentally feasible.

The effective QFI for the globally catalyzed  W-type squeezed state $\left|\Psi_{\mathrm{cws}}\right\rangle$ given by Eq. (\ref{eq:30})  is found to be  the following expression  (see Eq. (\ref{eq:C10}) in the Appendix \ref{C_HEP})
\begin{equation}\label{eq.34}
	\begin{aligned}
		H_{\mathrm{cws}}=&4\mathcal{N}_{4}^{2}d\left[\widetilde{N}^{2}\mathcal{D}_{m}\left(\frac{\triangle_{t,\tau}h_{t,\tau}}{g_{t,\tau}^{5/2}}\right)-3\widetilde{N}_{m}^{2}\mathcal{D}_{m}\left(\frac{\triangle_{t,\tau}}{g_{t,\tau}^{3/2}}\right)+1\right],\\
		&-4\mathcal{N}_{4}^{4}d^{2}\left[\widetilde{N}_{m}^{2}\mathcal{D}_{m}\left(\frac{\triangle_{t,\tau}}{g_{t,\tau}^{3/2}}\right)-1\right]^{2},
	\end{aligned}
\end{equation}
where we have defined $h_{t,\tau}=2+B^{2}C_{\tau}^{2}A_{t}^{2}$ and $g_{t,\tau}=1-B^{2}A_{t}^{2}C_{\tau}^{2}$. In addition to tuning the input resource $N$ and the network size $d$, the catalysis-related parameters $\theta$ and $m$ can also be adjusted to optimize the estimation precision. As shown in Fig.~\ref{fig:9} (a-b) for fixed $N=1$ and $\theta=\pi/3$, the effective QFI $H_{\mathrm{cws}}$ increases with the network size $d$. This indicates that $\left|\Psi_{\mathrm{cws}}\right\rangle$ is particularly advantageous for large-scale networks,  and implies that more quantum resource synergies are required for larger networked systems. Meanwhile, for each $d$, there exists an optimal catalytic photon number $m_{\mathrm{opt}}$, which is found to be $8$ for all considered cases, and $H_{\mathrm{cws}}$ remains above the HL for all considered $d$ and $m$. For fixed $d=5$ and $N=1$, $H_{\mathrm{cws}}$ exhibits an oscillatory dependence on $\theta$, and its maximum value increases with $m$, while staying above the HL throughout the full $\theta$ range. In addition, for fixed $d=5$ and $\theta=\pi/3$, $H_{\mathrm{cws}}$ increases with both $N$ and $m$, and multiphoton catalysis ($m>1$) enables HL-beating performance over the entire $N$ range.

The variations of $H_{\mathrm{cws}}$ with $N$, $\theta$, and $m$ are found to be largely consistent with those of $N_{\mathrm{cws}}$, indicating that the behavior of $H_{\mathrm{cws}}$ is predominantly determined by $N_{\mathrm{cws}}$. The small residual differences between them may be attributed to the effect of the weighting factor.

According to weak QCRB, the higher the effective QFI, the better the sennsitivity. So we compare the effective \textcolor{red}{QFI} with and without quantum catalysis through the gain factor
\begin{equation}
	\begin{aligned}
		G_{\mathrm{cws-ws}}=10\log\left(\frac{H_{\mathrm{cws}}}{H_{\mathrm{ws}}}\right),
	\end{aligned}
\end{equation}
which indicates that if the gain greater than zero, we can say that the $\left|\Psi_{\mathrm{cws}}\right\rangle$ has better sensing performance for simultaneous estimation of multiple phases than $\left|\Psi_{\mathrm{ws}}\right\rangle$. This also implies that the sensing scheme using three quantum resources (catalysis, entanglement, squeezing) has better sensitivity than that using two quantum resources (entanglement, squeezing).
%(Note that, the same squeezed state is employed in this case, except that the former has been catalyzed).

Fig.~\ref{fig:9} (d)–(f) show the gain factor $G_{\mathrm{cws-ws}}$ as functions of the catalytic parameters ($m$, $\theta$) and the input resource $N$, respectively. As shown in Fig.~\ref{fig:9} (d), for all considered network sizes $d$, the gain first increases with $m$ and then decreases slightly, reaching its maximum at $m=8$. The corresponding peak gains are $4.57$~dB, $4.74$~dB, $4.33$~dB, and $3.35$~dB for $d=1,5,10,$ and $20$, respectively. As $d$ increases, the overall gain decreases, and the threshold catalytic photon number required to obtain a positive gain also becomes larger. 
Fig.~\ref{fig:9}(e) shows that, for fixed $d=5$ and $N=1$, $G_{\mathrm{cws-ws}}$ oscillates with the BS parameter $\theta$, with positive gain appearing only in suitable $\theta$ windows and maxima at $(\theta_{\mathrm{opt}},G_{\max})=(0.94,1.78~\mathrm{dB})$, $(1.12,3.44~\mathrm{dB})$, and $(1.02,5.23~\mathrm{dB})$ for $m=4,6,8$, respectively; moreover, increasing $m$ generally enlarges the positive-gain region and the optimal gain.
In Fig.~\ref{fig:9} (f), with $d=5$ and $\theta=\pi/3$ fixed, the gain decreases with increasing $N$, while the largest gain appears in the low-energy regime. In particular, for $m=8$, the gain reaches as high as $33.77$~dB. Moreover, as $m$ increases, the region where positive gain exists becomes larger.

%These results indicate that the sensitivity advantage introduced by quantum catalysis based on compression and entanglement is significant

These results show that the sensitivity advantage of the three-resource DQN phase-sensing scheme over the two-resource scheme can be enhanced by tuning the catalysis-related parameters, with the largest enhancement occurring for an optimal $m$, a suitable $\theta$, and low input energy $N$. 
The decrease of the gain with increasing $d$ reflects the distribution of available resources over more network nodes, which weakens the relative catalytic advantage. For fixed $N$ and $d$, since $H_{\mathrm{ws}}$ is independent of $m$ and $\theta$, the optimal-$m$ and $\theta$-window behaviors of $G_{\mathrm{cws-ws}}$ directly follow from the enhancement of $H_{\mathrm{cws}}$, arising from increased effective photon-number resources and catalysis-induced non-Gaussianity.
As $N$ increases, the saturation of $\bar N_{\mathrm{cws}}$ relative to the growing $\bar N_{\mathrm{ws}}$ reduces the catalytic resource advantage. Thus, the decrease of $G_{\mathrm{cws-ws}}$ reflects a crossover from catalysis-induced non-Gaussian enhancement at low $N$ to the intrinsic squeezing, and entanglement of the original W-type squeezed probe at high $N$.

%%%%%%%%%%%%%%%%%%%%%%%%%%%%%%
\begin{figure}[t]
	\centering
	\includegraphics[width=1.0\columnwidth]{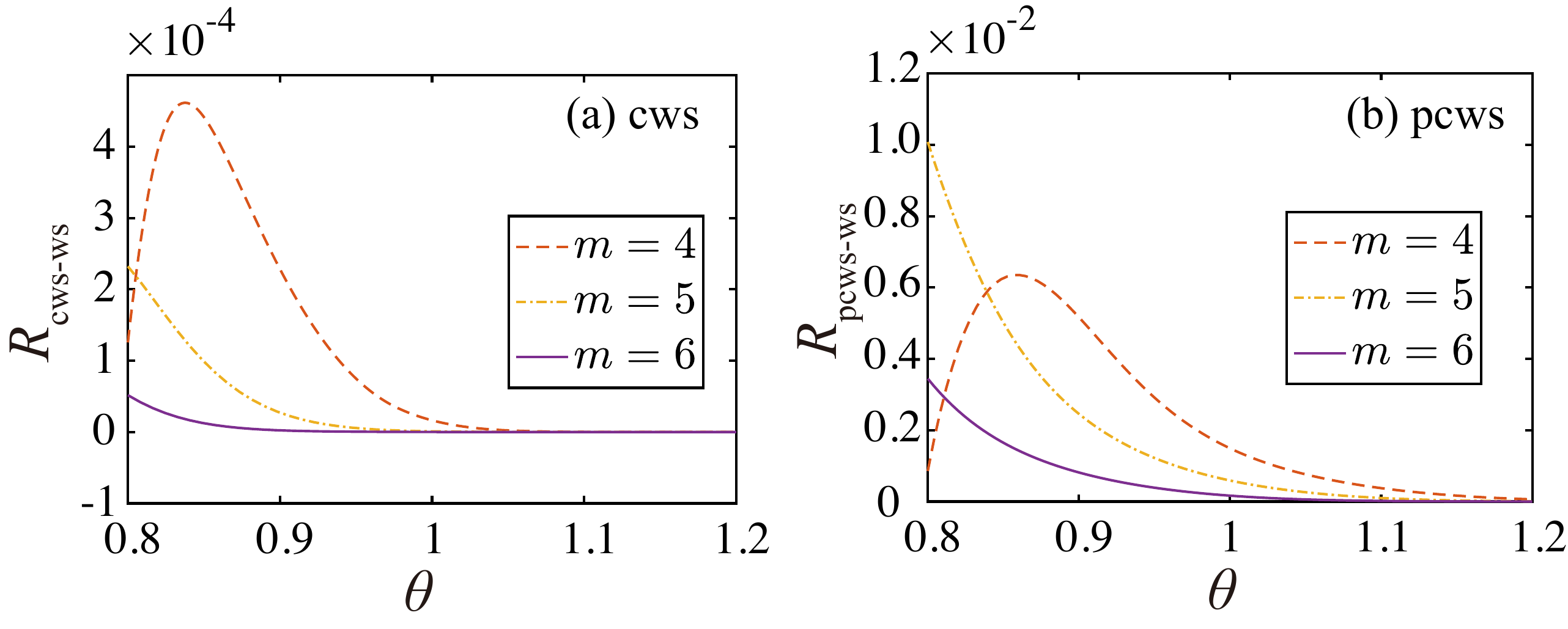}
	\caption{(a) The cooperation factor for global catalysis $R_{\mathrm{cws-ws}}$  as a function of the transmissivity coefficient $\theta$ for different catalytic photons $m=4,5,6$ for given $N=1$ and $d=2$ and (b) The cooperation factor for partial catalysis $R_{\mathrm{pcws-ws}}$ as a function of $\theta$  for $m=4,5,6$ when $N=1$, $d=2$, and $s=1$.\label{fig:10}}
\end{figure}
%%%%%%%%%%%%%%%%%%%%%%%%%%%%%%%

Then, we caculate the cooperation factor to measure  the joint effect of gain and success probability, as follows
\begin{equation}
	\begin{aligned}
		R_{\mathrm{cws-ws}}&=\bigtriangleup H_{\mathrm{cws-ws}}\times P_{\mathrm{cws}}.
	\end{aligned}
\end{equation} 
where $\bigtriangleup H_{\mathrm{cws-ws}}=H_{\mathrm{cws}}-H_{\mathrm{ws}}$ and $P_{\mathrm{cws}}$ is the success probability. Fig.~\ref{fig:10} (a) shows the cooperation factor $R_{\mathrm{cws-ws}}$ under global catalysis as a function of the beam-splitter parameter $\theta$ for different catalytic photon numbers $m$, with $N=1$ and $d=2$. It is found that $R_{\mathrm{cws-ws}}$ remains positive over the plotted range, indicating that $\left|\Psi_{\mathrm{cws}}\right\rangle$ still retains an advantage over $\left|\Psi_{\mathrm{ws}}\right\rangle$ when the success probability is taken into account. For all considered $m$, $R_{\mathrm{cws-ws}}$ first increases and then decreases with $\theta$, reaching a maximum at an optimal $\theta$, which indicates the existence of an optimal parameter window that balances metrological enhancement and success probability. As $m$ increases, catalysis yields a larger QFI enhancement, but at the cost of a faster reduction in the success probability, thereby suppressing the $R_{\mathrm{cwc-wc}}$ performance. This reveals a trade-off between sensing enhancement and success probability, and suggests that effective sensing gain can be achieved by properly choosing $m$ and $\theta$ while maintaining a relatively high success probability. Comparing Fig.~\ref{fig:10} (a) with  Fig.~\ref{fig:5} (a), it can be seen that the DQN sensing scheme using three quantum resources improves the cooperation coefficient of global catalysis by four orders of magnitude compared with that using two quantum resources.

We now discuss sensing advantage of quantum squeezing through making a comparison of the sensing performance between $\left|\Psi_{\mathrm{cws}}\right\rangle$ with $\left|\Psi_{\mathrm{cwc}}\right\rangle$.
Under the same input quantum resources $\bar{N}_{\mathrm{wc}}=\bar{N}_{\mathrm{ws}}$, we can also compare the effective QFI of $\left|\Psi_{\mathrm{cws}}\right\rangle$ with $\left|\Psi_{\mathrm{cwc}}\right\rangle$ to evaluate whether a quantum probe utilizing three quantum resources (catalysis, entanglement, and squeezing) outperforms one utilizing only two resources (catalysis and entanglement) for distributed phase sensing in DQN. So we compare the gain factor  $G_{\mathrm{cws-cwc}}=10\log\left(\frac{H_{\mathrm{cws}}}{H_{\mathrm{cwc}}}\right)$, if the gain greater than $0$, we can say that using three quantum resources have better sensing performance than probes using two quantum resources.

%%%%%%%%%%%%%%%%%%%%%%%%%%%%%%
\begin{figure}[t]
	\centering
	\includegraphics[width=0.78\columnwidth]{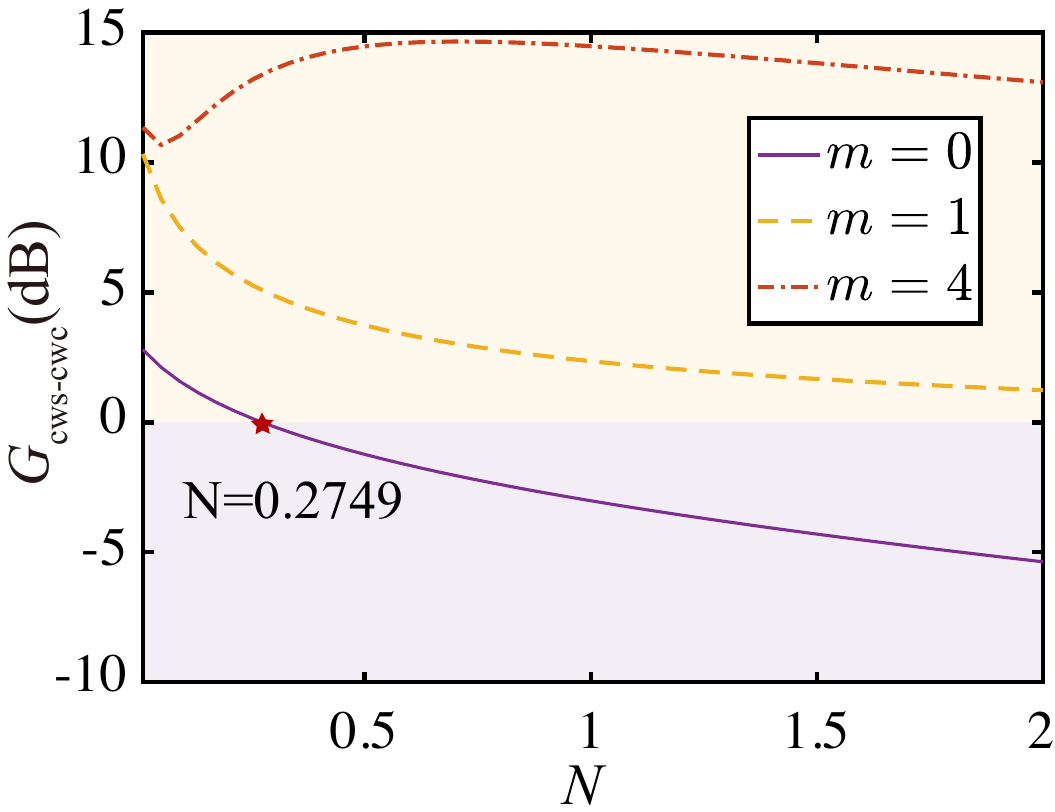}
	\caption {Sensitivity gain factor $G_{\mathrm{cws-cwc}}$ as a function of the input resource number $N$ for different catalytic photons $m=0,1,4$ when the BS transmittance $\theta=\pi/3$ and the network node number $d=5$. The yellow shaded area indicates $\left|\Psi_{\mathrm{cws}}\right\rangle$ exhibits better estimation precision than $\left|\Psi_{\mathrm{cwc}}\right\rangle$. \label{fig:11}} 
\end{figure}
%%%%%%%%%%%%%%%%%%%%%%%%%%%%%%%
Figure~\ref{fig:11} shows that the gain factor $G_{\mathrm{cws-cwc}}$ depends strongly on both the input resource $N$ and the catalytic photon number $m$. For fixed $\theta=\pi/3$ and $d=5$, vacuum-photon catalysis ($m=0$) leads to a monotonic decrease of $G_{\mathrm{cws-cwc}}$ with increasing $N$, and the gain becomes negative when $N\geq 0.2749$, indicating that $\left|\Psi_{\mathrm{cws}}\right\rangle$ performs worse than $\left|\Psi_{\mathrm{cwc}}\right\rangle$ in this regime. By contrast, for $m\geq 1$, $G_{\mathrm{cws-cwc}}$ remains positive throughout the plotted range of $N$, showing that multiphoton catalysis enables $\left|\Psi_{\mathrm{cws}}\right\rangle$ to consistently outperform $\left|\Psi_{\mathrm{cwc}}\right\rangle$. Moreover, the gain is significantly enhanced as $m$ increases. The maximum gain also increases with the catalytic photon number, reaching $10.33$~dB, and $14.47$~dB, for $m=1,4$ respectively. Overall, these results demonstrate that, over most of the considered parameter range, the $\left|\Psi_{\mathrm{cws}}\right\rangle$ scheme achieves a clear sensitivity advantage over the $\left|\Psi_{\mathrm{cwc}}\right\rangle$ scheme.

Physically, this sensitivity advantage is consistent with the enhancement mechanisms revealed in Fig.~\ref{fig:4} (d-f) and Fig.~\ref{fig:9} (d-f). Since $\left|\Psi_{\mathrm{cws}}\right\rangle$ combines squeezing, catalysis, and entanglement, while $\left|\Psi_{\mathrm{cwc}}\right\rangle$ only involves catalysis and entanglement, the additional squeezing resource in $\left|\Psi_{\mathrm{cws}}\right\rangle$ can cooperate with the catalysis-induced non-Gaussianity under multiphoton catalysis, leading to a stronger cooperative enhancement and hence a larger sensitivity gain over $\left|\Psi_{\mathrm{cwc}}\right\rangle$.

\subsection{The case of partial Quantum Catalysis}

In this subsection, we study the sensing performance of of the DQN phase sensing scheme using three quantum resources under the condition of partial quantum catalysis. We assume that  only first $s+1$ modes of $d+1$ modes in the quanutm probe are catalyzed,  the remaining $d-s$ modes are not catalyzed. In this case,  the quantum probe state  after partial quantum catalysis can be written as follows
\begin{equation}\label{eq:39}
	\begin{aligned}
		\left|\Psi_{\mathrm{pcws}}\right\rangle=&\mathcal{N}_{3}'\big(\sum_{m=0}^{s}\left|0\right\rangle _{0}\left|0\right\rangle _{1}...\left|\psi_{s}'\right\rangle _{m}...\left|0\right\rangle _{d},\\
		&+\sum_{m=s+1}^{d}\left|0\right\rangle _{0}\left|0\right\rangle _{1}...\left|r\right\rangle _{m}...\left|0\right\rangle _{d}\big),
	\end{aligned}
\end{equation} 
where $\left|\psi_{s}'\right\rangle$ is defined by Eq. (\ref{eq:31}), $\mathcal{N}_{3}'$ is the normalization constant given by
\begin{equation}
	\begin{aligned}
		\mathcal{N}_{3}^{'-2}=&\left(s+1\right)\left[1+s\widetilde{N}_{m}^{2}m!^{2}+\left(d-s\right)\widetilde{N}_{m}m!\sqrt{r}\right]\\
		&+\left(d-s\right)\left[1+\left(d-s-1\right)\operatorname{sech}r+\left(s+1\right)\widetilde{N}_{m}m!\sqrt{\operatorname{sech}r}\right].
	\end{aligned}
\end{equation} 

Making use of Eq.~(\ref{eq:27}), we can obtain the quantum resources in the quantum probe state given by Eq. (\ref{eq:39})  as
\begin{equation}
	\begin{aligned}
		\bar{N}_{\mathrm{pcws}}=&\mathcal{N}_{3}^{'2}\left(s+1\right)\left(\widetilde{N}_{m}^{2}\mathcal{D}_{m}\left[\frac{\triangle_{t,\tau}}{\left(1-B^{2}A_{t}^{2}C_{\tau}^{2}\right)^{3/2}}\right]-1\right)\\&+\mathcal{N}_{3}^{'2}\left(d-s\right)\sinh^{2}r.
	\end{aligned}
\end{equation} 

Starting with the quantum probe state given by Eq.~(\ref{eq:39}),   we obtain the effective QFI under the condition of partial catalysis as follows
\begin{equation}
	\begin{aligned}
		H_{\mathrm{pcws}}=&4\mathcal{N}_{3}^{'2}s\left[\widetilde{N}_{m}^{2}\mathcal{D}_{m}\left(\frac{\triangle_{t,\tau}h_{t,\tau}}{g_{t,\tau}^{5/2}}\right)-3\widetilde{N}_{m}^{2}\mathcal{D}_{m}\left(\frac{\triangle_{t,\tau}}{g_{t,\tau}^{3/2}}\right)+1\right]\\&+4\mathcal{N}_{3}^{'2}\left(d-s\right)\sinh^{2}r\left(3\sinh^{2}r+2\right)\\&-4\mathcal{N}_{3}^{'4}\left\{ s\left[\widetilde{N}_{m}^{2}\mathcal{D}_{m}\left(\frac{\triangle_{t,\tau}}{g_{t,\tau}^{3/2}}\right)-1\right]+\left(d-s\right)\sinh^{2}r\right\} ^{2}.
	\end{aligned}
\end{equation} 

\begin{figure}[t]
	\centering
	\includegraphics[width=0.78\columnwidth]{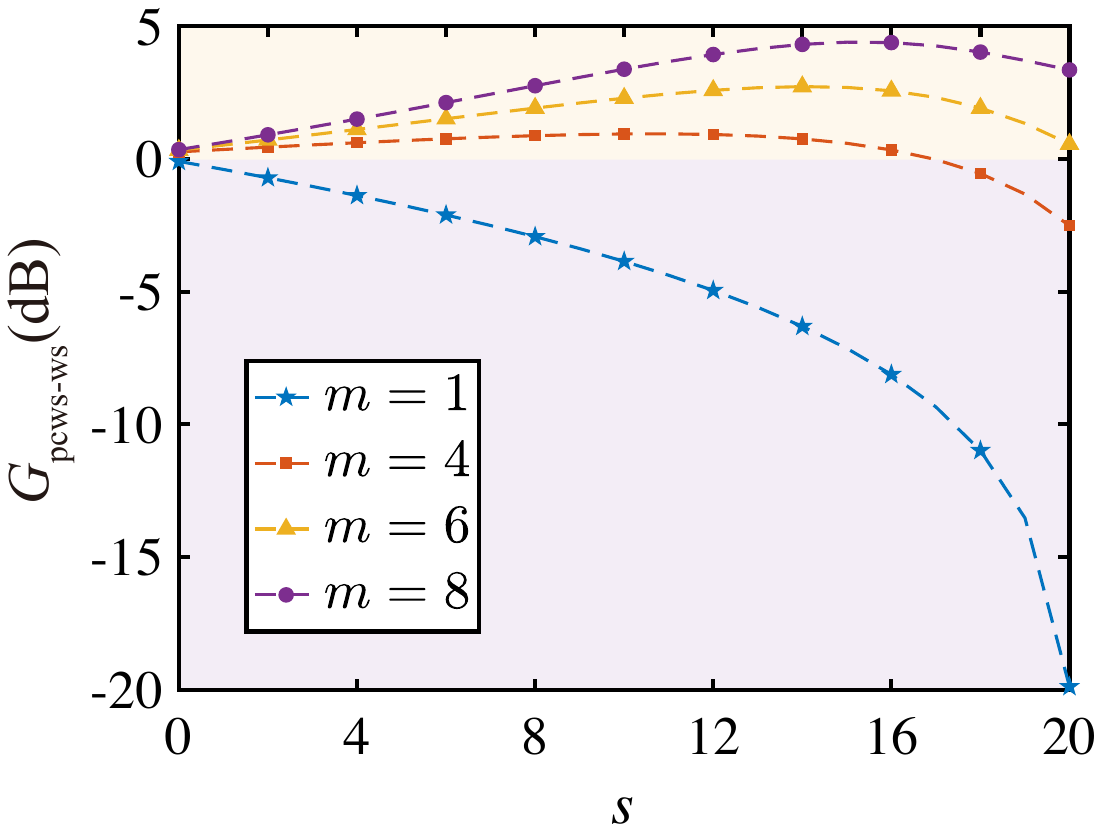}
	\caption{Sensitivity gain factor $G_{\mathrm{pcws-ws}}$ as a funcation of the number of catalytic modes $s$ for different catalytic photons $m=1,4,6,8$ when the BS transmittance $\theta=\pi/3$, the input resource number $N=1$ and the network node number  $d=20$. The yellow shaded area indicates $\left|\Psi_{\mathrm{pcws}}\right\rangle$ exhibits better estimation precision than $\left|\Psi_{\mathrm{ws}}\right\rangle$.\label{fig:12}}
\end{figure}

In order to evaluate the quantum advantage of the partial analysis scheme over the scheme  without catalysis, we discuss the gain factor $G_{\mathrm{pcws-ws}}=10\log\left(\frac{H_{\mathrm{pcws}}}{H_{\mathrm{ws}}}\right)$. If the ratio is greater than zero, it means that only $s$ modes catalyzed can obtain a better sensitivity than without catalyzing. Note that, to ensure the same input resources, when we make the comparison, we use the same squeezed state. In Fig.~\ref{fig:12}, we plot the gain factor $G_{\mathrm{pcws-ws}}$ for $d=20$ with respect to the modes of catalysis $s$ for given $N=1$ and $\theta=\frac{\pi}{3}$. The results show that for multiphoton catalysis scenario ($m>1$), there exists an optimal catalytic mode in which the gain factor reaches maximum. When $m=4$, $m=6$ and $m=8$, the optimal catalytic modes are $11$, $14$ and $15$, respectively, which implies that only a specific number of modes need to be catalyzed to achieve the maximum gain.  Additionally, we observe that as the number of catalyzing photons increases, the improvement in sensitivity at optimal catalyzed mode number continues to enhance. Furthermore, it is observed that as the number of modes increases further, the catalytic case instead performs worse, which indicates that partial catalysis in the optimal catalytic mode is superior to global catalysis.

%%%%%%%%%%%%%%%%%%%%%%%%%%%%%%
%%%%%%%%%%%%%%%%%%%%%%%%%%%%%%%
In order to solve the defect of low cooperation coefficient caused by global catalysis, we can further improve the cooperation coefficient by truncating the quantum state of $d+1$-mode at $s=1$ in catalytic DQN for given $d=2$. Based on this,  we discuss the trade-off between "existence gain" and "success probability" in partial catalysis
\begin{equation}
	\begin{aligned}
		R_{\mathrm{pcws-ws}}&=\Delta{H}_{\mathrm{pcws-ws}}\times P_{\mathrm{pcws}} ,
		\end{aligned}
\end{equation}
where $\Delta{H}_{\mathrm{pcws-ws}}=(H_{\mathrm{pcws}}-H_{\mathrm{ws}})$ represents the sensitivity gain and $P_{\mathrm{pcws}}$ represents the probability of successful preparation, which can be obtain as followed (see Eq. (\ref{eq:B9}) in Appendix \ref{B_HEP})
\begin{equation}\label{eq:44}
	\begin{aligned}
		P_{\mathrm{pcws}}=&\mathcal{N}_{3}^{2}\Big\{\left(s+1\right)\left[\left(N_{m}'\right)^{-2}+d\operatorname{sech} r\left(\cos\theta\right)^{2m}\right]\\&+\left(d-s\right)\left(\cos\theta\right)^{2m}\left(1+d\operatorname{sech} r\right)\Big\}.
		\end{aligned}
\end{equation}

Fig.~\ref{fig:10} (b), the cooperation coefficient under partial catalysis is significantly larger than that under global catalysis over the whole plotted range, indicating that partial catalysis is much more effective in enhancing the cooperative performance of the network. Similar to the global-catalysis case, $R_{\mathrm{pcws-ws}}$ generally decreases with increasing $\theta$.
%and the $m=5$ and $m=6$ curves are monotonic. For $m=4$, the curve again shows a pronounced maximum, revealing an optimal $\theta$ for achieving the strongest cooperative enhancement. 
Moreover, although the cooperation coefficient still decreases with increasing $m$, its value under partial catalysis remains two orders of magnitude higher than that under global catalysis, which clearly demonstrates the superiority of partial catalysis for distributed phase estimation in catalyzed DQNs.
\section {The practical measurement scheme}\label{section 4}
%%%%%%%%%%%%%%%%

For our ($d+1$)-mode DQN sensing scheme shown in Fig.~\ref{fig:1}, an additional linear optical network $\hat{V}$ is introduced before the measurement stage in order to efficiently extract the phase information to be estimated. Here we consider one reference mode and 
$d$ signal modes: after encoding phase the reference mode $\hat{a}_{0}'=\hat{a}_{0}$, while the remaining $d$-modes are signal modes $\hat{a}_{j}'=\hat{a}_{j}e^{-i\phi_{j}}$   $(j=1,...,d)$. Accordingly, the target parameter is defined as the average phase $\bar{\phi}=\frac{1}{d}\sum_{j=1}^{d}\phi_{j}$. To estimate this quantity, the phase information distributed among the signal modes must be recombined into a single output mode for subsequent readout.

For this purpose, we choose a linear network with $d+1$ ports as the readout network described by the unitary matrix~\cite{Vourdas2005Fourier}
\begin{equation}
	\begin{aligned}
		V_{kj}=\frac{1}{\sqrt{d+1}}\exp \left(\frac{2\pi i}{d+1}kj\right), ~~ ~~~~~k,j=0,1,...,d.
	\end{aligned}
\end{equation} 
which leads to the corresponding mode transformation   $\hat{c}_{k}=\frac{1}{\sqrt{d+1}}\sum_{j=0}^{d}\exp\left(\frac{2\pi i}{d+1}kj\right)\hat{a}_{j}'$.
This transformation is lossless and preserves the total photon number due to  $V^{\dagger}V=I$. Experimentally, a ($d+1$)-port linear network is a standard lossless multiport linear optical network that can be realized by cascading beam splitters and phase shifters. In particular, when $d+1=2$ and $d+1=3$, they  reduce to the standard $50:50$ beam splitter and the tritter, respectively. The first output mode of the readout network is the $\hat{c}_{0}$ mode  given by
\begin{equation}
	\begin{aligned}
		\hat{c}_{0}&=\frac{1}{\sqrt{d+1}}\left(\hat{a}_{0}+\sum_{j=1}^{d}e^{-i\phi_{j}}\hat{a}_{j}\right),
	\end{aligned}
\end{equation} 
which is the equal-weight superposition of all input modes. In the symmetric estimation case, where all signal arms acquire the same phase $\phi_{j}=\phi$ for $j=1,...,d$.  The above expression can be written as  
\begin{equation}
	\begin{aligned}
		\hat{c}_{0}&=\frac{1}{\sqrt{d+1}}\left(\hat{a}_{0}+e^{-i\bar{\phi}}\sum_{j=1}^{d}\hat{a}_{j}\right),
	\end{aligned}
\end{equation} 
where $\bar{\phi}$ is  the average phase information defined by Eq. (\ref{eq:5}).
Therefore, the average phase information is concentrated into the single output mode $\hat{c}_{0}$, so that an effective estimation of $\bar{\phi}$ can be achieved by performing a subsequent measurement on this mode alone.

In what follows we study the  sensitivity of the average phase in the DQN phase sensing scheme using the homodyne measurement method. As an example to evaluate the  sensitivity in a practical measurement scheme, we calculate the  sensitivity of the average phase in the DQN phase sensing scheme involving two types of qauntum resources (quantum catalysis and entanglement).   In order to do this, we need to  know  the expectation value and fluctuation of the quadrature operator
\begin{equation}
	\begin{aligned}
		\hat{X}_{c_{0}}=\frac{1}{\sqrt{2}}\left(\hat{c}_{0}+\hat{c}_{0}^{\dagger}\right).
	\end{aligned}
\end{equation} 

For the DQN phase sensing scheme with the global quantum catalysis, the quantum probe state after quantum catalysis is  $\left|\Psi_{\mathrm{cwc}}\right\rangle$ given by Eq. (\ref{eq:10}). It is straightforward to calculate the expectation value of the quadrature operator $\hat{X}_{c_{0}}$ in the  global catalyzed quantum state  with the following result
\begin{equation}
	\begin{aligned}
		\left\langle \hat{X}^{(g)}_{c_{0}}\right\rangle =2\mathcal{N}_{d}\mathrm{Re}\left[(A_{10}+d\varepsilon_{c}\lambda_{c}^{\ast})\left(1+de^{-i\bar{\phi}}\right)\right]
	\end{aligned}
\end{equation} 
where we have set $\mathcal{N}_{d}=\frac{\mathcal{N}_{2}^{2}}{\sqrt{2\left(d+1\right)}}$, and $A_{10}=\left\langle \psi^{'}\right|\hat{a}\left|\psi^{'}\right\rangle$ which can be found in Appendix~\ref{C_HEP}.  $\lambda_{c}$ and $\varepsilon_{c}$ are given by
\begin{equation}
	\begin{aligned}
	\lambda_{c}&=\langle 0|\psi^{'}\rangle =N_{m}\left(\cos\theta\right)^{m}e^{-\frac{\left|\alpha\right|^{2}}{2}}, \\
	\varepsilon_{c}&=\langle 1|\psi^{'}\rangle=N_{m}\mathcal{L}_{m}\left(\tan^{2}\theta\right)\left(\cos\theta\right)^{1+m}e^{-\frac{\left|\alpha\right|^{2}}{2}}\alpha.
	\end{aligned}
\end{equation} 
Simmilarly, the expectation value of the operator $\left\langle \left(\hat{X}^{(g)}_{c_{0}}\right)^{2}\right\rangle$ in the $\left|\Psi_{cwc}\right\rangle$ can be obtained with the following expression
\begin{equation}
	\begin{aligned}
	\left\langle \left(\hat{X}^{(g)}_{c_{0}}\right)^{2}\right\rangle =\frac{1}{2}\left[\left\langle \hat{c}_{0}^{2}\right\rangle +\left\langle \hat{c}_{0}^{\dagger2}\right\rangle +\left\langle \hat{c}_{0}\hat{c}_{0}^{\dagger}\right\rangle +\left\langle \hat{c}_{0}^{\dagger}\hat{c}_{0}\right\rangle \right],
	\end{aligned}
\end{equation} 
where the relevant expectation value on the right-hand side are given by
\begin{equation}
	\begin{aligned}
			\left\langle \hat{c}_{0}^{2}\right\rangle &=\frac{\mathcal{N}_{2}^{2}\left(1+de^{-2i\bar{\phi}}\right)}{d+1}\left(A_{20}+\sqrt{2}d\lambda_{c}^{\ast}\omega_{c}\right),\\
			\left\langle \hat{c}_{0}^{\dagger2}\right\rangle &=\left(\left\langle \hat{c}_{0}^{2}\right\rangle\right)^{\ast},\\ 
			\left\langle \hat{c}_{0}^{\dagger}\hat{c}_{0}\right\rangle &=\mathcal{N}_{2}^{2}\left[A_{11}-1+\frac{d\left|\varepsilon_{c}\right|^{2}}{d+1}\left(2\cos\bar{\phi}+d-1\right)\right],\\
			\left\langle \hat{c}_{0}\hat{c}_{0}^{\dagger}\right\rangle &=1+\left\langle \hat{c}_{0}^{\dagger}\hat{c}_{0}\right\rangle
		\end{aligned}
	\end{equation} 
where $A_{20}=\left\langle \psi^{'}\right|\hat{a}^{2}\left|\psi^{'}\right\rangle$ which can be found in Appendix~\ref{C_HEP}. And $\omega_{c}$ is given by
\begin{equation}
	\begin{aligned}
		\omega_{c}=&\langle 2|\psi^{'}\rangle =\frac{N_{m}}{\sqrt{2}}\alpha^{2}e^{-\frac{\left|\alpha\right|^{2}}{2}}\mathcal{L}_{m}\left(2\tan^{2}\theta\right)\left(\cos\theta\right)^{2+m}.
	\end{aligned}
\end{equation} 

For the DQN phase sensing scheme with the partial quantum catalysis, the quantum probe state after quantum catalysis is  $\left|\Psi_{\mathrm{pcwc}}\right\rangle$ given by Eq. (\ref{eq:23}). The  relevant expectation values of the quadrature operator $\hat{X}_{c_{0}}$ in the  partial catalysis case can be obtained as follows
\begin{equation}
	\begin{aligned}
		\left\langle \hat{X}^{(p)}_{c_{0}}\right\rangle =&2\mathcal{N}'_{d}\mathrm{Re}\left[A_1+e^{-i\bar{\phi}}\left(sA_1+\left(d-s\right)A_2\right)\right],\\
	\end{aligned}
\end{equation} 
and
\begin{equation}
	\begin{aligned}
		\left\langle\left(\hat{X}^{(p)}_{c_{0}}\right)^{2}\right\rangle=&1+2\mathcal{N}'_{d}\Bigg\{
		\left(A_{6}+A_{7}\right)+\mathrm{Re}\Big[A_{3}+2e^{-i\bar{\phi}}A_{5}\\
		&+e^{-2i\bar{\phi}}\left(sA_{3}+\left(d-s\right)A_{4}\right)\Big]\Bigg\},
	\end{aligned}
\end{equation} 
where we we have set $\mathcal{N}'_{d}=\frac{\mathcal{N}_{1}^{'2}}{\sqrt{2\left(d+1\right)}}$ and introduced the notations
\begin{equation}
	\begin{aligned}
		A_{1}&=A_{10}+s\varepsilon_{c}\lambda_{c}^{\ast}+\left(d-s\right)\varepsilon_{c}e^{-\left|\alpha\right|^{2}/2},\\
		A_{2}&=\alpha\left[1+\left(d-s-1\right)e^{-\left|\alpha\right|^{2}}+\left(s+1\right)\lambda_{c}e^{-\left|\alpha\right|^{2}/2}\right],\\
		A_{3}&=A_{20}+\sqrt{2}s\lambda_{c}^{\ast}\omega_{c}+\sqrt{2}\left(d-s\right)\omega_{c}e^{-\left|\alpha\right|^{2}/2},\\
		A_{4}&=\alpha^{2}\left(1+\left(d-s-1\right)e^{-\left|\alpha\right|^{2}}+\left(s+1\right)\lambda_{c}^{\ast}e^{-\left|\alpha\right|^{2}/2}\right),\\
		A_{5}&=s\left|\varepsilon_{c}\right|^{2}+\left(d-s\right)\varepsilon_{c}^{\ast}\alpha e^{-\left|\alpha\right|^{2}/2},\\
		A_{6}&=\left(s+1\right)\left(A_{11}-1\right)+\left(d-s\right)\left|\alpha\right|^{2}+s\left(s-1\right)\left|\varepsilon_{c}\right|^{2},\\
		A_{7}&=\left(d-s\right)e^{-\left|\alpha\right|^{2}/2}\left[s\left(\alpha^{\ast}\varepsilon_{c}+\alpha\varepsilon_{c}^{\ast}\right)+\left|\alpha\right|^{2}\left(d-s-1\right)e^{-\left|\alpha\right|^{2}/2}\right].
	\end{aligned}
\end{equation} 

The average phase sensitivity in the homodyne measurement scheme for global and partial catalysis can be given by the error transfer equation
\begin{equation}
	\begin{aligned}
	\Delta \bar{\phi}=\frac{\sqrt{\left\langle \hat{X}_{c_{0}}^{2}\right\rangle -\left\langle \hat{X}_{c_{0}}\right\rangle ^{2}}}{\left|\partial\left\langle \hat{X}_{c_{0}}\right\rangle /\partial\bar{\phi}\right|}.
	\end{aligned}
\end{equation} 
%%%%%%%%%%%%%
\begin{figure}[t]
	\centering
	\includegraphics[width=1.0\columnwidth]{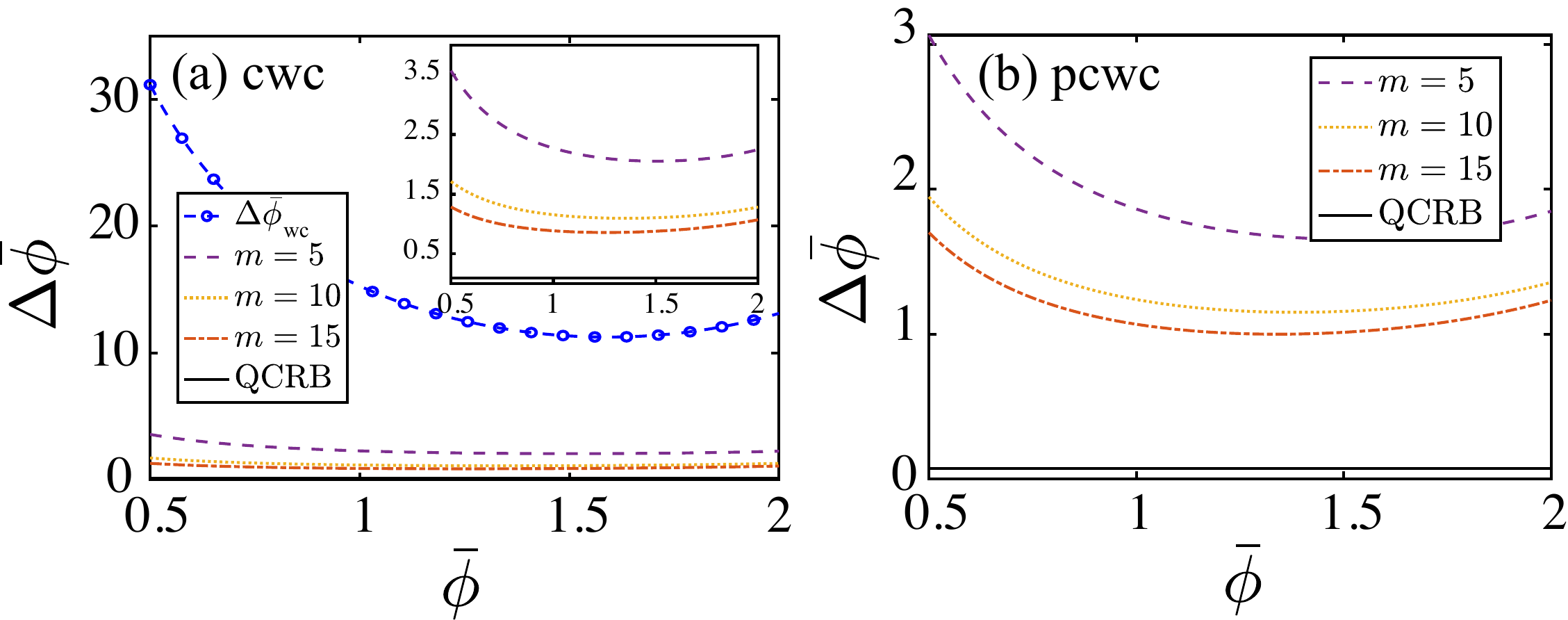}
	\caption{(a) The average phase measurement sensitivity in the homodyne measurement scheme under global catalysis for different values of catalytic photon number $m=5,10,15$ when the BS transmittance $\theta=5\pi /12$, the network node number  $d=5$ and the input resource number $N=1$, and the black solid line denotes the corresponding QCRB for global catalysis when $m=15$. The inset provides a magnified view of the catalyzed cases for clearer comparison. (b) The average phase measurement sensitivity in the homodyne measurement scheme under partial catalysis for $m=5,10,15$ when $\theta=5\pi /12$, $d=5$, $s=3$ and $N=1$. and the black line represents the weak QCRB for partial catalysis when $m=15$. \label{fig:13}}
\end{figure}

In Fig.~\ref{fig:13} (a), we plot the practical measurement sensitivity of the  average phase  in the  global quantum catalysis scheme as funcation of the average phase for different values of catalytic photon number $m$ when $\theta=5\pi /12$, $d=5$ and $N=1$. The solid line in figure denotes the QCRB given by the previous section. From Fig.~\ref{fig:13} (a), we found that  with the same input resources, the catalytic operation can also enhance the practical
	sensitivity of the average phase in the homodyne measurement.  It can further be seen that the practical measurement sensitivity described by $\Delta \bar{\phi}$ becomes better with increasing the catalytic photon number $m$, and it more closely approaches to the QCRB, as shown more clearly in the inset. This implies that stronger catalysis can achieve better measurement performance. Moreover, for the case of the global quantum catalysis, the practical measurement sensitivity obtained by this measurement scheme is already very close to the QCRB, which shows that the adopted homodyne measurement method is close to be optimal.

In Fig.~\ref{fig:13} (b), we plot the practical measurement sensitivity of the  average phase  in the  partial quantum catalysis scheme. In this case,
the partially catalyzed state $\left|\Psi_{\mathrm{pcwc}}\right\rangle$, which contains $d+1$ modes, with partial catalysis applied to the optimal subset of modes. For the catalytic DQN with $d=5$, we find that the optimal number of catalyzed modes is $s=3$. We then show the measurement error in the homodyne detection scheme as a function of the mean phase for different catalytic photon numbers $m=5,10,15$, with $\theta=5\pi /12$, $d=5$, $s=3$ and $N=1$. We find that, when catalysis is truncated to the optimal mode, the measurement sensitivity under partial catalysis not only outperforms the non-catalytic case, but also surpasses that achieved under global catalysis with the same parameter settings. This indicates that partial catalysis can further improve the sensitivity in the practical measurement scheme. As the number of catalytic photons increases, the measurement error decreases and gradually approaches the corresponding QCRB.

%%%%%%%%%%%%
\section{The Impact Of Photon Loss}\label{section 5}
%%%%%%%%%%%%%

In any practical  scenario, noise cannot be avoided due to influence of environment. Therefore, it is essential to investigate the behavior of the effective QFI under photon-loss conditions to evaluate the robustness of sensing performance for the DQN phase sensing scheme.
The BS is generally used to model photon losses through introducing an auxiliary mode. We consider one input mode of the BS as the system mode with input state $\left|\Psi\right\rangle_{S}$, and the other input port mode  as the environmental mode  in the vacuum  state $\left|0\right\rangle_{E}$. After the BS transformation, the output state becomes
\begin{equation}
	\begin{aligned}
		\left|\Psi\right\rangle_{SE} =\hat{B}_{SE}\left(\eta\right)\left|\Psi\right\rangle _{S}\left|0\right\rangle _{E},
	\end{aligned}
\end{equation} 
where $\hat{B}_{SE}$ is the BS operator, coupling the environment to the system and $\eta$ is transmittance, $\eta=1$ indicates no photon loss while $\eta=0$ means photon complete loss. 

The density matrix of the output state is $\rho_{SE}=\hat{B}_{SE}\left(\eta\right)\left|\Psi_{S}\right\rangle \left|0\right\rangle _{E}\left\langle 0\right|_{E}\left\langle \Psi_{S}\right|\hat{B}_{SE}^{\dagger}\left(\eta\right)$. By tracing out the environmental mode, we obtain the reduced density matrix of the system under noise
\begin{equation}\label{eq:43}
	\begin{aligned}
		\rho_{S}\left(\eta\right)&=\sum_{l}\hat{\Pi}_{l}\left(\eta\right)\left|\Psi\right\rangle _{S}\left\langle \Psi\right|_{S}\hat{\Pi}_{l}^{\dagger}\left(\eta\right),
	\end{aligned}
\end{equation} 
where $\hat{\Pi}_{l}\left(\eta\right)=\left\langle l\right|_{E}\hat{B}_{S,E}\left(\eta\right)\left|0\right\rangle _{E}$  are $\eta$-dependent Kraus operators. A possible set of Kraus operators in each mode is given by
\begin{equation}
	\begin{aligned}
		\hat{\Pi}_{l}\left(\eta\right)=\frac{\sqrt{\left(1-\eta\right)^{l}}}{\sqrt{l!}}\eta^{\frac{\hat{n}}{2}}\hat{a}^{l},
	\end{aligned}
\end{equation} 
where $\hat{a}$ represents the annihilation operator of the system mode.

To address the non-unitary dynamics of open quantum systems, purification provides a framework in which  the system-environment composite evolves unitarily. By extending the Hilbert space through an auxiliary environment, the system’s noisy dynamics (e.g., decoherence or dissipation) are encoded in the unitary evolution of the global state  for the system-environment (S+E) composite,  which evolves to  the entangled state
\begin{equation}
	\begin{aligned}
		\left|\Psi\right\rangle _{SE}&=\sum_{l}\hat{\Pi}_{l}\left(\eta\right)\left|\Psi\right\rangle _{S}\left|l\right\rangle _{E},
	\end{aligned}
\end{equation} 
where $\left|l\right\rangle _{E}$ belongs to the environment, and we are summing over a complete set of environment states. Indeed, by takingthe trace of the density operator corresponding to the state $\left|\Psi\right\rangle _{SE}$ with respect to the environment degrees offreedom, one recovers Eq~(\ref{eq:43}).

This is a Kraus operator representation of the lossy evolution of system  without considering the phase shift. Similarly, we can introduce the phase coding operator $e^{-i\hat{n}\phi}$ and write out the Kraus operator in the presence of phase shift to explain the phase shift of $N-l$ photons passing through dispersive media. The Kraus operator has  the following form \cite{escher2011general}
\begin{equation}
	\begin{aligned}
		\hat{\Pi}_{l}\left(\eta\right)=\frac{\sqrt{\left(1-\eta\right)^{l}}}{\sqrt{l!}}e^{-i\left(\hat{n}-\gamma l\right)\phi}\eta^{\frac{\hat{n}}{2}}\hat{a}^{l},
	\end{aligned}
\end{equation} 
where $\gamma$ is a constant. Parameter $\gamma=0$ and $\gamma=-1$ correspond to photon losses occurring before and after the phase shifter.

Generally, the QFI of the noisy system will be smaller than that of the S+E, because the additional freedom offered by the environment should increase the QFI in principle. Therefore, we  can define the bound  $\mathcal{F}\left(\rho_{S,loss}\right)\leq\mathcal{F}\left(\rho_{SE}\right)$.

%%%%%%%%%%
\begin{figure*}[t]
	\centering
	\includegraphics[width=0.9\textwidth]{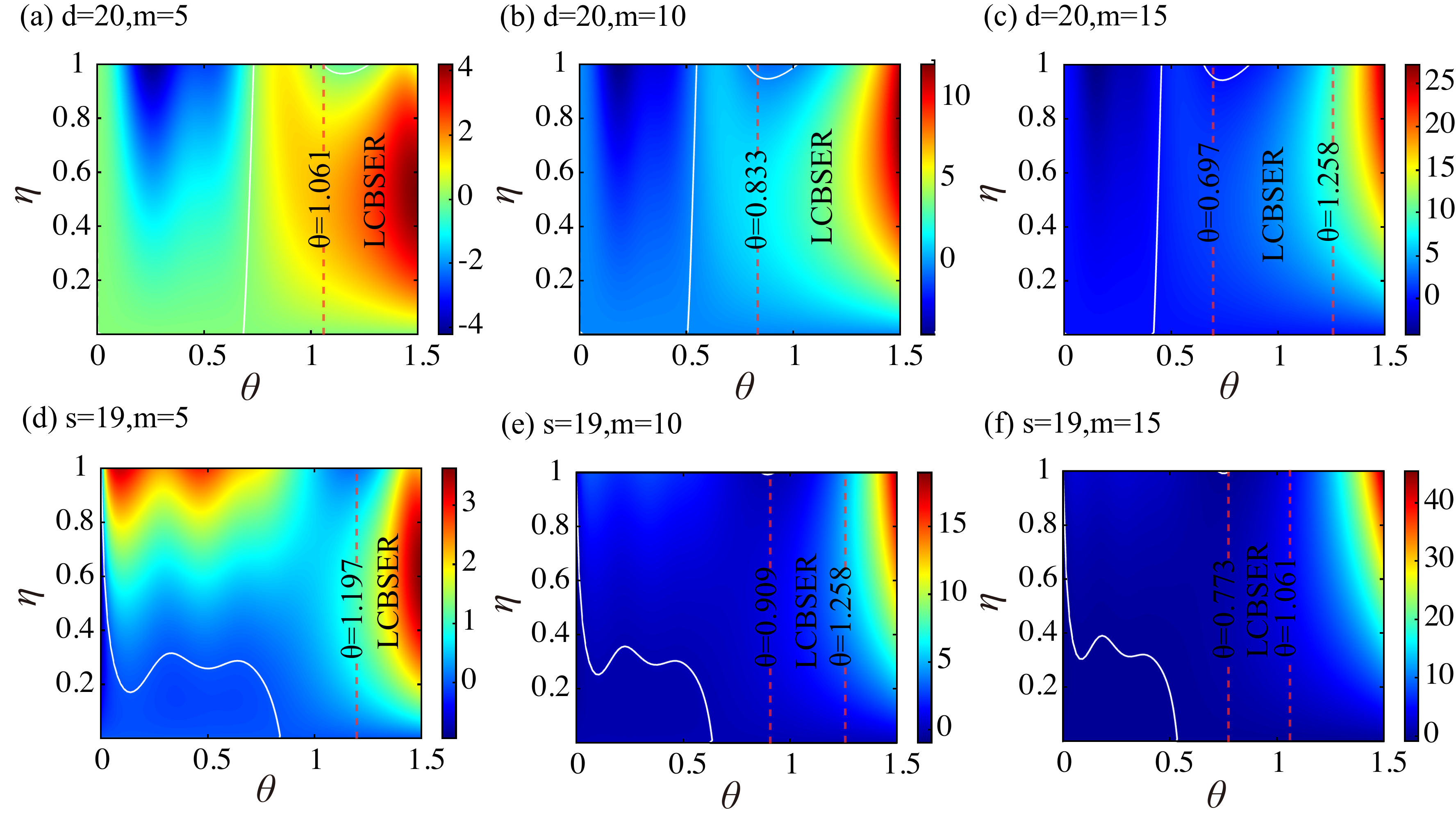}
	\caption{Sensitivity enhancement induced by global and partial catalysis relative to the uncatalyzed scheme, shown as a function of $\theta$ and $\eta$. The top row corresponds to the global-catalysis case and shows the distribution of $H_{\mathrm{cwc},l}-H_{\mathrm{wc},l}$ for $d=20$ and $N=1/2$, where panels (a)–(c) correspond to $m=5$, $10$, and $15$, respectively. The bottom row corresponds to the partial-catalysis case and shows the distribution of $H_{\mathrm{pcwc},l}-H_{\mathrm{wc},l}$ for $d=20$, $N=1/2$, and $s=19$, where panels (d)–(f) correspond to $m=5$, $10$, and $15$, respectively. The red dash line indicate the corresponding $\theta$-constraints for the LESR region at each $m$. The white curves indicate the zero-contour lines: $H_{\mathrm{cwc},l}-H_{\mathrm{wc},l}=0$ for global catalysis and $H_{\mathrm{pcwc},l}-H_{\mathrm{wc},l}=0$ for partial catalysis. \label{fig:14}} 
\end{figure*}
%%%%%%%%%%

In what follows, we shall investigate the influence of photon losses on the sensing performance of the DQN phase sensing by using the  Kraus-operator method. For the DQN sensing scheme  described by Fig.~\ref{fig:1}, in the preparation stage, a probe state is created  in the form \textcolor{red}{$\left|\Psi_{\mathrm{w}}\right\rangle$}. Although the probe state may be correlated, the evolution is separated for different modes. Therefore, in the mode $i$, evolution is determined by the parameter $\phi_{i}$, expressed in terms of Kraus operators $\hat{\Pi}_{l_{i}}^{\left(i\right)}\left(\phi_{i}\right)$, which satisfies the normalization condition $\sum_{l_{i}}\hat{\Pi}_{l_{i}}^{\left(i\right)\dagger}\left(\phi_{i}\right)\hat{\Pi}_{l_{i}}^{\left(i\right)}\left(\phi_{i}\right)=I$. The evolved state is then given by ( 0-mode is the reference mode)
\begin{equation}
	\begin{aligned}
		\left|\Psi_{SE}\left(\boldsymbol{\phi}\right)\right\rangle =\sum_{\boldsymbol{l}}\hat{\Pi}_{\boldsymbol{l}}\left(\boldsymbol{\phi}\right)\left|\Psi\right\rangle _{S}^{\left(d+1\right)}\left|l\right\rangle _{E}^{\left(d+1\right)},
	\end{aligned}
\end{equation}
where $\boldsymbol{\phi}=\left(\phi_{1},...,\phi_{d}\right)$, $\boldsymbol{l}=\left(l_{0},l_{1},...,l_{d}\right)$, $\left|0\right\rangle _{E}^{\left(d+1\right)}=\otimes_{i=0}^{d}\left|0\right\rangle _{E}$, $\left|l\right\rangle _{E}^{\left(d+1\right)}=\otimes_{i=0}^{d}\left|l_{i}\right\rangle _{E}$, and $\hat{\Pi}_{\boldsymbol{l}}\left(\boldsymbol{\phi}\right)$ is the set of Kraus operators describing $d+1$ quantum channels, which only acts on the system state and can be decomposed
\begin{equation}
	\begin{aligned}
		\hat{\Pi}_{\boldsymbol{l}}\left(\boldsymbol{\phi}\right)&=\hat{\Pi}_{l_{0}}^{\left(0\right)}\left(0\right)\otimes\hat{\Pi}_{l_{1}}^{\left(1\right)}\left(\phi_{1}\right)\otimes...\otimes\hat{\Pi}_{l_{d}}^{\left(d\right)}\left(\phi_{d}\right), \\
		&=\otimes_{i=0}^{d}\sqrt{\frac{\left(1-\eta_{i}\right)^{l_{i}}}{l_{i}!}}e^{-i\phi_{i}\cdot\left(\hat{n}_{i}-\gamma l_{i}\right)}\eta^{\frac{\hat{n}_{i}}{2}}\hat{a}_{i}^{l}.
	\end{aligned}
\end{equation}

It is known that, the precision  estimation of $\boldsymbol{\phi}$, described by its covariance matrix $Cov(\boldsymbol{\Phi})$, is limited by the QCRB, described by Eq.~\ref{eq:3}. For multi-parameter estimation problems with losses, we have the following  inequality \cite{Jie2014Quantum}
\begin{equation}
	\begin{aligned}
		Cov(\boldsymbol{\Phi})\geq\boldsymbol{\mathcal{F}}^{-1}\geq \boldsymbol{\mathcal{F}}\left(\hat{\Pi}_{\boldsymbol{l}}\left(\boldsymbol{\phi}\right)\right)^{-1},
	\end{aligned}
\end{equation}
where $\boldsymbol{\mathcal{F}}$ is the QFI matrix of the noisy system QFI and $\boldsymbol{\mathcal{F}}\left(\hat{\Pi}_{\boldsymbol{l}}\left(\boldsymbol{\phi}\right)\right)$ is the QFI matrix of the large Hilbert space $S+E$.
In the DQN phase sensing scheme under our consideration, the parameter focused on is the multiphase linear combination \textcolor{red}{$\bar{\phi}$}. Therefore, we can obtain the weak QCRB for estimating the global parameter \textcolor{red}{$\bar{\phi}$} in the presence of photon losses
\begin{equation}
	\bigtriangleup^{2} \bar{\phi}\ge\frac{1}{H}\ge\frac{1}{H_{l}},
\end{equation}
where  $H=\sum_{ij}\boldsymbol{\mathcal{F}}_{ij}$ and $H_{l}=\sum_{ij}\boldsymbol{\mathcal{F}}_{ij}\left(\hat{\Pi}_{\boldsymbol{l}}\left(\boldsymbol{\phi}\right)\right)$, called the effective QFI in the presence of losses, which provides a metric to evaluate the robustness of the sensing performance in distributed parameter estimation under the noise condition.

Because $\left|\Psi_{SE}\left(\boldsymbol{\phi}\right)\right\rangle$ is pure state, we can obtain the matrix elements of the effective QFI matrix in the presence of losses
\begin{equation}
	\begin{aligned}
		\boldsymbol{\mathcal{F}}_{ij}\left(\hat{\Pi}_{\boldsymbol{l}}\left(\boldsymbol{\phi}\right)\right)&=4\left[\left\langle\hat{H}_{2}^{\left(ij\right)}\right\rangle _{S}-\left\langle\hat{H}_{1}^{\left(i\right)}\right\rangle _{S}\left\langle \hat{H}_{1}^{\dagger\left(j\right)}\right\rangle _{S}\right], 
	\end{aligned}
\end{equation}
where all average values are taken to the system state $\left|\Psi\right\rangle_{S}$, and 
\begin{equation}
	\begin{aligned}
		\hat{H}_{1}^{(i)}
		&=
		\sum_{l_i}
		\frac{\partial \left(\hat{\Pi}_{l_i}^{(i)}(\phi_i)\right)^{\dagger}}
		{\partial \phi_i}
		\hat{\Pi}_{l_i}^{(i)}(\phi_i),
		\qquad i=1,\ldots,d, \\
		\hat{H}_{2}^{(ij)}
		&=
		\begin{cases}
			\hat{H}_{1}^{(i)}
			\left(\hat{H}_{1}^{(j)}\right)^{\dagger},
			& i\neq j,\quad i,j=1,\ldots,d, \\[1.0ex]
			\displaystyle
			\sum_{l_i}
			\frac{\partial \left(\hat{\Pi}_{l_i}^{(i)}(\phi_i)\right)^{\dagger}}
			{\partial \phi_i}
			\frac{\partial \hat{\Pi}_{l_i}^{(i)}(\phi_i)}
			{\partial \phi_i},
			& i=j,\quad i,j=1,\ldots,d .
		\end{cases}
	\end{aligned}
\end{equation}
By using the Kraus operator, we can find that
\begin{equation}
	\begin{aligned}
		\hat{H}_{1}^{(i)}
		&= h_{1}\hat{n}_{i},
		\qquad i=1,\ldots,d, \\
		\hat{H}_{2}^{(ij)}
		&=
		\begin{cases}
			\hat{H}_{1}^{(i)}
			\left(\hat{H}_{1}^{(j)}\right)^{\dagger},
			& i\neq j,\quad i,j=1,\ldots,d, \\[1.0ex]
			\left|h_{1}\right|^{2}\hat{n}_{i}^{2}
			+h_{2}\hat{n}_{i},
			& i=j,\quad i,j=1,\ldots,d .
		\end{cases}
	\end{aligned}
\end{equation}
where $h_{1}=i\left[\left(\gamma+1\right)\left(1-\eta\right)-1\right]$, and $h_{2}=\left(\gamma+1\right)^{2}\left(1-\eta\right)\eta$. For simplicity of calculation, we suppose that $\eta_{i}=\eta$ for all $i$ and $\gamma=0$.
Thus we can obtain the effective QFI in the presence of loss as follows

\begin{equation}
	\begin{aligned}
		H_{l}=&\sum_{i,j=1,i\neq j}^{d}4\left|h_{1}\right|^{2}\left(\left\langle \hat{n}_{i}\hat{n}_{j}\right\rangle _{S}-\left\langle \hat{n}_{i}\right\rangle _{S}\left\langle \hat{n}_{j}\right\rangle _{S}\right)\\&+\sum_{i=1}^{d}4\left[\left|h_{1}\right|^{2}\left(\left\langle \hat{n}_{i}^{2}\right\rangle _{S}-\left\langle \hat{n}_{i}\right\rangle _{S}^{2}\right)+h_{2}\left\langle \hat{n}_{i}\right\rangle _{S}\right],
	\end{aligned}
\end{equation}
where $\left\langle \hat{n}_{i}\right\rangle _{S}$ denotes the average photon number in the 
$i$-th mode of the system's initial state.
Given the initial state of the system, the effective QFI under loss can be calculated. For instance, the effective QFI of a multimode W-type state in the presence of losses is given by
\begin{equation}\label{eq:54}
	\begin{aligned}
		H_{\mathrm{w},l}&=\eta^{2}H_{w}+4\left(1-\eta\right)\eta\mathcal{N}^{2}\sum_{i=1}^{d}\left\langle \Psi_{\mathrm{w}}\right|\hat{n}_{i}\left|\Psi_{\mathrm{w}}\right\rangle . 
	\end{aligned}
\end{equation}
%where $\left\langle \hat{n}\right\rangle$ express as the average number of photons per mode.
where $\mathcal{N}$ represent the normalization constant of the corresponding multimode W-type state. The first term $H_{\mathrm{w}}$ reflects the degradation of the ideal effective QFI  of the system due to loss, while the second term $4\left(1-\eta\right)\eta\mathcal{N}^{2}\sum_{i=1}^{d}\left\langle \Psi_{\mathrm{w}}\right|\hat{n}_{i}\left|\Psi_{\mathrm{w}}\right\rangle$ arises from the interaction between the system and the environment, enhancing sensitivity in certain regimes. 
 
 Based on the analytical results obtained in the previous sections, we can directly derive the effective QFI  of uncatalyzed, globally catalyzed, and partially catalyzed schemes for multimode W-type state in the presence of losses. Below, we present the explicit expressions of the effective QFI for the quantum probe states $\left|\Psi_{\mathrm{wc}}\right\rangle$, $\left|\Psi_{\mathrm{cwc}}\right\rangle$, $\left|\Psi_{\mathrm{pcwc}}\right\rangle$, $\left|\Psi_{\mathrm{ws}}\right\rangle$, $\left|\Psi_{\mathrm{cws}}\right\rangle$, and $\left|\Psi_{\mathrm{pcws}}\right\rangle$ under lossy conditions
\begin{equation}
	\begin{aligned}
		&H_{\mathrm{wc},l}=\eta^{2}H_{\mathrm{wc}}+4\left(1-\eta\right)\eta\mathcal{N}_{1}^{2}d\left|\alpha\right|^{2},\\
		&H_{\mathrm{cwc},l}=\eta^{2}H_{\mathrm{cwc}}+4\left(1-\eta\right)\eta\mathcal{N}_{3}^{2}d\left\langle \hat{n}'\right\rangle,\\
		&H_{\mathrm{pcwc},l}=\eta^{2}H_{\mathrm{pwc}}+4\left(1-\eta\right)\eta\mathcal{N}_{3}^{'2}\left[s\left\langle \hat{n}^{'}\right\rangle +\left(d-s\right)\left|\alpha\right|^{2} \right],\\
		&H_{\mathrm{ws},l}=\eta^{2}H_{\mathrm{ws}}+4\left(1-\eta\right)\eta\mathcal{N}_{2}^{2}d\sinh^{2}r,\\
		&H_{\mathrm{cws},l}=\eta^{2}H_{\mathrm{cws}}+4\left(1-\eta\right)\eta\mathcal{N}_{4}^{2}d\left\langle \hat{n}'\right\rangle _{\left|\psi_{s}'\right\rangle }, \\
		&H_{\mathrm{pcws},l}=\eta^{2}H_{\mathrm{pcws}}+4\left(1-\eta\right)\eta\mathcal{N}_{4}^{'2}\left[s\left\langle \hat{n}'\right\rangle_{\left|\psi_{s}'\right\rangle } +\left(d-s\right)\sinh^{2}r\right],
	\end{aligned}
\end{equation}
where $s$ denotes the truncated modulus of partial catalysis, that is, only the first $s+1$ modes are catalysed.

In order to discuss the effect of losses on the sensing performance of different quantum probes in DQN phase sensing scheme, we rewrite Eq. (\ref{eq:54}) as a quadratic function of the dissipation intensity $H_{\mathrm{w},l}=\left(H_{\mathrm{w}}-4\mathcal{N}^{2}\sum_{i=1}^{d}\left\langle \Psi_{\mathrm{w}}\right|\hat{n}_{i}\left|\Psi_{\mathrm{w}}\right\rangle \right)\eta^{2}+4\mathcal{N}^{2}\sum_{i=1}^{d}\left\langle \Psi_{\mathrm{w}}\right|\hat{n}_{i}\left|\Psi_{\mathrm{w}}\right\rangle\eta$. 
It is interesting to note that, according to the properties of quadratic functions, we can find that there exists a loss-enhanced sensitivity region under the following condition 
\begin{equation}\label{eq:72}
	H_{\mathrm{w}}<2\left(\bar{N}_{\mathrm{w}}-\mathcal{N}^{2}\left\langle\psi_{0}\right| \hat{n}\left|\psi_{0}\right\rangle\right), 
\end{equation}
where $\left\langle\psi_{0}\right| \hat{n}\left|\psi_{0}\right\rangle$ express as the average number of photons in the $0$-th mode and $\bar{N}_{\mathrm{w}}$ represent the average number of photons corresponding to the quantum probe state. Eq.~(\ref{eq:72}) can be described as the discriminant of the region whether the loss-enhanced sensitivity exists. 

We  can numerically analyzed the catalytic conditions under which there exists a loss-enhanced sensitivity region for $\left|\Psi_{\mathrm{wc}}\right\rangle$, $\left|\Psi_{\mathrm{cwc}}\right\rangle$, $\left|\Psi_{\mathrm{pcwc}}\right\rangle$, $\left|\Psi_{\mathrm{ws}}\right\rangle$, $\left|\Psi_{\mathrm{cws}}\right\rangle$ and $\left|\Psi_{\mathrm{pcws}}\right\rangle$, respectively, for a fixed number of input resources $N=1/2$ and the network nodes $d=20$. It is found that loss-enhanced sensitivity arises only for the states $\left|\Psi_{\mathrm{cwc}}\right\rangle$ and $\left|\Psi_{\mathrm{pcwc}}\right\rangle$. For a fixed number of catalytic photons, this enhancement occurs only when the transmission coefficient $\theta$ lies within specific parameter regimes, indicating that these conditions are necessary for achieving loss-enhanced sensitivity. Specifically, for the  W-type coherent state under global catalysis $\left|\Psi_{\mathrm{cwc}}\right\rangle$, loss-enhanced sensitivity appears when $\theta \ge 1.061$ for $m=5$, $\theta \ge 0.833$ for $m=10$, and $0.697 \le \theta \le 1.258$ for $m=15$. For the  W-type coherent state under partial catalysis $\left|\Psi_{\mathrm{pcwc}}\right\rangle$ where $s=19$, the corresponding regions are $\theta \ge 1.197$ for $m=5$, $0.901 \le \theta \le 1.258$ for $m=10$, and $0.773 \le \theta \le 1.061$ for $m=15$.

To further identify parameter regimes where catalytic enhancement and loss-enhanced sensitivity region coexist. 
Fig.~\ref{fig:14}  shows the sensitivity enhancement induced by global and partial catalysis relative to the uncatalyzed scheme in the presence of losses, as functions of $\theta$ and $\eta$. The calculations are performed for $d=20$ and $N=1/2$, with catalytic photon numbers $m=5,10,15$ in panels (a)–(c) for global catalysis and (d)–(f) for partial catalysis, where the latter is implemented with truncation parameter $s=19$. The red dashed lines indicate the corresponding $\theta$-constraints, while the white curves denote the zero-contour lines of the enhancement. It can be seen that loss-enhanced sensitivity appears only within the specific parameter intervals constrained by $\theta$, and these intervals vary significantly with the catalytic photon number $m$.
 Furthermore, the overlap between the catalysis-enhanced region (corresponding to the area with color-scale values greater than 1) and the loss-enhanced sensitivity region (bounded by the red dashed lines) defines the loss-catalysis dual-enhanced sensitivity region (LCBESR), indicating that both catalysis and loss contribute positively to the sensitivity in this regime. A comparison between the top and bottom rows shows that global catalysis generally yields a broader LCBESR, and this advantage becomes more pronounced for larger $m$; however, in terms of enhancement magnitude, partial catalysis typically provides a larger sensitivity gain and a wider positive gain region than global catalysis in the presence of losses.
%%%%%%
\begin{figure}[t]
	\centering
	\includegraphics[width=1.0\columnwidth]{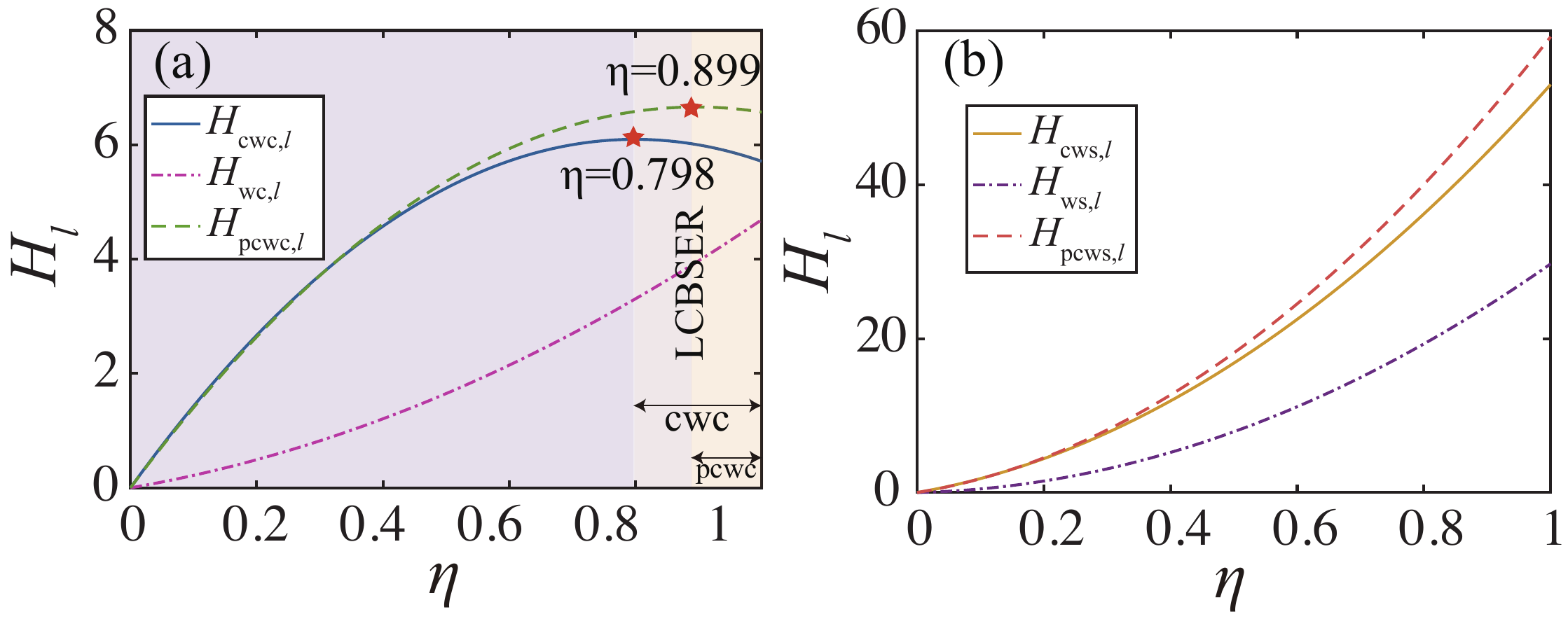}%\textwidth
	\caption{ Sensitivity in the presence of losses for the globally catalyzed, uncatalyzed, and partially catalyzed schemes versus $\eta$. (a) $H_{\mathrm{cwc},l}$, $H_{\mathrm{wc},l}$, and $H_{\mathrm{pcwc},l}$ for $d=20$, $m=5$, $N=1/2$, and $\theta=11\pi/25$, with partial catalysis truncated at $s=19$. The yellow region denotes the corresponding LCBESR, and the stars mark the critical values $\eta=0.798$ and $\eta=0.899$ for the global and partial catalysis cases, respectively. (b) $H_{\mathrm{cws},l}$, $H_{\mathrm{ws},l}$, and $H_{\mathrm{pcws},l}$ for $d=20$, $m=5$, $N=1/2$, and $\theta=\pi/3$, with partial catalysis truncated at $s=19$. \label{fig:15}}
\end{figure}
%%%%%%%

Fig.~\ref{fig:15} (a) shows $H_{\mathrm{wc},l}$, $H_{\mathrm{cwc},l}$, and $H_{\mathrm{pcwc},l}$ as functions of the loss parameter $\eta$ for $d=20$, $m=5$, $N=1/2$, $\theta=11\pi/25$, and $s=19$ in the partial-catalysis scheme, respectively. For global catalysis, $H_{\mathrm{cwc},l}$ increases with $\eta$ and satisfies $H_{\mathrm{cwc},l}>H_{\mathrm{wc},l}$ in the range $0.798<\eta<1$, indicating the existence of a LCBESR, while for $0<\eta<0.798$ the sensitivity decreases with increasing loss, yielding an optimal loss rate $\eta_{\mathrm{opt}}=0.798$ for $\left|\Psi_{\mathrm{cwc}}\right\rangle$. For partial catalysis, a LCBESR appears in the range $0.899<\eta<1$, where $H_{\mathrm{pcwc},l}$ increases with $\eta$ and remains larger than $H_{\mathrm{wc},l}$. When $0<\eta<0.899$, $H_{\mathrm{pcwc},l}$ decreases with increasing loss, implying an optimal loss rate $\eta_{\mathrm{opt}}=0.899$ for $\left|\Psi_{\mathrm{pcwc}}\right\rangle$. Moreover, $H_{\mathrm{cwc},l}$ remains larger than $H_{\mathrm{wc},l}$, while $H_{\mathrm{pcwc},l}$ remains larger than $H_{\mathrm{cwc},l}$, indicating that catalytic operations enhance the sensing performance in the presence of loss, and that partial catalysis yields a further improvement compared with global catalysis. In Fig.~\ref{fig:15} (b), we discuss the influence of loss on $H_{\mathrm{ws},l}$, $H_{\mathrm{cws},l}$ and $H_{\mathrm{pcws},l}$ for given $d=20$, $m=5$, $N=1/2$, $\theta=\pi/3$, and $s=19$ in the partial-catalysis scheme. Our analysis reveals that loss will reduce the sensitivity, and the greater the loss, the smaller the sensitivity, but $H_{\mathrm{pcws},l}>H_{\mathrm{cws},l}>H_{\mathrm{ws},l}$ still meet, which also confirms that for $\left|\Psi_{\mathrm{ws}}\right\rangle$, catalysis can improve sensing performance even in the case of loss and that partial catalysis provides a further enhancement over global catalysis.  

We also can find that under lossy conditions, the effective QFI hierarchy follows $H_{\mathrm{pcws},l}>H_{\mathrm{cws},l}>H_{\mathrm{ws},l}>H_{\mathrm{pcwc},l}>H_{\mathrm{cwc},l}>H_{\mathrm{wc},l}$ for identical system parameters. 
This shows that WS-type probes consistently achieve higher effective QFI than WC-type probes. Moreover, both global and partial catalysis further enhance the sensitivity, with partial catalysis providing the strongest improvement. Although the WS-type probes are more sensitive to environmental noise, they still outperform the corresponding WC-type probes under loss. In contrast, the WC-type probes are less affected by decoherence, indicating greater robustness but lower sensitivity. Notably, for the WC-type probes, both global and partial catalysis give rise to a loss-catalysis dual-enhanced sensitivity region (LCBESR) under appropriate parameter regimes, where catalysis and loss simultaneously enhance the sensitivity. Overall, these results confirm that catalytic operations, especially partial catalysis, significantly improve the sensing performance of both multimode W-type quantum probes under realistic lossy conditions. 
%Remarkably, the $\left|\Psi_{cwc}\right\rangle$ exhibits a counterintuitive behmultimode W-type squeezed state (three resources) further enhances sensitivity by $22.37$~dB and $20.55$~dB over the two-resource squeezed state and catalyzed coherent state, respectively. Additionally, we reveal that the sensitivity of quantum network sensing depends on the number of catalytic modes: for catalyzed multimode W-type states, the sensitivity pavior: its effective QFI does not monotonically decrease with increasing loss. Instead, there exists an optimal loss rate maximizing the effective QFI. For our parameter regime, this maximum occurs at a transmittance of $\eta=0.5832$, as illustrated in Fig.~\ref{fig:14}. This phenomenon suggests a nontrivial interplay between catalytic operations and environmental dissipation, potentially enabling loss-optimized quantum sensing strategies.

%%%%%%%%%%%%%%%%%%
\section{Conclusions and discussion}\label{section 6}
%%%%%%%%%%%%%%%%%%

In this work, we have proposed and systematically analyzed a quantum-enhanced distributed network sensing scheme that integrates three TQRs  (quantum catalysis, entanglement, and squeezing)  for multiphase estimation. By introducing quantum catalysis into the DQN framework, we demonstrate that the sensing performance of hybrid quantum probes significantly outperforms that of quantum probes relying on fewer quantum resources, both in ideal and lossy environments.

Our key findings are threefold. First, quantum catalysis, whether applied globally or partially, markedly enhances the effective QFI and reduces the weak QCRB, enabling precision approaching the Heisenberg limit. Notably, partial catalysis—where only a subset of network modes are catalyzed—consistently outperforms global catalysis in terms of both theoretical precision bounds and practical homodyne measurement sensitivity. This advantage arises from a more concentrated distribution of non-Gaussian resources, which is more effective for multiphase estimation in W-type entangled networks.

 Second, the simultaneous use of all three TQRs  (quantum catalysis, entanglement, and squeezing) yields the highest sensitivity. Specifically, the multimode catalyzed W-type squeezed state achieves maximum gains of up to $33.77$dB over the multimode uncatalyzed W-type squeezed state and $14.47$dB over the multimode catalyzed W-type coherent state, showing the benefits of adding catalysis to a squeezed-entangled probe and adding squeezing to a catalyzed-entangled probe, respectively. These improvements arise from the cooperation among catalysis-induced non-Gaussian modification, squeezing-induced phase-sensitive fluctuations, and multimode entanglement. At low input resources, catalysis provides a strong relative enhancement, whereas at higher input resources the intrinsic squeezing, and entanglement of the original probe become more important. 
%Third, the proposed scheme exhibits nontrivial robustness under photon loss. For W-type coherent-state probes, both global and partial catalysis create a loss–catalysis dual-enhanced sensitivity region (LCBESR), where moderate loss counterintuitively improves sensitivity. In contrast, W-type squeezed-state probes, while more sensitive to loss, still maintain a precision hierarchy   $H_{pcws,l}>H_{cws,l}>H_{ws,l}$  under identical conditions, confirming that catalysis—especially partial catalysis remains beneficial in noisy environments.

Third, the proposed scheme exhibits nontrivial robustness under photon loss. For WC-type probes, both global and partial catalysis can produce a loss-catalysis dual-enhanced sensitivity region (LCBESR), where 
moderate loss counterintuitively improves the sensitivity. Under identical loss and parameter conditions, the WS-type probes mixing three TQRs retain an overall advantage over the WC-type probes using only two TQRs, and partial catalysis consistently gives the best performance within both the WC-type and WS-type probe families, with 
$H_{\mathrm{pcws},l}>H_{\mathrm{cws},l}>H_{\mathrm{ws},l}>H_{\mathrm{pcwc},l}>H_{\mathrm{cwc},l}>H_{\mathrm{wc},l}$. These results confirm that catalysis, especially partial catalysis, remains useful under 
loss, and that three TQRs continues to provide superior sensing performance in noisy environments. 

Our practical homodyne measurement scheme validates that the theoretical QCRB can be closely approached, and that increasing the catalytic photon number systematically reduces the estimation error of the average phase. Furthermore, the experimental feasibility of generating W-type catalyzed squeezed states is supported by a proposed optical network based on BSs, Kerr interactions, and conditional detection.

From a broader perspective, this work establishes that quantum-resource hybridization is a powerful strategy for distributed quantum metrology. It also raises several open questions. For instance, how does the optimal catalytic mode number scale with network size and topology? Can adaptive or feedback-enhanced catalysis further improve the trade-off between success probability and sensitivity? Moreover, extending the current scheme to nonlinear phase estimation or to other entangled graph states (e.g., GHZ or cluster states) may unlock new regimes of quantum-enhanced sensing.

In conclusion, our results provide both theoretical foundations and practical guidelines for designing high-sensitivity DQN sensors. The demonstrated advantages of partial catalysis and three TQRs hybridization mark a significant step toward real-world deployment of quantum networks for metrology, imaging, and synchronized clock networks.

%%%%%%%%%%%%%%%%%%
%Our scheme can be used to enhance the estimation precision of spatial distribution and thus improve the accuracy of sensing problems such as phase imaging~\cite{Humphreys2013Quantum} and globally synchronized clocks~\cite{K2014A}.

%\vspace{5mm}
\begin{acknowledgements}
This work is supported by  the Quantum Science and Technology-National Science and Technology Major Project (Grant No. 2024ZD0301000), the Natural Science Foundation of China (NSFC) (Grant Nos. 12247105, 12175060, 12421005,  252300421221), Hunan provincial sci-tech program (Grant No. 2023ZJ1010),  Henan Science and Technology Major Project (No. 241100210400),    XJ-Lab key project (Grant No. 23XJ02001), and the Natural Science Foundation of Henan Province (Grant No. 252300421221)
\end{acknowledgements}
%%%%%%%%%%%%%%%%%%%%%%%%%%%%%%%%%%%%%%%

\appendix

\section{Multi-photon catalytic quantum states}\label{A_HEP}

\begin{figure}[t]
	\centering
	\includegraphics[width=0.7\columnwidth]{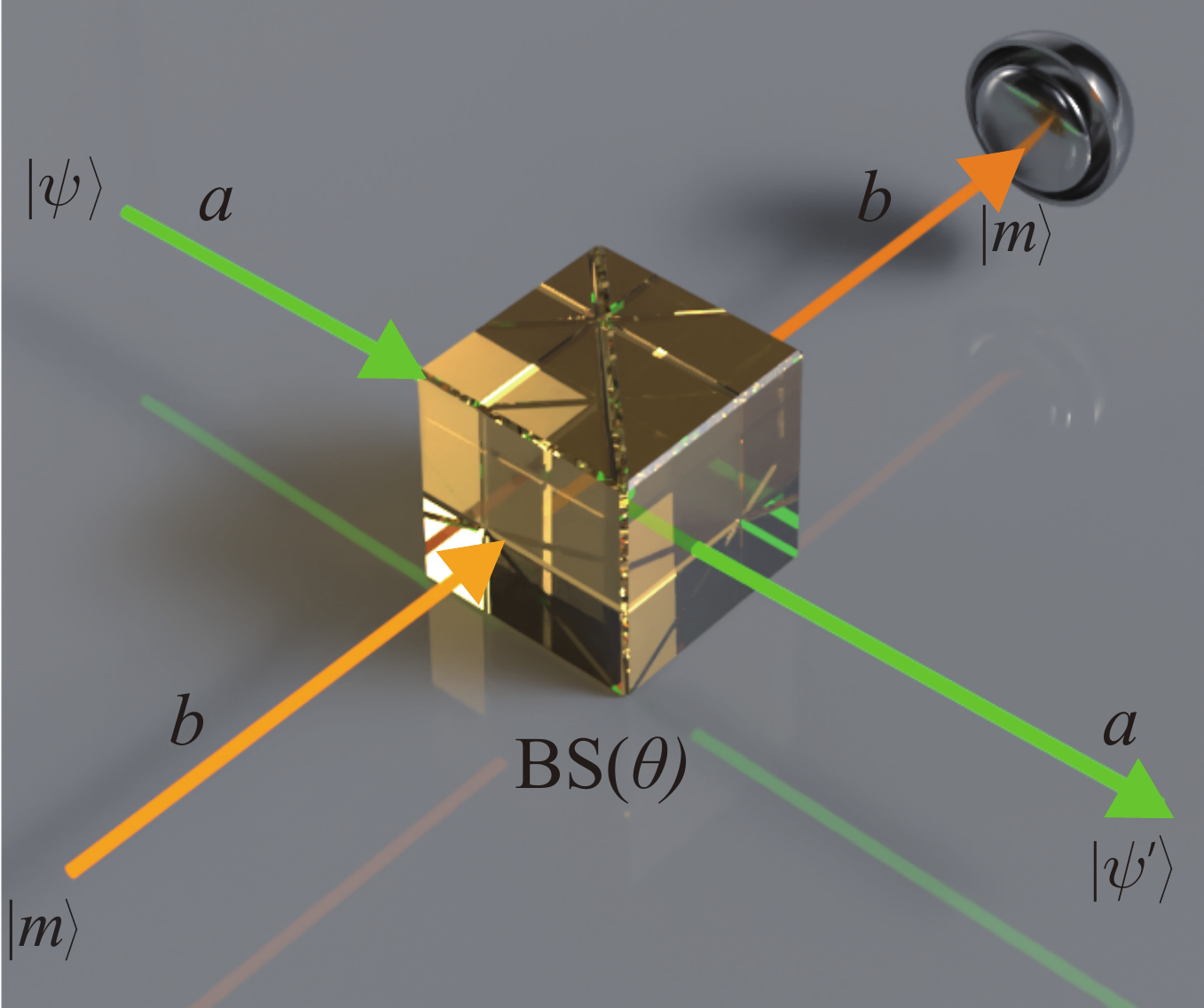}
	\caption{Theoretical model of multiphoton catalysis. There are two input ports $a$ and $b$ as well as two corresponding output ports, an arbitrary input pure state $\left|\psi\right\rangle$ at $a$ port and an m-photon Fock state at $b$ port (at a BS of transmission $\tau=\cos^{2}\theta$ ), 
		conditional measuring $m$ photons at output port corresponding to $b$, then $a$ corresponding output port generate the catalytic state $\left|\psi'\right\rangle$ of the input pure state. \label{fig:S1}}
\end{figure}

In this appendix, we derive expressions of multi-photon catalytic coherent state and multi-photon catalytic squeezed state.
The multi-photon quantum catalytic operation is physically accomplished by a BS as shown in Fig~\ref{fig:S1}. A pure state $\left|\psi\right\rangle$ and a multiphoton Fock state $\left|m\right\rangle$ are injected to the two input ports of the BS with transmittance $\tau=\cos^{2}\theta$. Multiphoton detection is performed at one of the output ports, if $m$ photons are detected, at this time  the output state at the other output port is the multiphoton catalytic state  noted as $\left|\psi'\right\rangle$ ~\cite{Bartley2012Multiphoton,Li2016Multiphoton,Hu2017Continuous}.
%%%%%%%%%%------------------------------------------%%%%%%%%%%%%
%%%%%%%%%%%%------------------------------------------%%%%%%%%%%%

In general, the output state of the BS can be expressed as
\begin{equation}
	\begin{aligned}
		\left|\psi\right\rangle_{out}=\hat{B}(\theta)\left|\psi\right\rangle_{a}\left|m\right\rangle_{b},
	\end{aligned}
\end{equation}
where $\hat{B}\left(\theta\right)=:\exp\left[\left(\cos\theta-1\right)\left(\hat{a}^{\dagger}\hat{a}+\hat{b}^{\dagger}\hat{b}\right)+\left(\hat{a}^{\dagger}\hat{b}-\hat{a}\hat{b}^{\dagger}\right)\right]:$ is the normal ordering form of the BS operator, $\left|\psi\right\rangle$ and $|m\rangle$ are the input states of the catalyzed and catalytic  modes, respectively. 

According to the BS catalytic model, in order to successfully generate the catalytic quantum state at port $a$, it is necessary to perform projection measurements $\hat{M}=\left| m\right\rangle_{bb}\left\langle m\right|$ at the $b$-mode corresponding output port with the following output state
\begin{equation}
	\begin{aligned}
		\left|\psi\right\rangle_{out}&=\hat{M}\hat{B}\left(\theta\right)\left|\psi\right\rangle_{a}\left|m\right\rangle_{b}, \\
		&=\left| m\right\rangle_{b}\left\langle m\right|_{b}\hat{B}(\theta )\left|\psi\right\rangle_{a}\left|m\right\rangle_{b}, \\
		&=\hat{C}\left|\psi\right\rangle_{a}\left|m\right\rangle_{b},
	\end{aligned}
\end{equation}
where $\hat{C}$ is  called as the catalytic operator ~\cite{Li2016Multiphoton,Hu2017Continuous,Zhang2021Improved}
\begin{equation}\label{eq:A3}
	\begin{aligned}
		\hat{C}&=\left\langle m\right|_{b}\hat{B}\left(\theta\right)\left|m\right\rangle_{b}, \\
		&=:\mathcal{L}_{m}\left(\hat{a}^{\dagger}\hat{a}\tan^{2}\theta\right):\left(\cos\theta\right)^{\hat{a}^{\dagger}\hat{a}+m}, \\ 
	\end{aligned}
\end{equation}
where $\mathcal{L}_{m}\left(.\right)$ is Laguerre polynomial. Utilizing the generating function of Laguerre polynomials $\mathcal{L}_{m}\left(x\right)=\frac{1}{m!}\frac{\partial^{m}}{\partial\tau^{m}}\left(\frac{e^{-\frac{x\tau}{1-\tau}}}{1-\tau}\right)|_{\tau=0}$ and the normal ordering transformation $e^{x\hat{a}^{\dagger}\hat{a}}=:\exp\left[\left(e^{x}-1\right)\hat{a}^{\dagger}\hat{a}\right]:$, we can obtain
\begin{equation}
	\begin{aligned}
		:\mathcal{L}_{m}\left(\hat{a}^{\dagger}\hat{a}tan^{2}\theta\right):&=\frac{1}{m!}\frac{\partial^{m}}{\partial\tau^{m}}\left\{ \frac{1}{1-\tau}\left(\frac{1-\tau\sec^{2}\theta}{1-\tau}\right)^{\hat{a}^{\dagger}\hat{a}}\right\} |_{\tau=0}.
	\end{aligned}
\end{equation}

Therefore, performing a quantum catalytic operation  on a quantum state of the   $a$ mode $|\psi\rangle_{a}$, the catalyzed state can be written as $\left|\psi'\right\rangle_{a}=\hat{C}\left|\psi\right\rangle_{a}$.

If we input a single-mode coherent state, i.e $\left|\psi\right\rangle_{a}=\left|\alpha\right\rangle$, then we obtain the normalized catalytic coherent state as~\cite{Bartley2012Multiphoton,Li2016Multiphoton}
\begin{equation}\label{eq:A5}
	\begin{aligned}
		\left|\psi'\right\rangle&=N_{m}:\mathcal{L}_{m}\left(\hat{a}^{\dagger}\hat{a}\tan^{2}\theta\right):\left(\cos\theta\right)^{\hat{a}^{\dagger}\hat{a}+m}\left|\alpha\right\rangle, \\
		&=\bar{N}_{m}\mathcal{L}_{m}\left(\hat{a}^{\dagger}\mu\right)\left|\alpha_{\theta}\right\rangle,
	\end{aligned}
\end{equation}
where we have  introduced  $\alpha_{\theta}=\alpha\cos\theta$ and $\bar{N}_{m}=\left(\cos\theta\right)^m e^{-\frac{\left|\alpha\right|^{2}\sin^{2}\theta}{2}}N_{m}$, with $\mu=\alpha\cos\theta\tan^{2}\theta$ and $N_{m}$ being the normalization constant. From $\left\langle\psi'|\psi'\right\rangle=1$, we can obtain
\begin{equation}
	\begin{aligned}\label{ES.A3}
		\bar{N}^{-2}_{m}=&\left\langle\alpha_{\theta}\right|\mathcal{L}_{m}\left(\hat{a}\mu^{\ast}\right)\mathcal{L}_{m}\left(\hat{a}^{\dagger}\mu\right)\left|\alpha_{\theta}\right\rangle \\
		=&\frac{1}{\pi}\int d^2z \left|\mathcal{L}_{m}\left(z\mu^{\ast}\right)\right|^{2}\\
		&\times\exp\left[-\left|\alpha\cos\theta\right|^{2}-\left|z\right|^{2}+\left(z^{\ast}\alpha+\alpha^{\ast}z\right)\cos\theta\right] \\
		=&\sum_{n,k}^{m}\Pi_{n,k}^{m}\left(-\mu,\mu^{\ast}\right)H_{n,k}\left(\alpha_{\theta}^{\ast},-\alpha_{\theta}\right),
	\end{aligned}
\end{equation}
where $\Pi_{n,k}^{m}\left(x,y\right)=\binom{m}{n}\binom{m}{k}\frac{x^{n}y^{k}}{n!k!}$ and we have used the expansion of the Laguerre polynomial $\mathcal{L}_{m}\left(x\right)=\sum_{k=0}^{m}\frac{\left(-1\right)^{k}m!x^{k}}{k!k!\left(m-k\right)!}$ and the integral form of two-variable Hermite polynomials $H_{m,n}\left(\xi,\eta\right)=\left(-1\right)^{n}e^{\xi\eta}\int\frac{dz^{2}}{\pi}z^{n}z^{\ast m}e^{-\left|z\right|^{2}+\xi z-\eta z^{\ast}}$. 

If we input a single-mode squeezed state, i.e~$\left|\psi\right\rangle_{a}=\left|r\right\rangle$ with $r$ being the squeezing parameter, after performing $m$-photon catalytic operation, we can obtain the normalized catalytic squeezed state as
\begin{equation}\label{eq:A7}
	\begin{aligned}
		\left|\psi_{s}'\right\rangle =&N_{m}':\mathcal{L}_{m}\left(\hat{a}^{\dagger}\hat{a}\tan^{2}\theta\right):\left(\cos\theta\right)^{\hat{a}^{\dagger}\hat{a}+m}\left|r\right\rangle _{a}\\
		=&\frac{1}{m!}\frac{\partial^{m}}{\partial\tau^{m}}\left\{ \frac{1}{1-\tau}\left(\frac{1-\tau \sec^{2}\theta}{1-\tau}\right)^{\hat{a}_{i}^{\dagger}\hat{a}_{i}}\right\} \left(\cos\theta\right)^{\hat{a}_{i}^{\dagger}\hat{a}_{i}+m}\left|r\right\rangle _{a}|_{\tau=0} \\
		=&N_{m}'\left(\cos\theta\right)^{m}\frac{1}{m!}\frac{\partial^{m}}{\partial\tau^{m}}\left\{ \frac{1}{1-\tau}\left(\frac{1-\tau \sec^{2}\theta}{1-\tau}\right)^{\hat{a}^{\dagger}\hat{a}}\right\} \\
		&\times\sqrt{\operatorname{sech}r}e^{\hat{a}^{\dagger}\hat{a}ln\left(\cos\theta\right)}\exp\left(\frac{1}{2}\hat{a}^{\dagger2}\tanh r\right)\left|0\right\rangle _{a}|_{\tau=0}\\
		=&N_{m}'\left(\cos\theta\right)^{m}\frac{1}{m!}\frac{\partial^{m}}{\partial\tau^{m}}\left\{ \frac{1}{1-\tau}\left(\frac{1-\tau \sec^{2}\theta}{1-\tau}\right)^{\hat{a}^{\dagger}\hat{a}}\right\} \\&\times\sqrt{\operatorname{sech}r}\exp\left(\frac{1}{2}\hat{a}^{\dagger2}\cos^{2}\theta \tanh r\right)\left|0\right\rangle _{a}|_{\tau=0} \\
	\end{aligned}
\end{equation}

Substituting  the transformation relation $e^{\hat{a}^{\dagger}\hat{a}\chi}\hat{a}^{\dagger}e^{-\hat{a}^{\dagger}\hat{a}\chi}=e^{\chi}\hat{a}^{\dagger}$ into (\ref{eq:A7}), we can obtain
\begin{equation}\label{eq:A8}
	\begin{aligned}
		\left|\psi_{s}'\right\rangle
		=&N_{m}'\frac{\sqrt{\operatorname{sech}r}\left(\cos\theta\right)^{m}}{m!}\frac{\partial^{m}}{\partial\tau^{m}}\left\{ \frac{e^{\hat{a}^{\dagger}\hat{a}\lambda}}{1-\tau}\right\}e^{\frac{1}{2}\hat{a}^{\dagger2}\cos^{2}\theta \tanh r}\left|0\right\rangle _{a}|_{\tau=0}\\
		=&N_{m}'\frac{\sqrt{\operatorname{sech}r}\left(\cos\theta\right)^{m}}{m!}\frac{\partial^{m}}{\partial\tau^{m}}\left\{ \frac{e^{\frac{1}{2}\left(\hat{a}^{\dagger}e^{\lambda}\right)^{2}\cos^{2}\theta \tanh r}}{1-\tau}\right\} \left|0\right\rangle _{a}|_{\tau=0}\\
		=&N_{m}'\frac{\sqrt{\operatorname{sech}}\left(\cos\theta\right)^{m}}{m!} \frac{\partial^{m}}{\partial\tau^{m}}\left\{ \frac{e^{\frac{1}{2}\cos^{2}\theta \tanh r C_{\tau}^{2}\hat{a}^{\dagger2}}}{1-\tau}\right\} \left|0\right\rangle _{a}|_{\tau=0} \\
		=&\widetilde{N}_{m}\frac{\partial^{m}}{\partial\tau^{m}}\left\{ \frac{e^{\frac{1}{2}\cos^{2}\theta \tanh r C_{\tau}^{2}\hat{a}^{\dagger2}}}{1-\tau}\right\} \left|0\right\rangle _{a}|_{\tau=0}
	\end{aligned}
\end{equation}
where $\lambda=\ln\left(\frac{1-\tau \sec^{2}\theta}{1-\tau}\right)$, $C_{\tau}=\frac{1-\tau \sec^{2}\theta}{1-\tau}$, $\widetilde{N}_{m}=N_{m}'\frac{\sqrt{\operatorname{sech}}\left(\cos\theta\right)^{m}}{m!}$ with $N_{m}'$ is the normalization constant. Making use of $\left\langle \psi_{s}'\right.\left|\psi_{s}'\right\rangle=1$, we can obtain the coefficient associated with the normalization constant
\begin{equation}
	\begin{aligned}
		 \widetilde{N}_{m}^{-2}&=\left\langle 0\right|\frac{\partial^{m}}{\partial t^{m}}\frac{\partial^{m}}{\partial\tau^{m}}\frac{e^{\left(\frac{1}{2}B\hat{a}^{2}A_{t}^{2}\right)}}{1-t}\frac{e^{\left(\frac{1}{2}B\hat{a}^{\dagger2}C_{\tau}^{2}\right)}}{1-\tau}\left|0\right\rangle _{a}|_{t,\tau=0}\\
		&=\widetilde{N}_{m}^{2}\mathcal{D}_{m}\Delta_{t,\tau}\frac{1}{\pi}\int d^2z  e^{-\left|z\right|^{2}+\frac{1}{2}Bz^{2}A_{t}^{2}+\frac{1}{2}Bz^{\ast2}C_{\tau}^{2}}\\
		&=\mathcal{D}_{m}\left(\frac{\Delta_{t,\tau}}{\sqrt{1-B^{2}A_{t}^{2}C_{\tau}^{2}}}\right)\\
		&=\mathcal{D}_{m}\left(\frac{\Delta_{t,\tau}}{\sqrt{g_{t,\tau}}}\right)|_{t,\tau=0}=1
	\end{aligned}
\end{equation}
where $\mathcal{D}_{m}=\frac{\partial^{m}}{\partial t^{m}}\frac{\partial^{m}}{\partial\tau^{m}}\left(.\right)|_{t,\tau=0}$, $A_{t}=\frac{1-t\sec^{2}\theta}{1-t}$, $B=\cos^{2}\theta \tanh r$, $\Delta_{t,\tau}=\frac{1}{1-t}\frac{1}{1-\tau}$, and $g_{t,\tau}=1-B^{2}A_{t}^{2}C_{\tau}^{2}$.

\section{Successful probability of generating catalytic entangled states}\label{B_HEP}

In this appendix, we calculate successful probability of generating catalytic W-type coherent  and squeezd states.
In our DQN sensing scheme, multi-photon catalysis is realized by  $d+1$ independent BSs. If the input quantum state of the $d+1$ catalyzed modes $a_{j}$ are $d+1$-mode entangled states $\left|\Psi_{\mathrm{w}}\right\rangle$, and the input quantum state of the $d+1$ ancillary catalysis modes $b_{j}$ are Fock states $\otimes_{j=0}^{d}\left|m\right\rangle _{b_{j}}$, so the input state can be written as $\left|\Psi_{in}\right\rangle =\left|\Psi_{\mathrm{w}}\right\rangle \otimes_{j=0}^{d}\left|m\right\rangle _{b_{j}}$, the corresponding output state of the catalytic network can be written as follows
\begin{equation}\label{eq:B1}
	\left|\Psi_{out}\right\rangle =\otimes_{j=0}^{d}\hat{B}_{j}\left(\theta\right)\left|\Psi_{\mathrm{w}}\right\rangle \left|m\right\rangle _{b_{j}}.
\end{equation}
 
Then $d+1$ local $m$-photon number measurements are performed on the $b_{j}$ mode, which can be represented by the $d+1$ independent positive operator-value measure (POVM) $\hat{M}=\otimes_{j=0}^{d}\left|m\right\rangle_{b_{j}b_{j}} \left\langle m\right|$. After carry out the POVM,  the output state given by Eq. (\ref{eq:B1}) becomes
\begin{equation}
	\begin{aligned}
		\left|\Psi_{out}'\right\rangle &=\otimes_{j=0}^{d}\left|m\right\rangle_{b_{j}}  \left\langle m\right|_{b_{j}} \hat{B}_{j}\left(\theta\right)\left|\Psi_{\mathrm{w}}\right\rangle \left|m\right\rangle _{b_{j}}, \\
		&=\otimes_{j=0}^{d}\hat{C}_{j}\left(\theta\right)\left|\Psi_{\mathrm{w}}\right\rangle \left|m\right\rangle _{b_{j}}.
	\end{aligned}
\end{equation}

When the measurement on the output catalysis modes $b_{j}$ yields the result $\left|m\right\rangle_{b_{j}}$, then the reduced state of the output signal mode becomes:
\begin{equation}
	\begin{aligned}
		\left|\Psi_{\mathrm{c}}\right\rangle &=\otimes_{j=0}^{d}\frac{1}{\sqrt{P_{c}}}\hat{C}_{j}\left(\theta\right)\left|\Psi_{\mathrm{w}}\right\rangle,
	\end{aligned}
\end{equation}
where the probability of detecting $m$ photons is given by
\begin{equation}
	\begin{aligned}
		P_{c}=&\left\langle\Psi_{out}\right|\hat{M}\left|\Psi_{out}\right\rangle,\\
		=&\otimes_{j=0}^{d}\left\langle\Psi_{\mathrm{w}}\right| \left\langle m\right| _{b_{j}}\hat{B}^{\dagger}_{j}\left(\theta\right)\left|m\right\rangle_{b_{j}} \\
		&\times \left\langle m\right|_{b_{j}} \hat{B}_{j}\left(\theta\right)\left|\Psi_{\mathrm{w}}\right\rangle \left|m\right\rangle _{b_{j}}, \\
		=&\otimes_{j=0}^{d}\left\langle\Psi_{\mathrm{w}}\right| \hat{C}^{\dagger}_{j}(\theta)\hat{C}_{j}(\theta)\left|\Psi_{\mathrm{w}}\right\rangle \\
	\end{aligned}
\end{equation}
which is the successful probability of generating a catalytic $d+1$-mode entangled quantum state.

For the globally catalytic W-type coherent state $\left|\Psi_{\mathrm{cwc}}\right\rangle$ given by Eq.~(\ref{eq:10}), making use of Eq. (\ref{eq:A3}) and  Eq. (\ref{eq:A5}), we can calculate  the successful probability  
\begin{equation}\label{eq:B5}
	\begin{aligned}
		P_{\mathrm{cwc}}=&\left\langle \Psi_{\mathrm{wc}}\right|\otimes_{j=0}^{d}\hat{C}_{j}^{\dagger}\left(\theta\right)\hat{C}_{j}\left(\theta\right)\left|\Psi_{\mathrm{wc}}\right\rangle ,\\
		=&\frac{\left(d+1\right)\mathcal{N}_{1}^{2}}{N_{m}^{2}}\left(\cos\theta\right)^{2md}\Big[1+d\left|\left\langle \psi'\right|\left.0\right\rangle \right|^{2}\Big],\\
		=&\frac{\left(d+1\right)\mathcal{N}_{1}^{2}}{N_{m}^{2}}\left(\cos\theta\right)^{2md}\Big[1+dN_{m}^{2}\left(\cos\theta\right)^{2m}e^{-\left|\alpha\right|^{2}}\Big],\\
		=&\mathcal{N}_{1}^{2}N_{m}^{-2}\mathcal{N}_{2}^{-2}\left(\cos\theta\right)^{2md},
	\end{aligned}
\end{equation}
which is the expression of  the successful probability given by Eq. (\ref{eq:17}) in Sec. \ref{section 2}.

For the partially  catalytic W-type coherent state $\left|\Psi_{\mathrm{pcwc}}\right\rangle$ given by Eq. (\ref{eq:23}),  making use of Eq.~(\ref{eq:A3}) and  Eq. (\ref{eq:A5}), we can obtain the successful probability as follows
\begin{equation}\label{eq:B6}
	\begin{aligned}
		P_{\mathrm{pcwc}}=&\left\langle \Psi_{\mathrm{wc}}\right|\otimes_{j=0}^{s}\hat{C}_{j}^{\dagger}\left(\theta\right)\hat{C}_{j}\left(\theta\right)\left|\Psi_{\mathrm{wc}}\right\rangle \\
		=&\mathcal{N}_{1}^{2}\Biggl\{\left(s+1\right)\Big[\frac{(\cos\theta)^{2ms}}{N_{m}^{2}}(\left\langle \psi'|\psi'\right\rangle +s|\left\langle \psi'|0\right\rangle|^{2})\\
		&+(d-s)\frac{(\cos\theta)^{2ms+m}}{N_{m}}\left\langle \psi'|0\right\rangle\left\langle 0|\alpha\right\rangle\Big]\\
		&+(d-s)\big[\frac{(s+1)(\cos\theta)^{2ms+m}}{N_{m}}\left\langle 0|\psi'\right\rangle \left\langle \alpha|0\right\rangle\\
		&+(\cos\theta)^{2m(s+1)}(1+(d-s-1)|\left\langle\alpha|0\right\rangle|^{2})\big]\Biggr\},\\
		=&\mathcal{N}_{1}^{2}\left(\cos\theta\right)^{2ms}\big\{(s+1)\left[\left(N_{m}\right)^{-2}+d(\cos\theta)^{2m}\exp(-|\alpha|^{2})\right]\\&+(d-s)(\cos\theta)^{2m}\left[1+d\exp(-|\alpha|^{2})\right]\big\},
	\end{aligned}
\end{equation}
which is the expression of  the successful probability given by Eq. (\ref{eq:28}) in Sec. \ref{section 2}.
In above derivation we have used  
\begin{equation}
	\begin{aligned}
		\left|\psi'\right\rangle&=N_m\hat{C}\left(\theta\right)\left|\alpha\right\rangle,  \\
		\mathcal{N}_{2}^{-2}&=\left(d+1\right)\left[1+dN_{m}^{2}\left(\cos\theta\right)^{2m}e^{-\left|\alpha\right|^{2}}\right].\\
	\end{aligned}
\end{equation}

For the globally catalytic W-type squeezed state given by Eq. (\ref{eq:30}),  making use of Eq. (\ref{eq:A3}) and  Eq. (\ref{eq:A5}), we can obtain the successful probability as follows
\begin{equation}\label{eq:B8}
	\begin{aligned}
		P_{\mathrm{cws}}&=\left\langle \Psi_{\mathrm{ws}}\right|\otimes_{j=0}^{d}\hat{C}_{j}^{\dagger}\left(\theta\right)\hat{C}_{j}\left(\theta\right)\left|\Psi_{\mathrm{ws}}\right\rangle\\
		&=\frac{\left(d+1\right)\mathcal{N}_{3}^{2}}{N_{m}^{'2}}\left(\cos\theta\right)^{2md}\left[1+d\left|\left\langle \psi_{s}'\right|\left|0\right\rangle \right|^{2}\right]\\
		&=\frac{\left(d+1\right)\mathcal{N}_{3}^{2}}{N_{m}^{'2}}\left(\cos\theta\right)^{2md}\left(1+d\widetilde{N}_{m}^{2}m!^{2}\right)\\
		&=\mathcal{N}_{3}^{2}N_{m}^{'-2}\mathcal{N}_{4}^{-2}\left(\cos\theta\right)^{2md},
	\end{aligned}
\end{equation}
which is the expression of  the successful probability given by Eq. (\ref{eq:33}) in Sec. \ref{section 3}.

Simmilarly, for the partially catalytic W-type squeezed state $\left|\Psi_{\mathrm{pcws}}\right\rangle$ given by Eq. (\ref{eq:39}), we can obtain the successful probability as follows
\begin{equation}\label{eq:B9}
	\begin{aligned}
		P_{\mathrm{pcws}}=&\left\langle \Psi_{\mathrm{ws}}\right|\otimes_{j=0}^{s}\hat{C}_{j}^{\dagger}\left(\theta\right)\hat{C}_{j}\left(\theta\right)\left|\Psi_{\mathrm{ws}}\right\rangle ,\\=&\mathcal{N}_{3}^{2}\Biggl\{\left(s+1\right)\Big[\frac{(\cos\theta)^{2ms}}{N_{m}'^{2}}(\left\langle \psi_{s}'|\psi_{s}'\right\rangle +s|\left\langle \psi_{s}'|0\right\rangle|^{2})\\&+(d-s)\frac{(\cos\theta)^{2ms+m}}{N_{m}'}\left\langle \psi'_{s}|0\right\rangle\left\langle 0|r\right\rangle\Big]\\&+(d-s)\big[\frac{(s+1)(\cos\theta)^{2ms+m}}{N_{m}'}\left\langle 0|\psi'_{s}\right\rangle \left\langle r|0\right\rangle\\&+(\cos\theta)^{2m(s+1)}(1+(d-s-1)|\left\langle r|0\right\rangle|^{2})\big]\Biggr\},\\
		=&\mathcal{N}_{3}^{2}\Big\{\left(s+1\right)\left[\left(N_{m}'\right)^{-2}+d\operatorname{sech} r\left(\cos\theta\right)^{2m}\right]\\&+\left(d-s\right)\left(\cos\theta\right)^{2m}\left(1+d\operatorname{sech} r\right)\Big\}.
	\end{aligned}
\end{equation}
which is the expression of  the successful probability given by Eq. (\ref{eq:44}) in Sec. \ref{section 3}.
In above derivation we have used  
\begin{equation}
	\begin{aligned}
		\left|\psi_{s}'\right\rangle&=N_{m}'\hat{C}\left(\theta\right)\left|r\right\rangle, \\
		\mathcal{N}_{4}^{-2}&=\left[\left(d+1\right)\left(1+d\widetilde{N}_{m}^{2}m!^{2}\right)\right]. \\
	\end{aligned}
\end{equation}

\section{QFI for multi-photon catalytic quantum states}\label{C_HEP}

In this appendix, we present calculations of the QFI for globally and partially catalytic W-type of coherent and squeezed states.
We now discuss the effect of catalytic operations on QFI. In our schemes, the multi-photon catalytic pure quantum state is allowed to pass through an optical network, which is subsequently phase-encoded, and eventually, its phase is estimated. The encoded quantum state as
\begin{equation}
	\left|\psi_{\mathrm{c}}\left(\phi\right)\right\rangle=e^{-i\hat{a}^{\dagger}\hat{a}\phi}\left|\psi_{\mathrm{c}}\right\rangle, 
\end{equation}

According to the formula for the QFI of the pure quantum state $F=4\left[\left\langle \partial_{\phi}\psi\left(\phi\right)|\partial_{\phi}\psi\left(\phi\right)\right\rangle-\left|\left\langle \partial_{\phi}\psi\left(\phi\right)|\psi\left(\phi\right)\right\rangle\right|^{2}\right]$, we can obtain the QFI of the multi-photon catalytic pure quantum state, as followed
\begin{equation}\label{eq:C2}
	\begin{aligned}
		\mathcal{F}'&=4\left[\left\langle \psi_{\mathrm{c}}\right|\hat{n}^{2}\left|\psi_{\mathrm{c}}\right\rangle-\left|\left\langle \psi_{\mathrm{c}}\right|\hat{n}\left|\psi_{\mathrm{c}}\right\rangle\right|^{2}\right]
	\end{aligned}
\end{equation}

If the input pure state is a coherent state, the quantum state after multiphoton catalysis becomes $\left|\psi_{\mathrm{c}}\right\rangle=\left|\psi'\right\rangle$. In order to calculate the QFI, we need the following average value
\begin{equation}\label{eq:C3}
	\begin{aligned}
		A_{qp}=&\left\langle \psi'\right|\hat{a}^{q}\hat{a}^{\dagger p}\left|\psi'\right\rangle\\ =&\left|\bar{N}_{m}\right|^{2}\left\langle \alpha_{\theta}\right|\mathcal{L}_{m}\left(\mu^{\ast}\hat{a}\right)\hat{a}^{q}\hat{a}^{\dagger p}\mathcal{L}_{m}\left(\mu\hat{a}^{\dagger}\right)\left|\alpha_{\theta}\right\rangle \\
		=&\frac{1}{\pi}\int\left|\bar{N}_{m}\right|^{2}\left\langle \alpha_{\theta}\right|\mathcal{L}_{m}\left(\mu^{\ast}\hat{a}\right)\hat{a}^{q}
		\left|z\right\rangle  \left\langle z\right|\hat{a}^{\dagger p}\mathcal{L}_{m}\left(\mu\hat{a}^{\dagger}\right)\left|\alpha_{\theta}\right\rangle dz^{2}\\
		=&\frac{1}{\pi}\int\left|\bar{N}_{m}\right|^{2}\mathcal{L}_{m}\left(\mu^{\ast}z\right)z^{q}z^{\ast p}\mathcal{L}_{m}\left(\mu z^{\ast}\right)\left|\left\langle\alpha_{\theta}|z\right\rangle \right|^{2}dz^{2}\\=&\left|\bar{N}_{m}\right|^{2}e^{-\left|\alpha\right|^{2}\cos^{2}\theta}\sum_{n,k=0}^{m}\left(\begin{array}{c}
			m\\
			n
		\end{array}\right)\left(\begin{array}{c}
			m\\
			k
		\end{array}\right)\frac{\left(-1\right)^{n+k}}{n!k!}\mu^{n}\mu^{\ast k}\\
		&\times\int\frac{dz^{2}}{\pi}z^{\left(q+k\right)}(z^{\ast})^{\left(p+n\right)}}e^{-\left|z\right|^{2}+z\alpha_{\theta}^{\ast}+z^{\ast}\alpha_{\theta}\\
		=&\left|\bar{N}_{m}\right|^{2}\sum_{n,k}^{m}\Pi_{n,k}^{m}\left(-\mu,\mu^{\ast}\right)\left(-1\right)^{q}H_{p+n,q+k}\left(\alpha^{\ast}_{\theta},-\alpha_{\theta}\right)
	\end{aligned}
\end{equation}
 where we have used the relationship of $\left|\left\langle\alpha_{\theta}|z\right\rangle\right|^{2}=e^{-\left|\alpha_{\theta}\right|^{2}-\left|z\right|^{2}+(z\alpha^{*}_{\theta}+z^{*}\alpha_{\theta})}$. Substituting Eq. (\ref{eq:C3}) into Eq. (\ref{eq:C2}), we can obtain  the QFI of the multi-photon catalytic coherent state 
\begin{equation}\label{eq:C4}
	\begin{aligned}
		\mathcal{F}'=&4\bar{N}_{m}^{2}\sum_{n,k}^{m}\Pi_{n,k}^{m}\left(-\mu,\mu^{*}\right)H_{n+2,k+2}\left(\alpha^{\ast}_{\theta},-\alpha_{\theta}\right)\\&-4\left[\bar{N}_{m}^{2}\sum_{n,k}^{m}\Pi_{n,k}^{m}\left(-\mu,\mu^{*}\right)H_{n+1,k+1}\left(\alpha^{\ast}_{\theta},-\alpha_{\theta}\right)\right]^{2}\\
		&+4\bar{N}_{m}^{2}\sum_{n,k}^{m}\Pi_{n,k}^{m}\left(-\mu,\mu^{*}\right)H_{n+1,k+1}\left(\alpha^{\ast}_{\theta},-\alpha_{\theta}\right),
	\end{aligned}
\end{equation}

\begin{figure}[t!]
	\centering
	\includegraphics[width=0.45\textwidth]{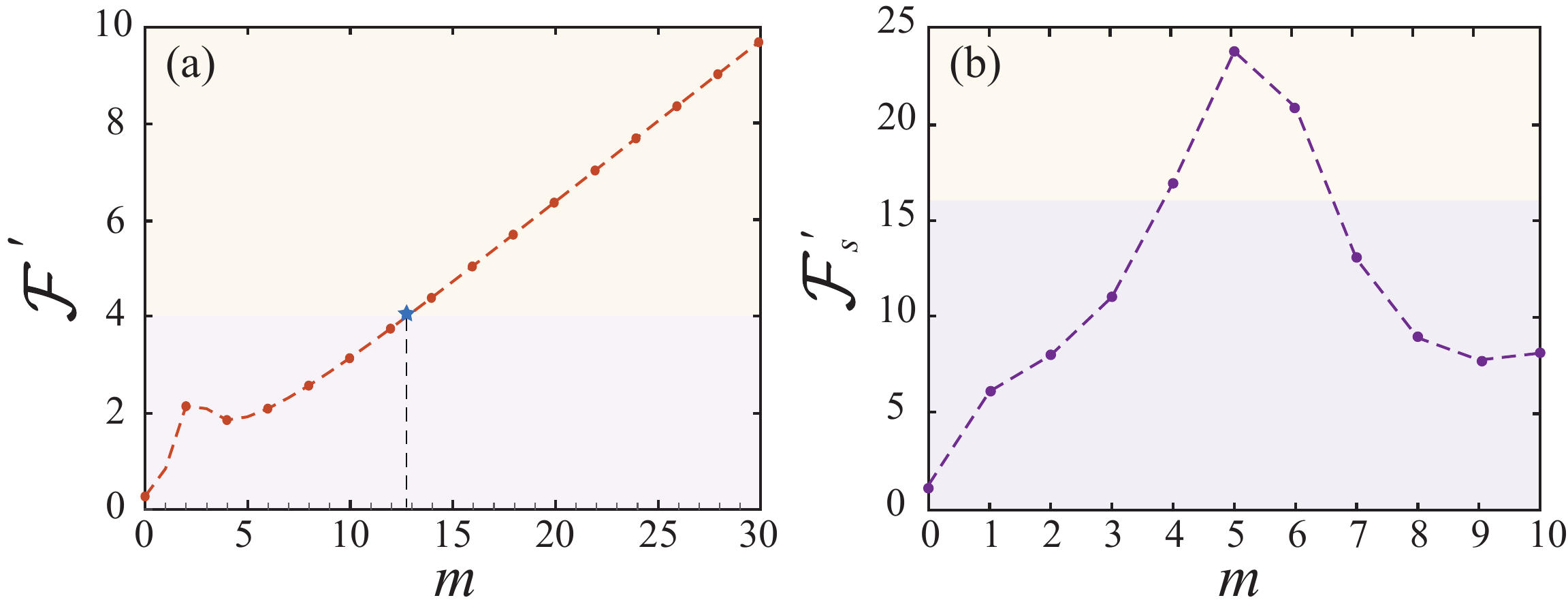}
	\caption{(a) QFI for multi-photon catalytic coherent states with different catalytic photon numbers $m$ with $\alpha=1$, and $50:50$ BS. (b) QFI for multi-photon catalytic squeezed states with different catalytic photon numbers $m$ with $r=0.8814$, and $50:50$ BS. The yellow shading indicates that the catalytic operation improves the estimation precision, while the purple shading indicates the opposite. \label{fig:S2}}
\end{figure}

When $m=0$ and $\theta=0$, Eq. (\ref{eq:C4}) corresponds to the QFI of the coherent state, i.e., $\mathcal{F}_{\mathrm{c}}=4\left|\alpha\right|^{2}$ which reached the standard quantum limit ($\mathcal{F}_{\mathrm{c}}\propto n $), and this result is consistent with Ref.~\cite{ozdemir2019Parity}. In Fig.~\ref{fig:S2} (a), we plot the QFI of the catalytic squeezed state as a funcation of the number of catalytic photons, given an input resource number is $1~(\alpha=1)$ and a transmittance of $1/2$. As shown in Fig.~\ref{fig:S2} (a), we find the higher the number of catalytic photons, the greater the QFI. And with $m>12$, the catalytic coherent state provides better sensitivity than the coherent state, i.e., $\mathcal{F}'\ge4\left|\alpha\right|^{2}$. In other words, if the number of catalytic photons reaches a certain value, the catalytic quantum state can show better performance as a quantum probe for parameter estimation.

According Eq.~(\ref{eq:13}),obtain we can  obtain the effective QFI of multiphoton globally catalyzed multimode W-type coherent state $\left|\Psi_{\mathrm{cwc}}\right\rangle$  as follows
\begin{equation}\label{eq:C5}
	\begin{aligned}
		H_{\mathrm{cwc}}=&4\mathcal{N}_{2}^{2}\left(d\left\langle\hat{n}'^{2}\right\rangle-\mathcal{N}_{2}^{2}d^{2}\left\langle\hat{n}'\right\rangle^{2}\right) \\
		=&4d\mathcal{N}_{2}^{2}\Big[\bar{N}_{m}^{2}\sum_{n,k}^{m}\Pi_{n,k}^{m}\left(-\mu,\mu^{*}\right)H_{n+2,k+2}\left(\alpha^{\ast}_{\theta},-\alpha_{\theta}\right)\\
		&+3\bar{N}_{m}^{2}\sum_{n,k}^{m}\Pi_{n,k}^{m}\left(-\mu,\mu^{*}\right)H_{n+1,k+1}\left(\alpha^{\ast}_{\theta},-\alpha_{\theta}\right)+1\Big] \\
		&-4d^{2}\mathcal{N}_{2}^{4}\Big[\bar{N}^{2}_{m}\sum_{n,k}^{m}\Pi_{n,k}^{m}\left(-\mu,\mu^{*}\right)H_{n+1,k+1}\left(\alpha^{\ast}_{\theta},-\alpha_{\theta}\right)+1\Big]^{2}\\
		=&4d\mathcal{N}_{2}^{2}\bar{N}_{m}^{2}\Big[\sum_{n,k}^{m}\Pi_{n,k}^{m}\left(-\mu,\mu^{*}\right)H_{n+2,k+2}\left(\alpha^{\ast}_{\theta},-\alpha_{\theta}\right) \\
		&+\left(3-2d\mathcal{N}_{2}^{2}\right)\sum_{n,k}^{m}\Pi_{n,k}^{m}\left(-\mu,\mu^{*}\right)H_{n+1,k+1}\left(\alpha^{\ast}_{\theta},-\alpha_{\theta}\right) \\
		&-\mathcal{N}_{2}^{2}\bar{N}_{m}^{2}d\left(\sum_{n,k}^{m}\Pi_{n,k}^{m}\left(-\mu,\mu^{*}\right)H_{n+1,k+1}\left(\alpha^{\ast}_{\theta},-\alpha_{\theta}\right)\right)^{2}\Big] \\
		&+4d\mathcal{N}_{2}^{2}\left(1-d\mathcal{N}_{2}^{2}\right).
	\end{aligned}
\end{equation}

If the input pure state is a squeezed state, the quantum state after multiphoton catalysis becomes $\left|\psi_{\mathrm{c}}\right\rangle=\left|\psi_{s}'\right\rangle$. After phase encoding, and we can similarly compute the QFI after phase encoding using the the Eq.~(\ref{eq:C2}) with the following result
\begin{equation}
	\begin{aligned}
		\mathcal{F}_{s}'=&4(\left\langle \hat{n}_{s}'^{2}\right\rangle-\left\langle \hat{n}_{s}'\right\rangle^{2}),\\ 
		=&4\left[\widetilde{N}_{m}^{2}\mathcal{D}_{m}\left(\frac{\triangle_{t,\tau}h_{t,\tau}}{g_{t,\tau}^{5/2}}\right)-3\widetilde{N}_{m}^{2}\mathcal{D}_{m}\left(\frac{\triangle_{t,\tau}}{g_{t,\tau}^{3/2}}\right)+1\right]\\&-4\left[\widetilde{N}_{m}^{2}\mathcal{D}_{m}\left(\frac{\triangle_{t,\tau}}{g_{t,\tau}^{3/2}}\right)-1\right]^{2}.
	\end{aligned}
\end{equation}
In the derivation of above equation we have used the following average value
\begin{equation}
	\begin{aligned}
		B_{pq}=&\left\langle \psi_{s}'\right|\hat{a}^{p}\hat{a}^{\dagger q}\left|\psi_{s}'\right\rangle \\
		=&\frac{\partial^{p+q}}{\partial s^{p}\partial k^{q}}\left\langle \psi_{s}'\right|e^{s\hat{a}}e^{k\hat{a}^{\dagger}}\left|\psi_{s}'\right\rangle |_{s=k=0},\\=&\widetilde{N}_{m}^{2}\frac{\partial^{p+q}}{\partial s^{p}\partial k^{q}}\frac{\partial^{m}}{\partial t^{m}}\frac{\partial^{m}}{\partial\tau^{m}}\left\langle 0\right|\frac{e^{\frac{1}{2}B\hat{a}^{2}A_{t}^{2}} }{1-t}e^{s\hat{a}}e^{k\hat{a}^{\dagger}}\frac{e^{\frac{1}{2}B\hat{a}^{\dagger2}C_{\tau}^{2}}}{1-\tau}
		\left|0\right\rangle |_{t=\tau=0}{}_{s=k=0},\\=&\widetilde{N}_{m}^{2}\frac{\partial^{p+q}}{\partial s^{p}\partial k^{q}}\mathcal{D}_{m}\frac{\Delta_{t,\tau}}{\pi}\int\left\langle 0\right|\exp(\frac{1}{2}B\hat{a}^{2}A_{t}^{2}+s\hat{a})\left|z\right\rangle \\
		&\times\left\langle z\right|\exp(k\hat{a}^{\dagger}+\frac{1}{2}B\hat{a}^{\dagger2}C_{\tau}^{2})\left|0\right\rangle d^{2}z|_{s=k=0},\\=&\widetilde{N}_{m}^{2}\frac{\partial^{p+q}}{\partial s^{p}\partial k^{q}}\mathcal{D}_{m}\frac{\Delta_{t,\tau}}{\pi}\int e^{-\left|z\right|^{2}+sz+\frac{1}{2}BA_{t}^{2}z^{2}+kz^{\ast}+\frac{1}{2}BC_{\tau}^{2}z^{\ast2}}d^{2}z|_{s=k=0},\\=&\widetilde{N}_{m}^{2}\mathcal{D}_{m}\frac{\Delta_{t,\tau}}{\sqrt{g_{t,\tau}}}\frac{\partial^{p+q}}{\partial s^{p}\partial k^{q}}\exp\left\{ \frac{sk+\frac{1}{2}B\left(s^{2}C_{\tau}^{2}+k^{2}A_{t}^{2}\right)}{g_{t,\tau}}\right\} |_{s=k=0},
	\end{aligned}
\end{equation}

Specifically, when $s=k=1$, we obtain
\begin{equation}
	\begin{aligned}
		B_{11}&=\left\langle \psi_{s}'\right|\hat{a}\hat{a}^{\dagger}\left|\psi_{s}'\right\rangle\\ &=\widetilde{N}_{m}^{2}\mathcal{D}_{m}\frac{\Delta_{t,\tau}}{\sqrt{g_{t,\tau}}}\frac{\partial^{2}}{\partial s\partial k}\exp\left\{ \frac{sk+\frac{1}{2}B\left(s^{2}C_{\tau}^{2}+k^{2}A_{t}^{2}\right)}{g_{t,\tau}}\right\} |_{s=k=0},\\&=\widetilde{N}_{m}^{2}\mathcal{D}_{m}\left(\frac{\Delta_{t,\tau}}{g_{t,\tau}^{3/2}}\right),
	\end{aligned}
\end{equation}
when $s=k=2$, we have
\begin{equation}
	\begin{aligned}
		B_{22}&=\left\langle \psi_{s}'\right|\hat{a}^{2}\hat{a}^{\dagger2}\left|\psi_{s}'\right\rangle\\ &=\widetilde{N}_{m}^{2}\mathcal{D}_{m}\frac{\Delta_{t,\tau}}{\sqrt{g_{t,\tau}}}\frac{\partial^{4}}{\partial s^{2}\partial k^{2}}\exp\left\{ \frac{sk+\frac{1}{2}B\left(s^{2}C_{\tau}^{2}+k^{2}A_{t}^{2}\right)}{g_{t,\tau}}\right\} |_{s=k=0},\\&=\widetilde{N}_{m}^{2}\mathcal{D}_{m}\left(\frac{\Delta_{t,\tau}h_{t,\tau}}{g_{t,\tau}^{5/2}}\right),
	\end{aligned}
\end{equation}
in which $h_{t,\tau}=2+B^{2}C_{\tau}^{2}A_{t}^{2}$.

When $m=0$ and $\theta=0$, it corresponds to the QFI of the squeezed state, i.e., $\mathcal{F}_{s}=8\sinh^{2}r\left(\sinh^{2}r+1\right)$ which reached the Heisenberg quantum limit ($\mathcal{F}_{s}\propto n^{2} $). In Fig.~\ref{fig:S2} (b), we plot the QFI of the catalytic squeezed state as a funcation of the number of catalytic photons, given an input resource number is $1~(r=0.8814)$ and a transmittance of $1/2$. As shown in the figure, we find that the behavior of the QFI for the catalyzed squeezed state differs from that of the catalyzed coherent state as the number of catalytic photons increases. In this case, there exists an optimal number of catalytic photons that maximizes the QFI, beyond which further increases in the number of catalytic photons lead to a reduction in the QFI. Moreover, the QFI of the catalyzed squeezed state exceeds that of the squeezed state with the same input resource number only when the number of catalytic photons falls within a specific range$(4\le m\le6)$.

By comparing Figure (a) and Figure (b), we observe that for the same input resource, the QFI of the squeezed state exceeds that of the coherent state (i.e., $\mathcal{F}_{c}<\mathcal{F}_{s}$). This confirms that quantum probes utilizing quantum resources exhibit superior sensing performance compared to those without, a conclusion already supported by Ref.~\cite{abbott2016observation}. Additionally, we find that for the same number of catalytic photons, the QFI of the catalyzed squeezed state is greater than that of the catalyzed coherent state, demonstrating that hybridizing two quantum resources provides better sensing performance than using only one quantum resource.

Meanwhile, according Eq.~(\ref{eq:13}), the effective QFI of multiphoton globally catalyzed multimode W-type squzeed state $\left|\Psi_{\mathrm{cws}}\right\rangle$ can be obtained
\begin{equation}\label{eq:C10}
	\begin{aligned}
		H_{\mathrm{cws}}=&4\mathcal{N}_{4}^{2}\left(d\left\langle\hat{n}_{s}'^{2}\right\rangle-\mathcal{N}_{2}^{2}d^{2}\left\langle\hat{n}_{s}'\right\rangle^{2}\right), \\
		=&4\mathcal{N}_{4}^{2}d\left[\widetilde{N}^{2}\mathcal{D}_{m}\left(\frac{\triangle_{t,\tau}h_{t,\tau}}{g_{t,\tau}^{5/2}}\right)-3\widetilde{N}_{m}^{2}\mathcal{D}_{m}\left(\frac{\triangle_{t,\tau}}{g_{t,\tau}^{3/2}}\right)+1\right],\\
		&-4\mathcal{N}_{4}^{4}d^{2}\left[\widetilde{N}_{m}^{2}\mathcal{D}_{m}\left(\frac{\triangle_{t,\tau}}{g_{t,\tau}^{3/2}}\right)-1\right]^{2}.
	\end{aligned}
\end{equation}

\begin{figure}[t!]
	\centering
	\includegraphics[width=0.45\textwidth]{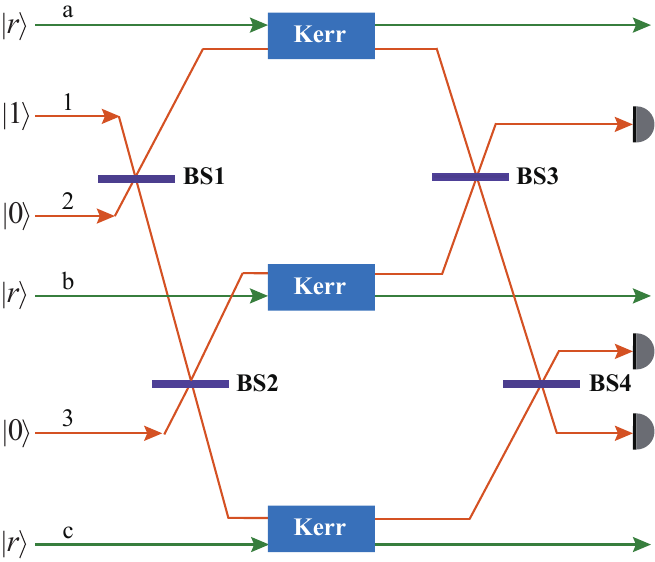}
	\caption{Scheme for generating three-mode catalytically entangled squeezed states, in which there exist $a$,$1$,$2$,$b$,$3$,$c$ paths from top to bottom. BS1, BS2, BS3, and BS4 are the BSs for forming entanglement states. $\left| r\right\rangle$ is a squeezed state, and Kerr denotes the Kerr interaction of two associated paths.} \label{fig:S3}
\end{figure}

\section{Experimental protocol for generating W-type squeezed state }\label{F_HEP} 
In this appendix, we present an experimental scheme to produce  W-type squeezed state.
One of the concerns of networked quantum sensing is how to design a network to generate multimode quantum probes for quantum metrology. In Ref.~\cite{Maleki2022Distributed}, a feasible scheme for generating multi-mode W-type coherent states in experiments is proposed. Here we extend this scheme to  generating multi-mode W-type squeezed  state.

In this scheme, we first prepare the state $\left|1\right\rangle$ $\left|0\right\rangle$ as the input state of the 50:50  BS1 and BS2, in which  a single
photon $\left|1\right\rangle$ is injected into the beam-splitter setup in path 1, however, path 2 and path 3 are kept in vacuum states $\left|0\right\rangle$, as shown in Fig.\ref{fig:S3}. The reflectance and transmittance of the $i$-th BS are respectively $(1-\tau_{i})$, $\tau_{i}$. Therefore, the $i$-th beam BS is transformed as follows
\begin{equation}
	\begin{aligned}
		\left|1\right\rangle\left|0\right\rangle\longrightarrow \tau_{i}\left|1\right\rangle\left|0\right\rangle+(1-\tau_{i})\left|0\right\rangle\left|1\right\rangle, \\
		\left|0\right\rangle\left|1\right\rangle\longrightarrow \tau_{i}\left|0\right\rangle\left|1\right\rangle-(1-\tau_{i})\left|1\right\rangle\left|0\right\rangle,
	\end{aligned}
\end{equation}
Thus after the BS1, the system  becomes
\begin{equation}
	\begin{aligned}
		\left|\psi_{0}\right\rangle=&\left(\tau_{1}\left|1\right\rangle_{1}\left|0\right\rangle_{2}\left|0\right\rangle_{3}+(1-\tau_{1})\left|0\right\rangle_{1}\left|1\right\rangle_{2}\left|0\right\rangle_{3}\right) \\
		&\otimes\left| r\right\rangle_{a}\left| r\right\rangle_{b}\left| r\right\rangle_{c},
	\end{aligned}
\end{equation}
Similarly, after the BS2, the system becomes
\begin{equation}
	\begin{aligned}
		\left|\psi_{1}\right\rangle=&\Big(\tau_{1}\tau_{2}\left|1\right\rangle_{1}\left|0\right\rangle_{2}\left|0\right\rangle_{3}+\tau_{1}(1-\tau_{2})\left|0\right\rangle_{1}\left|0\right\rangle_{2}\left|1\right\rangle_{3} \\
		&+(1-\tau_{1})\left|0\right\rangle_{1}\left|1\right\rangle_{2}\left|0\right\rangle_{3}\Big)\otimes\left| r\right\rangle_{a}\left| r\right\rangle_{b}\left| r\right\rangle_{c}.
	\end{aligned}
\end{equation}
Next, the single photon of each path is coupled to an external mode of squeezed state $\left|  r\right\rangle$ through cross-Kerr interactions $\hat{U}\left(t\right)=e^{-ikt\hat{b}^{\dagger}\hat{b}\hat{a}^{\dagger}\hat{a}}$ formed by a nonlinear Kerr medium, where $\hat{a}$($\hat{a}^{\dagger}$) is the annihilation (creation) operator
of photons in the one arm of the interferometer and $\hat{b}$($\hat{b}^{\dagger}$) is
the annihilation (creation) operator of the interacting squeezed
state~\cite{Maleki2022Distributed}. We make the Kerr interaction time $t$, then have $kt=\frac{\pi}{2}$. Then after three Kerr mediums, the whole system becomes
\begin{equation}
	\begin{aligned}
		\left|\psi_{2}\right\rangle=&\tau_{1}\tau_{2}\left|1\right\rangle_{1}\left|0\right\rangle_{2}\left|0\right\rangle_{3}\otimes\left| r\right\rangle_{a}\left| r\right\rangle_{b}\left|-r\right\rangle_{c} \\
		&+\tau_{1}(1-\tau_{2})\left|0\right\rangle_{1}\left|0\right\rangle_{2}\left|1\right\rangle_{3}\otimes\left| r\right\rangle_{a}\left|-r\right\rangle_{b}\left| r\right\rangle_{c} \\
		&+(1-\tau_{1})\left|0\right\rangle_{1}\left|1\right\rangle_{2}\left|0\right\rangle_{3}\otimes\left|-r\right\rangle_{a}\left| r\right\rangle_{b}\left| r\right\rangle_{c}, \\
		=&a\left|1,0,0\right\rangle\left|-r,r,r\right\rangle+b\left|0,1,0\right\rangle\left|r,-r,r\right\rangle \\
		&+c\left|0,0,1\right\rangle\left|r,r,-r\right\rangle, \\
	\end{aligned}
\end{equation}
where $a=\tau_{1}\tau_{2}$, $b=\tau_{1}(1-\tau_{2})$ and $c=(1-\tau_{1})$. Next, we then perform an anti-squeezing transformation on each mode, with the anti-squeezing parameter set equal in magnitude to the squeezing strength of the input state. This operation is locally performed on
each squeezed state mode. %similar to the displacement oper-
%ation demonstrated in [40–42].
In other words, we apply the anti-squeezing operator $\hat{S}\left(-r\right)\hat{S}\left(-r\right)\hat{S}\left(-r\right)$ to the state and attain
\begin{equation}
	\begin{aligned}
		\left|\psi_{3}\right\rangle=&a\left|1,0,0\right\rangle\left|-2r,0,0\right\rangle+b\left|0,1,0\right\rangle\left|0,-2r,0\right\rangle \\
		&+c\left|0,0,1\right\rangle\left|0,0,-2r\right\rangle
	\end{aligned}. \\
\end{equation}
Then, making $-2r=\xi$. Next, the system goes through BS3, which results in
\begin{equation}
	\begin{aligned}
		\left|\psi_{4}\right\rangle=&a\left|1,0,0\right\rangle \left|\xi,0,0\right\rangle\\
		&+\left|0,1,0\right\rangle\left(b\tau_{3}\left|0,\xi,0\right\rangle-cr_{3}\left|0,0,\psi'\right\rangle\right) \\
		&+\left|0,0,1\right\rangle\left(br_{3}\left|0,\xi,0\right\rangle+c\tau_{3}\left|0,0,\xi\right\rangle\right), \\
	\end{aligned}
\end{equation}
similarly, applying BS4 provides
\begin{equation}
	\begin{aligned}
		\left|\psi_{5}\right\rangle=&a\tau_{4}\left|\xi,0,0\right\rangle\left|1,0,0\right\rangle+c(1-\tau_{3})(1-\tau_{4})\left|0,0,\xi\right\rangle\left|1,0,0\right\rangle \\
		&-b\tau_{3}(1-\tau_{4})\left|0,\xi,0\right\rangle\left|1,0,0\right\rangle+a\tau_{4}\left|\xi,0,0\right\rangle\left|0,1,0\right\rangle \\
		&+b\tau_{3}\tau_{4}\left|0,\xi,0\right\rangle\left|0,1,0\right\rangle-cr_{3}\tau_{4}\left|0,0,\xi\right\rangle\left|0,1,0\right\rangle \\
		&+b(1-\tau_{3})\left|0,\xi,0\right\rangle\left|0,0,1\right\rangle +c\tau_{3}\left|0,0,\xi\right\rangle\left|0,0,1\right\rangle.
	\end{aligned}
\end{equation}
Finally, we post-select the state by performing the single-photon detection in 3 paths as a trigger. If we detect a single photon at the $1$st path, the corresponding output state at this time is
\begin{equation}
	at\tau_{4}\left|\xi,0,0\right\rangle+c(1-\tau_{3})(1-\tau_{4})\left|0,0,\xi\right\rangle-b\tau_{3}(1-\tau_{4})\left|0,\xi,0\right\rangle,
\end{equation}
in order to get the state we want, we make $a\tau_{4}=c(1-\tau_{3})(1-\tau_{4})=-b\tau_{3}(1-\tau_{4})$, i.e. $\tau_{1}\tau_{2}\tau_{4}=(1-\tau_{1})(1-\tau_{3})(1-\tau_{4})=-\tau_{1}(1-\tau_{2})\tau_{3}(1-\tau_{4})$. This limits the range of BSs used and sets the conditions to be met for the required quantum probes to be prepared. Therefore, in this case, we obtained the desired three-mode quantum probe
\begin{equation}
	\left|\psi_{6}\right\rangle=\mathcal{N}\left|\xi,0,0\right\rangle+\left|0,0,\xi\right\rangle+\left|0,\xi,0\right\rangle,
\end{equation}
where $\mathcal{N}$ is the normalization constant, so that we get a three-mode entangled squzeed state. Similarly, if we detect a single photon in the $2$nd path, we can also obtain a three-mode entangled squzeed state, but if we detect a single photon in the $3$rd path, we cannot generate a three-mode entangled squzeed state. It is also possible to extend the scheme to generate the multimode entangled squzeed state. However, the detection of the trigger of a single photon in the $3$rd path results in the generation of useless quantum states, so the method is rather inefficient~\cite{Zhang2017Quantum}.

\bibliography{HRNQS}	

@article{Maleki2022Distributed,
	title = {Distributed phase estimation and networked quantum sensors with {W}-type quantum probes}, 
	author = {Maleki, Yusef and Zubairy, M. Suhail},
	journal = {Phys. Rev. A},
	volume = {105},
	issue = {3},
	pages = {032428},
	numpages = {8},
	year = {2022},
	month = {Mar},
	publisher = {American Physical Society},
	doi = {10.1103/PhysRevA.105.032428},
}

@article{Ge2018Distributed,
	title = {Distributed Quantum Metrology with Linear Networks and Separable Inputs},
	author = {Ge, Wenchao and Jacobs, Kurt and Eldredge, Zachary and Gorshkov, Alexey V. and Foss-Feig, Michael},
	journal = {Phys. Rev. Lett.},
	volume = {121},
	issue = {4},
	pages = {043604},
	numpages = {6},
	year = {2018},
	month = {Jul},
	publisher = {American Physical Society},
	doi = {10.1103/PhysRevLett.121.043604},
}

@article{Proctor2018Multiparameter,   
	title = {Multiparameter Estimation in Networked Quantum Sensors},
	author = {Proctor, Timothy J. and Knott, Paul A. and Dunningham, Jacob A.},
	journal = {Phys. Rev. Lett.},
	volume = {120},
	issue = {8},
	pages = {080501},
	numpages = {6},
	year = {2018},
	month = {Feb},
	publisher = {American Physical Society},
	doi = {10.1103/PhysRevLett.120.080501},
}

@article{Gatto2019Distributed,
	title = {Distributed quantum metrology with a single squeezed-vacuum source},
	author = {Gatto, Dario and Facchi, Paolo and Narducci, Frank A. and Tamma, Vincenzo},
	journal = {Phys. Rev. Research},
	volume = {1},
	issue = {3},
	pages = {032024},
	numpages = {5},
	year = {2019},
	month = {Nov},
	publisher = {American Physical Society},
	doi = {10.1103/PhysRevResearch.1.032024},
}

@article{Li2016Multiphoton,
	author  = {Hu, Li-Yun and Wu, Jia-Ni and Liao, Zeyang and Zubairy, M. Suhail},
	title   = {Multiphoton catalysis with coherent state input: Nonclassicality and decoherence},
	journal = {J. Phys. B: At. Mol. Opt. Phys.},
	volume  = {49},
	pages   = {175504},
	year    = {2016},
	doi     = {10.1088/0953-4075/49/17/175504}
}

@article{lvovsky2002quantum,
	author  = {Lvovsky, A. I. and Mlynek, J.},
	title   = {Quantum-optical catalysis: Generating nonclassical states of light by means of linear optics},
	journal = {Phys. Rev. Lett.},
	volume  = {88},
	pages   = {250401},
	year    = {2002},
	doi     = {10.1103/PhysRevLett.88.250401}
}

@article{Bartley2012Multiphoton,
	title = {Multiphoton state engineering by heralded interference between single photons and coherent states},
	author = {Bartley, Tim J. and Donati, Gaia and Spring, Justin B. and Jin, Xian-Min and Barbieri, Marco and Datta, Animesh and Smith, Brian J. and Walmsley, Ian A.},
	journal = {Phys. Rev. A},
	volume = {86},
	issue = {4},
	pages = {043820},
	numpages = {6},
	year = {2012},
	month = {Oct},
	publisher = {American Physical Society},
	doi = {10.1103/PhysRevA.86.043820},
}

@article{Hu2017Continuous,
	title = {Continuous-variable entanglement via multiphoton catalysis},
	author = {Hu, Liyun and Liao, Zeyang and Zubairy, M. Suhail},
	journal = {Phys. Rev. A},
	volume = {95},
	issue = {1},
	pages = {012310},
	numpages = {15},
	year = {2017},
	month = {Jan},
	publisher = {American Physical Society},
	doi = {10.1103/PhysRevA.95.012310},
}

@article{jia2018comparison,
	author  = {Jia, Fang and Ye, Wei and Wang, Qi and Hu, Li-Yun and Fan, Hong-Yi},
	title   = {Comparison of nonclassical properties resulting from {non-Gaussian} operations},
	journal = {Laser Phys. Lett.},
	volume  = {16},
	pages   = {015201},
	year    = {2018},
	doi     = {10.1088/1612-202X/aaf1fb}
}

@article{zhang2021improved,
	author  = {Zhang, Huan and Ye, Wei and Wei, Chaoping and Xia, Ying and Chang, Shoukang and Liao, Zeyang and Hu, Liyun},
	title   = {Improved phase sensitivity in a quantum optical interferometer based on multiphoton catalytic two-mode squeezed vacuum states},
	journal = {Phys. Rev. A},
	volume  = {103},
	pages   = {013705},
	year    = {2021},
	doi     = {10.1103/PhysRevA.103.013705}
}

@article{zhao2024phase,
	author  = {Zhao, Zekun and Kang, Qingqian and Zhang, Huan and Zhao, Teng and Liu, Cunjin and Hu, Liyun},
	title   = {Phase estimation via coherent and photon-catalyzed squeezed vacuum states},
	journal = {Opt. Express},
	volume  = {32},
	pages   = {28267--28281},
	year    = {2024},
	doi     = {10.1364/OE.524262}
}

@article{J2015Quantum,
	author  = {Liu, Jing and Jing, Xiao-Xing and Wang, Xiaoguang},
	title   = {Quantum metrology with unitary parametrization processes},
	journal = {Sci. Rep.},
	volume  = {5},
	pages   = {8565},
	year    = {2015},
	doi     = {10.1038/srep08565}
}

@article{humphreys2013quantum,
	author  = {Humphreys, Peter C. and Barbieri, Marco and Datta, Animesh and Walmsley, Ian A.},
	title   = {Quantum enhanced multiple phase estimation},
	journal = {Phys. Rev. Lett.},
	volume  = {111},
	pages   = {070403},
	year    = {2013},
	doi     = {10.1103/PhysRevLett.111.070403}
}

@article{Jie2014Quantum,
	author  = {Yue, Jie-Dong and Zhang, Yu-Ran and Fan, Heng},
	title   = {Quantum-enhanced metrology for multiple phase estimation with noise},
	journal = {Sci. Rep.},
	volume  = {4},
	pages   = {5933},
	year    = {2014},
	doi     = {10.1038/srep05933}
}

@article{K2014A,
	author  = {K{\'o}m{\'a}r, P. and Kessler, E. M. and Bishof, M. and Jiang, L. and S{\o}rensen, A. S. and Ye, J. and Lukin, M. D.},
	title   = {A quantum network of clocks},
	journal = {Nat. Phys.},
	volume  = {10},
	pages   = {582--587},
	year    = {2014},
	doi     = {10.1038/nphys3000}
}

@article{pezze2017optimal,
	author  = {Pezz{\`e}, Luca and Ciampini, Mario A. and Spagnolo, Nicol{\`o} and Humphreys, Peter C. and Datta, Animesh and Walmsley, Ian A. and Barbieri, Marco and Sciarrino, Fabio and Smerzi, Augusto},
	title   = {Optimal measurements for simultaneous quantum estimation of multiple phases},
	journal = {Phys. Rev. Lett.},
	volume  = {119},
	pages   = {130504},
	year    = {2017},
	doi     = {10.1103/PhysRevLett.119.130504}
}

@article{zhang2017quantum,
	author  = {Zhang, Lu and Chan, Kam Wai Clifford},
	title   = {Quantum multiparameter estimation with generalized balanced multimode {NOON}-like states},
	journal = {Phys. Rev. A},
	volume  = {95},
	pages   = {032321},
	year    = {2017},
	doi     = {10.1103/PhysRevA.95.032321}
}

@article{liu2016quantum,
	author  = {Liu, Jing and Lu, Xiao-Ming and Sun, Zhe and Wang, Xiaoguang},
	title   = {Quantum multiparameter metrology with generalized entangled coherent state},
	journal = {J. Phys. A: Math. Theor.},
	volume  = {49},
	pages   = {115302},
	year    = {2016},
	doi     = {10.1088/1751-8113/49/11/115302}
}

@article{hong2021quantum,
	author  = {Hong, Seongjin and ur Rehman, Junaid and Kim, Yong-Su},
	title   = {Quantum enhanced multiple-phase estimation with multi-mode {N00N} states},
	journal = {Nat. Commun.},
	volume  = {12},
	pages   = {5211},
	year    = {2021},
	doi     = {10.1038/s41467-021-25451-4}
}

@article{Gessner2020Multiparameter,
	title = {Multiparameter squeezing for optimal quantum enhancements in sensor networks},
	author = {Gessner, M. and Smerzi, A. and Pezzè, L.},
	journal = {Nat. Commun.},
	volume = {11},
	pages = {3817},
	year = {2020},
	doi = {https://doi.org/10.1038/s41467-020-17471-3},
}

@article{Marco2021Distributed,
	author  = {Malitesta, Marco and Smerzi, Augusto and Pezz{\`e}, Luca},
	title   = {Distributed quantum sensing with squeezed-vacuum light in a configurable array of {Mach-Zehnder} interferometers},
	journal = {Phys. Rev. A},
	volume  = {108},
	pages   = {032621},
	year    = {2023},
	doi     = {10.1103/PhysRevA.108.032621}
}

@article{Gagatsos2016Gaussian,
	title = {Gaussian systems for quantum-enhanced multiple phase estimation},
	author = {Gagatsos, Christos N. and Branford, Dominic and Datta, Animesh},
	journal = {Phys. Rev. A},
	volume = {94},
	issue = {4},
	pages = {042342},
	numpages = {10},
	year = {2016},
	month = {Oct},
	publisher = {American Physical Society},
	doi = {10.1103/PhysRevA.94.042342},
}

@Article{abbott2016observation,
	author    = {Abbott, Benjamin P and Abbott, Richard and Abbott, Thomas D and Abernathy, Matthew R and Acernese, Fausto and Ackley, Kendall and Adams, Carl and Adams, Thomas and Addesso, Paolo and Adhikari, Rana X and others},
	journal   = {Phys. Rev. Lett.},
	title     = {{Observation of gravitational waves from a binary black hole merger}},
	year      = {2016},
	number    = {6},
	pages     = {061102},
	volume    = {116},
	doi       = {10.1103/PhysRevLett.116.061102},
	publisher = {APS},
}

@article{vahlbruch2016detection,
	author={Vahlbruch, Henning and Mehmet, Moritz and Danzmann, Karsten and Schnabel, Roman},
	journal={Phys. Rev. Lett.},
	title={Detection of 15 dB squeezed states of light and their application for the absolute calibration of photoelectric quantum efficiency},
	year={2016},
	number={11},
	pages={110801},
	volume={117},
	doi       = {10.1103/PhysRevLett.117.110801},
	publisher={APS},
}

@article{escher2011general,
	author={Escher, BM and de Matos Filho, Ruynet Lima and Davidovich, Luiz},
	journal={Nat. Phys.},
	title={General framework for estimating the ultimate precision limit in noisy quantum-enhanced metrology},
	year={2011},
	number={5},
	pages={406--411},
	volume={7},
	doi       = {https://doi.org/10.1038/nphys1958},
	publisher={Nature Publishing Group UK London}
}

@article{Giovannetti2006quantum,
	title = {Quantum Metrology},
	author = {Giovannetti, Vittorio and Lloyd, Seth and Maccone, Lorenzo},
	journal = {Phys. Rev. Lett.},
	volume = {96},
	issue = {1},
	pages = {010401},
	numpages = {4},
	year = {2006},
	month = {Jan},
	publisher = {American Physical Society},
	doi = {10.1103/PhysRevLett.96.010401},
}

@article{taylor2013biological,
	author  = {Taylor, Michael A. and Janousek, Jiri and Daria, Vincent and Knittel, Joachim and Hage, Boris and Bachor, Hans-A. and Bowen, Warwick P.},
	title   = {Biological measurement beyond the quantum limit},
	journal = {Nat. Photonics},
	volume  = {7},
	pages   = {229--233},
	year    = {2013},
	doi     = {10.1038/nphoton.2012.346}
}

@article{lai2019observation,
	author  = {Lai, Yu-Hung and Lu, Yu-Kun and Suh, Myoung-Gyun and Yuan, Zhiquan and Vahala, Kerry},
	title   = {Observation of the exceptional-point-enhanced {Sagnac} effect},
	journal = {Nature},
	volume  = {576},
	pages   = {65--69},
	year    = {2019},
	doi     = {10.1038/s41586-019-1777-z}
}

@article{helstrom1969quantum,
	author  = {Helstrom, Carl W.},
	title   = {Quantum detection and estimation theory},
	journal = {J. Stat. Phys.},
	volume  = {1},
	pages   = {231--252},
	year    = {1969},
	doi     = {10.1007/BF01007479}
}

@article{gerry2001generation,
	author  = {Gerry, Christopher C. and Campos, R. A.},
	title   = {Generation of maximally entangled photonic states with a quantum-optical {Fredkin} gate},
	journal = {Phys. Rev. A},
	volume  = {64},
	pages   = {063814},
	year    = {2001},
	doi     = {10.1103/PhysRevA.64.063814}
}

@article{joo2011quantum,
	author  = {Joo, Jaewoo and Munro, William J. and Spiller, Timothy P.},
	title   = {Quantum metrology with entangled coherent states},
	journal = {Phys. Rev. Lett.},
	volume  = {107},
	pages   = {083601},
	year    = {2011},
	doi     = {10.1103/PhysRevLett.107.083601}
}

@article{tan2014enhanced,
	author  = {Tan, Qing-Shou and Liao, Jie-Qiao and Wang, Xiaoguang and Nori, Franco},
	title   = {Enhanced interferometry using squeezed thermal states and even or odd states},
	journal = {Phys. Rev. A},
	volume  = {89},
	pages   = {053822},
	year    = {2014},
	doi     = {10.1103/PhysRevA.89.053822}
}

@article{lee2015quantum,
	author  = {Lee, Su-Yong and Lee, Chang-Woo and Nha, Hyunchul and Kaszlikowski, Dagomir},
	title   = {Quantum phase estimation using a multi-headed cat state},
	journal = {J. Opt. Soc. Am. B},
	volume  = {32},
	pages   = {1186--1192},
	year    = {2015},
	doi     = {10.1364/JOSAB.32.001186}
}

@article{chen2024asymmetry,
	author  = {Chen, Xiao-Tong and Zhang, Rui and Lu, Wang-Jun and Zuo, Yunlan and Jiao, Ya-Feng and Kuang, Le-Man},
	title   = {Asymmetry-enhanced phase sensing via asymmetric entangled coherent states},
	journal = {Phys. Rev. A},
	volume  = {109},
	pages   = {042609},
	year    = {2024},
	doi     = {10.1103/PhysRevA.109.042609}
}

@article{zhang2025enhancing,
	author  = {Zhang, Wen-Xun and Zhang, Rui and Zuo, Yunlan and Kuang, Le-Man},
	title   = {Enhancing the sensitivity of quantum fiber-optical gyroscope via a non-{Gaussian}-state probe},
	journal = {Adv. Quantum Technol.},
	volume  = {8},
	pages   = {2400270},
	year    = {2025},
	doi     = {10.1002/qute.202400270}
}

@article{dorner2009optimal,
	author  = {Dorner, Uwe and Demkowicz-Dobrza{\'n}ski, Rafa{\l} and Smith, Brian J. and Lundeen, Jeff S. and Wasilewski, Wojciech and Banaszek, Konrad and Walmsley, Ian A.},
	title   = {Optimal quantum phase estimation},
	journal = {Phys. Rev. Lett.},
	volume  = {102},
	pages   = {040403},
	year    = {2009},
	doi     = {10.1103/PhysRevLett.102.040403}
}

@article{joo2012quantum,
	author  = {Joo, Jaewoo and Park, Kimin and Jeong, Hyunseok and Munro, William J. and Nemoto, Kae and Spiller, Timothy P.},
	title   = {Quantum metrology for nonlinear phase shifts with entangled coherent states},
	journal = {Phys. Rev. A},
	volume  = {86},
	pages   = {043828},
	year    = {2012},
	doi     = {10.1103/PhysRevA.86.043828}
}

@article{krischek2011useful,
	author  = {Krischek, Roland and Schwemmer, Christian and Wieczorek, Witlef and Weinfurter, Harald and Hyllus, Philipp and Pezz{\'e}, Luca and Smerzi, Augusto},
	title   = {Useful multiparticle entanglement and sub-shot-noise sensitivity in experimental phase estimation},
	journal = {Phys. Rev. Lett.},
	volume  = {107},
	pages   = {080504},
	year    = {2011},
	doi     = {10.1103/PhysRevLett.107.080504}
}

@article{xia2020demonstration,
	author  = {Xia, Yi and Li, Wei and Clark, William and Hart, Darlene and Zhuang, Quntao and Zhang, Zheshen},
	title   = {Demonstration of a reconfigurable entangled radio-frequency photonic sensor network},
	journal = {Phys. Rev. Lett.},
	volume  = {124},
	pages   = {150502},
	year    = {2020},
	doi     = {10.1103/PhysRevLett.124.150502}
}

@article{sun2022quantum,
	author  = {Sun, Xiaocong and Li, Wei and Tian, Yuhang and Li, Fan and Tian, Long and Wang, Yajun and Zheng, Yaohui},
	title   = {Quantum positioning and ranging via a distributed sensor network},
	journal = {Photonics Res.},
	volume  = {10},
	pages   = {2886--2892},
	year    = {2022},
	doi     = {10.1364/PRJ.469166}
}

@article{wang2025exact,
	author  = {Wang, Jiaxuan and Agarwal, Girish S.},
	title   = {Exact quantum Fisher matrix results for distributed phases using multiphoton polarization {Greenberger-Horne-Zeilinger} states},
	journal = {Phys. Rev. A},
	volume  = {111},
	pages   = {012414},
	year    = {2025},
	doi     = {10.1103/PhysRevA.111.012414}
}

@article{caves1981quantum,
	author  = {Caves, Carlton M.},
	title   = {Quantum-mechanical noise in an interferometer},
	journal = {Phys. Rev. D},
	volume  = {23},
	pages   = {1693},
	year    = {1981},
	doi     = {10.1103/PhysRevD.23.1693}
}

@article{wineland1992spin,
	author  = {Wineland, David J. and Bollinger, John J. and Itano, Wayne M. and Moore, F. L. and Heinzen, Daniel J.},
	title   = {Spin squeezing and reduced quantum noise in spectroscopy},
	journal = {Phys. Rev. A},
	volume  = {46},
	pages   = {R6797},
	year    = {1992},
	doi     = {10.1103/PhysRevA.46.R6797}
}

@article{pezze2018quantum,
	author  = {Pezz{\`e}, Luca and Smerzi, Augusto and Oberthaler, Markus K. and Schmied, Roman and Treutlein, Philipp},
	title   = {Quantum metrology with nonclassical states of atomic ensembles},
	journal = {Rev. Mod. Phys.},
	volume  = {90},
	pages   = {035005},
	year    = {2018},
	doi     = {10.1103/RevModPhys.90.035005}
}

@article{toth2014quantum,
	author  = {T{\'o}th, G{\'e}za and Apellaniz, Iagoba},
	title   = {Quantum metrology from a quantum information science perspective},
	journal = {J. Phys. A: Math. Theor.},
	volume  = {47},
	pages   = {424006},
	year    = {2014},
	doi     = {10.1088/1751-8113/47/42/424006}
}

@article{ma2011quantum,
	author  = {Ma, Jian and Wang, Xiaoguang and Sun, Chang-Pu and Nori, Franco},
	title   = {Quantum spin squeezing},
	journal = {Phys. Rep.},
	volume  = {509},
	pages   = {89--165},
	year    = {2011},
	doi     = {10.1016/j.physrep.2011.08.003}
}

@article{zhuang2018distributed,
	author  = {Zhuang, Quntao and Zhang, Zheshen and Shapiro, Jeffrey H.},
	title   = {Distributed quantum sensing using continuous-variable multipartite entanglement},
	journal = {Phys. Rev. A},
	volume  = {97},
	pages   = {032329},
	year    = {2018},
	doi     = {10.1103/PhysRevA.97.032329}
}

@article{guo2020distributed,
	author  = {Guo, Xueshi and Breum, Casper R. and Borregaard, Johannes and Izumi, Shuro and Larsen, Mikkel V. and Gehring, Tobias and Christandl, Matthias and Neergaard-Nielsen, Jonas S. and Andersen, Ulrik L.},
	title   = {Distributed quantum sensing in a continuous-variable entangled network},
	journal = {Nat. Phys.},
	volume  = {16},
	pages   = {281--284},
	year    = {2020},
	doi     = {10.1038/s41567-019-0743-x}
}

@article{oh2020optimal,
	author  = {Oh, Changhun and Lee, Changhyoup and Lie, Seok Hyung and Jeong, Hyunseok},
	title   = {Optimal distributed quantum sensing using {Gaussian} states},
	journal = {Phys. Rev. Research},
	volume  = {2},
	pages   = {023030},
	year    = {2020},
	doi     = {10.1103/PhysRevResearch.2.023030}
}

@article{triggiani2021heisenberg,
	author  = {Triggiani, Danilo and Facchi, Paolo and Tamma, Vincenzo},
	title   = {Heisenberg scaling precision in the estimation of functions of parameters in linear optical networks},
	journal = {Phys. Rev. A},
	volume  = {104},
	pages   = {062603},
	year    = {2021},
	doi     = {10.1103/PhysRevA.104.062603}
}

@article{malitesta2023distributed,
	author  = {Malitesta, Marco and Smerzi, Augusto and Pezz{\`e}, Luca},
	title   = {Distributed quantum sensing with squeezed-vacuum light in a configurable array of {Mach-Zehnder} interferometers},
	journal = {Phys. Rev. A},
	volume  = {108},
	pages   = {032621},
	year    = {2023},
	doi     = {10.1103/PhysRevA.108.032621}
}

@article{liu2020quantum,
	author  = {Liu, Jing and Yuan, Haidong and Lu, Xiao-Ming and Wang, Xiaoguang},
	title   = {Quantum Fisher information matrix and multiparameter estimation},
	journal = {J. Phys. A: Math. Theor.},
	volume  = {53},
	pages   = {023001},
	year    = {2020},
	doi     = {10.1088/1751-8121/ab5d4d}
}

@article{albarelli2020perspective,
	author  = {Albarelli, Francesco and Barbieri, Marco and Genoni, Marco G. and Gianani, Ilaria},
	title   = {A perspective on multiparameter quantum metrology: From theoretical tools to applications in quantum imaging},
	journal = {Phys. Lett. A},
	volume  = {384},
	pages   = {126311},
	year    = {2020},
	doi     = {10.1016/j.physleta.2020.126311}
}

@article{fujiwara1995quantum,
	author  = {Fujiwara, Akio and Nagaoka, Hiroshi},
	title   = {Quantum Fisher metric and estimation for pure state models},
	journal = {Phys. Lett. A},
	volume  = {201},
	pages   = {119--124},
	year    = {1995},
	doi     = {10.1016/0375-9601(95)00279-N}
}

@article{petz1996geometries,
	author  = {Petz, D{\'e}nes and Sud{\'a}r, Csaba},
	title   = {Geometries of quantum states},
	journal = {J. Math. Phys.},
	volume  = {37},
	pages   = {2662--2673},
	year    = {1996},
	doi     = {10.1063/1.531535}
}

@article{petz1996monotone,
	author  = {Petz, D{\'e}nes},
	title   = {Monotone metrics on matrix spaces},
	journal = {Linear Algebra Appl.},
	volume  = {244},
	pages   = {81--96},
	year    = {1996},
	doi     = {10.1016/0024-3795(94)00211-8}
}

@article{matsumoto2002new,
	author  = {Matsumoto, Keiji},
	title   = {A new approach to the {Cram{\'e}r-Rao}-type bound of the pure-state model},
	journal = {J. Phys. A: Math. Gen.},
	volume  = {35},
	pages   = {3111},
	year    = {2002},
	doi     = {10.1088/0305-4470/35/13/307}
}

@article{ragy2016compatibility,
	author  = {Ragy, Sammy and Jarzyna, Marcin and Demkowicz-Dobrza{\'n}ski, Rafa{\l}},
	title   = {Compatibility in multiparameter quantum metrology},
	journal = {Phys. Rev. A},
	volume  = {94},
	pages   = {052108},
	year    = {2016},
	doi     = {10.1103/PhysRevA.94.052108}
}

@article{liu2021distributed,
	author  = {Liu, Li-Zheng and Zhang, Yu-Zhe and Li, Zheng-Da and Zhang, Rui and Yin, Xu-Fei and Fei, Yue-Yang and Li, Li and Liu, Nai-Le and Xu, Feihu and Chen, Yu-Ao and Pan, Jian-Wei},
	title   = {Distributed quantum phase estimation with entangled photons},
	journal = {Nat. Photonics},
	volume  = {15},
	pages   = {137--142},
	year    = {2021},
	doi     = {10.1038/s41566-020-00718-2}
}

@article{zhao2021field,
	author  = {Zhao, Si-Ran and Zhang, Yu-Zhe and Liu, Wen-Zhao and Guan, Jian-Yu and Zhang, Weijun and Li, Cheng-Long and Bai, Bing and Li, Ming-Han and Liu, Yang and You, Lixing and others},
	title   = {Field demonstration of distributed quantum sensing without post-selection},
	journal = {Phys. Rev. X},
	volume  = {11},
	pages   = {031009},
	year    = {2021},
	doi     = {10.1103/PhysRevX.11.031009}
}

@article{kim2024distributed,
	author  = {Kim, Dong-Hyun and Hong, Seongjin and Kim, Yong-Su and Kim, Yosep and Lee, Seung-Woo and Pooser, Raphael C. and Oh, Kyunghwan and Lee, Su-Yong and Lee, Changhyoup and Lim, Hyang-Tag},
	title   = {Distributed quantum sensing of multiple phases with fewer photons},
	journal = {Nat. Commun.},
	volume  = {15},
	pages   = {266},
	year    = {2024},
	doi     = {10.1038/s41467-023-44204-z}
}

@article{nagata2007beating,
	author  = {Nagata, Tomohisa and Okamoto, Ryo and O'Brien, Jeremy L. and Sasaki, Keiji and Takeuchi, Shigeki},
	title   = {Beating the standard quantum limit with four-entangled photons},
	journal = {Science},
	volume  = {316},
	pages   = {726--729},
	year    = {2007},
	doi     = {10.1126/science.1138007}
}

@article{xiang2011entanglement,
	author  = {Xiang, Guo-Yong and Higgins, Brendon L. and Berry, D. W. and Wiseman, Howard M. and Pryde, G. J.},
	title   = {Entanglement-enhanced measurement of a completely unknown optical phase},
	journal = {Nat. Photonics},
	volume  = {5},
	pages   = {43--47},
	year    = {2011},
	doi     = {10.1038/nphoton.2010.282}
}

@article{dowling2008quantum,
	author  = {Dowling, Jonathan P.},
	title   = {Quantum optical metrology---the lowdown on high-{N00N} states},
	journal = {Contemp. Phys.},
	volume  = {49},
	pages   = {125--143},
	year    = {2008},
	doi     = {10.1080/00107510802091298}
}

@article{sanders2012review,
	author  = {Sanders, Barry C.},
	title   = {Review of entangled coherent states},
	journal = {J. Phys. A: Math. Theor.},
	volume  = {45},
	pages   = {244002},
	year    = {2012},
	doi     = {10.1088/1751-8113/45/24/244002}
}

@Article{ozdemir2019Parity,
	author    = {{\"O}zdemir, {\c S}. K. and Rotter, S. and Nori, F. and Yang, L.},
	journal   = {Nat. Mater.},
	title     = {Parity\textendash Time Symmetry and Exceptional Points in Photonics},
	year      = {2019},
	pages     = {783--798},
	volume    = {18},
	doi       = {10.1038/s41563-019-0304-9},
}

@article{schnabel2010quantum,
	author  = {Schnabel, Roman and Mavalvala, Nergis and McClelland, David E. and Lam, Ping K.},
	title   = {Quantum metrology for gravitational wave astronomy},
	journal = {Nat. Commun.},
	volume  = {1},
	pages   = {121},
	year    = {2010},
	doi     = {10.1038/ncomms1122}
}

@article{fisher1925theory,
	author  = {Fisher, R. A.},
	title   = {Theory of statistical estimation},
	journal = {Math. Proc. Cambridge Philos. Soc.},
	volume  = {22},
	pages   = {700--725},
	year    = {1925},
	doi     = {10.1017/S0305004100009580}
}

@book{holevo2011probabilistic,
	author    = {Holevo, Alexander S.},
	title     = {Probabilistic and Statistical Aspects of Quantum Theory},
	series    = {Quaderni. Monographs},
	volume    = {1},
	publisher = {Springer},
	address   = {Pisa},
	year      = {2011}
}

@article{braunstein1996generalized,
	author  = {Braunstein, Samuel L. and Caves, Carlton M. and Milburn, Gerard J.},
	title   = {Generalized uncertainty relations: Theory, examples, and {Lorentz} invariance},
	journal = {Ann. Phys.},
	volume  = {247},
	pages   = {135--173},
	year    = {1996},
	doi     = {10.1006/aphy.1996.0040}
}

@book{cramer1999mathematical,
	author    = {Cram{\'e}r, Harald},
	title     = {Mathematical Methods of Statistics},
	series    = {Princeton Mathematical Series},
	volume    = {9},
	publisher = {Princeton University Press},
	address   = {Princeton, NJ},
	year      = {1999}
}

@article{helstrom1967minimum,
	author  = {Helstrom, Carl W.},
	title   = {Minimum mean-squared error of estimates in quantum statistics},
	journal = {Phys. Lett. A},
	volume  = {25},
	pages   = {101--102},
	year    = {1967},
	doi     = {10.1016/0375-9601(67)90366-0}
}

@article{qian2019heisenberg,
	author  = {Qian, Kevin and Eldredge, Zachary and Ge, Wenchao and Pagano, Guido and Monroe, Christopher and Porto, James V. and Gorshkov, Alexey V.},
	title   = {Heisenberg-scaling measurement protocol for analytic functions with quantum sensor networks},
	journal = {Phys. Rev. A},
	volume  = {100},
	pages   = {042304},
	year    = {2019},
	doi     = {10.1103/PhysRevA.100.042304}
}

@article{rubio2020quantum,
	author  = {Rubio, Jes{\'u}s and Knott, Paul A. and Proctor, Timothy J. and Dunningham, Jacob A.},
	title   = {Quantum sensing networks for the estimation of linear functions},
	journal = {J. Phys. A: Math. Theor.},
	volume  = {53},
	pages   = {344001},
	year    = {2020},
	doi     = {10.1088/1751-8121/ab9d46}
}

@article{bringewatt2021protocols,
	author  = {Bringewatt, Jacob and Boettcher, Igor and Niroula, Pradeep and Bienias, Przemyslaw and Gorshkov, Alexey V.},
	title   = {Protocols for estimating multiple functions with quantum sensor networks: Geometry and performance},
	journal = {Phys. Rev. Research},
	volume  = {3},
	pages   = {033011},
	year    = {2021},
	doi     = {10.1103/PhysRevResearch.3.033011}
}

@article{ehrenberg2023minimum,
	author  = {Ehrenberg, Adam and Bringewatt, Jacob and Gorshkov, Alexey V.},
	title   = {Minimum-entanglement protocols for function estimation},
	journal = {Phys. Rev. Research},
	volume  = {5},
	pages   = {033228},
	year    = {2023},
	doi     = {10.1103/PhysRevResearch.5.033228}
}

@article{ge2025heisenberg,
	author  = {Ge, Wenchao and Jacobs, Kurt},
	title   = {Heisenberg-limited continuous-variable distributed quantum metrology with arbitrary weights},
	journal = {Phys. Rev. Lett.},
	volume  = {135},
	pages   = {100801},
	year    = {2025},
	doi     = {10.1103/jkjj-3gvb}
}

@article{vourdas2005fourier,
	author  = {Vourdas, A. and Dunningham, J. A.},
	title   = {Fourier multiport devices},
	journal = {Phys. Rev. A},
	volume  = {71},
	pages   = {013809},
	year    = {2005},
	doi     = {10.1103/PhysRevA.71.013809}
}

@article{ma2026high,
	author  = {Ma, Jingxu and Yan, Jieli and Liang, Shaocong and Yan, Zhihui and Jia, Xiaojun and Xie, Changde and Peng, Kunchi},
	title   = {High-sensitivity distributed quantum sensing of multimodal parameters},
	journal = {Optica},
	volume  = {13},
	pages   = {541--547},
	year    = {2026},
	doi     = {10.1364/OPTICA.581723}
}

@article{lipka2024catalysis,
	author  = {Lipka-Bartosik, Patryk and Wilming, Henrik and Ng, Nelly H. Y.},
	title   = {Catalysis in quantum information theory},
	journal = {Rev. Mod. Phys.},
	volume  = {96},
	pages   = {025005},
	year    = {2024},
	doi     = {10.1103/RevModPhys.96.025005}
}

@article{datta2023catalysis,
	author  = {Datta, Chandan and Kondra, Tulja Varun and Miller, Marek and Streltsov, Alexander},
	title   = {Catalysis of entanglement and other quantum resources},
	journal = {Rep. Prog. Phys.},
	volume  = {86},
	pages   = {116002},
	year    = {2023},
	doi     = {10.1088/1361-6633/acfbec}
}

@article{zhang2023quantum,
	author  = {Zhang, Huan and Ye, Wei and Chang, Shoukang and Xia, Ying and Hu, Liyun and Liao, Zeyang},
	title   = {Quantum multiparameter estimation with multi-mode photon catalysis entangled squeezed state},
	journal = {Front. Phys.},
	volume  = {18},
	pages   = {42304},
	year    = {2023},
	doi     = {10.1007/s11467-023-1274-6}
}
\end{document}